\documentclass{pasa}

\usepackage[utf8]{inputenc}
\usepackage{graphicx}
\usepackage{float}
\usepackage{subcaption}
\usepackage{physics}
\usepackage[separate-uncertainty = true,multi-part-units=single]{siunitx}
\usepackage{hyperref}
\usepackage{booktabs}
\usepackage{multirow}
\usepackage{aas_macros}
\usepackage{natbib}
\usepackage{hypernat}
\usepackage{textcomp}
\usepackage{nicefrac}

\DeclareSIUnit\jansky{Jy}
\DeclareSIUnit\beam{beam}
\DeclareSIUnit\parsec{pc}

\newcommand*{\ra}[2][]{{
    \def\SIUnitSymbolDegree{\textsuperscript{h}}%
    \def\SIUnitSymbolArcminute{\textsuperscript{m}}%
    \def\SIUnitSymbolArcsecond{\textsuperscript{s}}%
    \ang[#1]{#2}}
}

\floatstyle{plaintop}
\restylefloat{table}

\hypersetup{colorlinks,citecolor=blue,linkcolor=blue,urlcolor=blue}

\def\clean{CLEAN} 

\title{Low(er) frequency follow-up of 28 candidate, large-scale synchrotron sources}
\author[Hodgson et al.]{Torrance Hodgson$^1$\thanks{torrance@pravic.xyz}, Melanie Johnston-Hollitt$^1$, Benjamin McKinley$^{1,2}$, Tessa Vernstrom$^3$, Valentina Vacca$^4$
\affil{$^1$International Centre for Radio Astronomy Research (ICRAR), Curtin University, 1 Turner Ave, Bentley, 6102, WA, Australia}%
\affil{$^2$ARC Centre of Excellence for All Sky Astrophysics in 3 Dimensions (ASTRO3D), Bentley, Australia}%
\affil{$^3$CSIRO Astronomy and Space Science, PO Box 1130, Bentley WA 6102, Australia}
\affil{$^4$INAF - Osservatorio Astronomico di Cagliari, Via della Scienza 5, I-09047 Selargius (CA), Italy}
}

\jid{PASA}
\doi{10.1017/pas.\the\year.xxx}
\jyear{\the\year}

\begin{document}

\begin{frontmatter}
\maketitle

\begin{abstract}
We follow up on a report by \citet{Vacca2018} of 28 candidate large-scale diffuse synchrotron sources in an \SI{8 x 8}{\degree} area of the sky (centred at RA \ra{5;00;00} Dec \ang{+5;48;00}). These sources were originally observed at 1.4 GHz using a combination of the single-dish Sardinia Radio Telescope (SRT) and archival NRAO VLA Sky Survey (NVSS) data. They are in an area with nine massive galaxy clusters at $z \approx 0.1$, and are candidates for the first detection of filaments of the synchrotron cosmic web. We attempt to verify these candidate sources with lower frequency observations at 154 MHz with the Murchison Widefield Array (MWA) and at 887 MHz with the Australian Square Kilometre Array Pathfinder (ASKAP). We use a novel technique to calculate the surface brightness sensitivity of these instruments to show that our lower frequency observations, and in particular those by ASKAP, are ideally suited to detect large-scale, extended synchrotron emission. Nonetheless, we are forced to conclude that none of these sources are likely to be synchrotron in origin or associated with the cosmic web.
\end{abstract}

\end{frontmatter}

\section{Introduction}

Up to half of the baryons in the present-day Universe are unaccounted for. We know how many baryons were present in the early Universe from fluctuations in the cosmic microwave background (CMB), and some 2 billion years later at redshift 3, the majority of the baryon budget of the Universe could be found in galaxies, proto-clusters and, mostly, in the Lyman-$\alpha$ forest. In the present-day Universe, however, if we take stock of the known baryon populations we come up short, and this has given rise to the `missing baryon problem' (e.g\@. see \citealp{Nicastro2017} for review). Cosmological simulations have long pointed to the likely explanation that these Baryons reside in a warm-hot intercluster medium (WHIM) that is distributed in a large-scale filamentary network, the so-called `cosmic web' (e.g\@. \citealp{Cen1999}). However, due to its extremely diffuse nature, intermediate temperature range ($10^5 - 10^7$ K), and highly ionised state, it is very difficult to detect. The low density of this medium and intermediate temperature result in only very weak X-ray emission via thermal free-free radiation; the highly ionised state makes detection via absorption/emission lines difficult; and the low mass, low density environment makes detection via the Sunyaev-Zel'dovich effect problematic (with the exception of bridges connecting close pairs of galaxy clusters).

Nonetheless, there have been early attempts to detect the cosmic web by way of some of these mechanisms. For example, \citet{Eckert2015} measured residual X-ray emission as large as \SI{8}{\mega \parsec} in scale around galaxy cluster Abell 2744, implying this existence of a large scale, energetic baryon population. \citet{Nicastro2018} claimed that Oxygen \textsc{vii} absorption features in a distant quaser pointed to the detection of an intervening overdense baryonic region. \citet{Tanimura2019} and \citet{deGraaff2019} have both independently claimed to have made statistical detections of the intercluster medium by way of the Sunyaev-Zel'dovich effect. Most recently, \citet{Macquart2020} have used the dispersion measure of a small number of localised fast radio bursts (FRBs) to measure the electron column density along the line of sight to these events, and have measured a value for the baryon count of the Universe that is consistent with those derived from CMB measurements.

Recently, there has been work to understand the radio emission properties of the cosmic web. Infall accretion shocks along the length of filaments and at the edge of clusters should have high Mach numbers ($\mathcal{M} \approx$ 10-100). These in turn are capable of producing relativistic electrons and---given the presence of background magnetic fields---associated synchrotron emission (e.g\@. \citealp{Wilcots2004}). Such emission would provide not only confirmation of the cosmic web but would also provide a probe into inter-cluster magnetic field strengths, which up till now are largely unknown. Early detection attempts such as \citet{Brown2017} and \citet{Vernstrom2017} have assumed synchrotron cosmic web emission to be spatially smooth and characteristically large in angular scale, in an effort to distinguish it from the more general extra-galactic synchrotron emission produced by radio galaxies. In \citet{Vernstrom2017}, for example, low frequency radio images were cross-correlated with galaxy density maps (as tracers of large scale structure), with the expectation that the synchrotron cosmic web would appear as excess radio emission with angular scales larger than the embedded radio galaxy population. More recent work utilising full magneto-hydrodynamic simulations has attempted to directly model the filamentary accretion shocks and from this derive values for their radio luminosity \citep{Vazza2015,Vazza2019}. As is typical of synchrotron shocked emission, these simulations suggest radio emission with steep spectral indices of approximately -1 to -1.5, as well as peak radio surface brightnesses on the order of \SI{E-6}{\jansky~ {arcsecond} \squared}. Such simulations, however, depend on assumptions about filamentary magnetic field strengths and electron acceleration efficiencies, which are poorly constrained or understood.

To date, these attempts at detecting the synchrotron cosmic web have been unsuccessful with two exceptions. In the first, a small `bridge' between two interacting clusters Abell 399 and 401 was recently reported to have been detected by \citet{Govoni2019}, however this emission is primarily the result of a pre-merger cluster-cluster interaction rather than the more general infall accretion shocks we expect to find in the cosmic web. The second, by \citealt{Vacca2018} (henceforth: VA18), is the focus of this current follow-up study.

VA18 reported the detection of 28 candidate, large-scale synchrotron radio sources using the single dish Sardinia Radio Telescope (SRT; \citealp{SRT}) and archival interferometric NRAO VLA\footnote{National Radio Astronomy Observatory Very Large Array} Sky Survey (NVSS; \citealp{Condon1998}) data observed at 1.4 GHz. These sources were observed in an \ang{8} $\times$ \ang{8} region of sky centred at RA \ra{5;0;0} and Dec \ang{5;48;0}. This region of sky contains 43 galaxy clusters, thirteen of which have spectroscopic redshifts, with nine being in the redshift range $0.08 \leq z \leq 0.15$ (see Tables 1 \& 2 in VA18 for full list). Additionally, some of these clusters have been identified as members of superclusters: \citet{Einasto2002} have catalogued superclusters SCL 061 and SCL 062, and \citet{ChowMartinez2014} have catalogued MSCC 145 which partially overlaps with SCL 062. However, VA18 exclude the possibility that these sources are associated with galaxy cluster cores due to the lack of associated X-ray emission typical of dense cluster environments; indeed, the sources populate a previously empty region of the X-ray luminosity / radio power space ($L_\text{X,0.1-2.4keV}-P_\text{1.4GHz}$). Instead, they have raised the possibility that these new-found synchrotron sources are in fact a detection of radio emission from the intercluster medium, that is, the synchrotron cosmic web.

Given the potential significance of these candidate sources and the new population of synchrotron sources they may represent, we here report on lower frequency observations using the Murchison Widefield Array (MWA; \citealp{Tingay2013, Wayth2018}) at 154 MHz and the Australian Square Kilometre Array Pathfinder (ASKAP; \citealp{Hotan2014}) at 887 MHz to verify the candidate sources and measure their spectral properties.

These lower frequencies are ideal for detecting synchrotron emission. The spectral energy distribution (SED) of synchrotron sources can usually be well approximated by a power law, where the spectral flux density $S$ is a function of frequency $\nu$ of the form:
\begin{equation}
    S\left(\nu\right) \propto \nu^\alpha
\end{equation}
The coefficient $\alpha$ is known as the spectral index. For astronomical synchrotron sources, this coefficient depends, amongst other things, on the electron injection power coupled with the aging dynamics of the electron population. Active radio galaxies (AGN) tend to have a shallower SED at around $\alpha \approx -0.7$, whilst as populations of relativistic electrons age, for example in AGN remnants, their SED tends to steepen. Synchrotron shocks tracing the cosmic web should have spectral indices of at least -0.7, and most likely -1 or steeper \citep{Vazza2015}. This typically negative spectral index ensures that synchrotron sources are brightest at lower radio frequencies. Thus, these lower frequency observations take advantage of the expected brighter emission to corroborate the detections in VA18, and additionally provide us with spectral information that can allow us to infer the emission mechanisms of any confirmed candidate sources.

This paper proceeds as follows: in Section 2 we briefly review the observations and data of VA18, before in Section 3 detailing our own observations with both MWA and ASKAP, which also includes our point source subtraction method. We measure our surface brightness sensitivity in Section 4, and in Section 5 we present the results of our observations. Finally, in Section 6 we discuss at length all potential corroborating detections as well as drawing from other extant surveys to help classify these emission sources.

\section{SRT+NVSS data}

VA18 fully document their observations and data processing methods, which we briefly summarize here. The SRT data consisted of 18 hours of observing in the L-band (1.3-1.8 GHz) using the `on-the-fly' mapping strategy, as well as some additional time on specific sub fields. The SRT has a beam size of \SI{13.9 x 12.4}{\arcminute} at 1550 MHz and the resulting images had a noise of \SI{20}{\milli \jansky \per \beam}. In addition to this low-resolution, single-dish data, VA18 also obtained archival NVSS observations of the field that were in two bands centred at 1.4~GHz, and which had a resolution of \SI{45}{\arcsecond} and an average noise of \SI{0.45}{\milli \jansky \per \beam}. The data sets were combined using SCUBE \citep{Murgia2016}, which performs a weighted sum in Fourier space of the power spectra of the single dish and the NVSS data after correcting for any misalignment of overall power on overlapping angular scales. To perform the combination, an SRT image was produced over the same frequency range as the NVSS image. The combined power spectrum was tapered with the NVSS beam and the data back-transformed to obtain the combined image. The resulting combined image was finally convolved to a resolution of \SI{3.5 x 3.5}{\arcminute} to accentuate large-scale emission, producing the `SRT+NVSS' combined map with a noise of \SI{3.7}{\milli \jansky \per \beam}.
    
To differentiate between compact emission and the presumed large-scale emission of the cosmic web, VA18 subtracted point sources from the `SRT+NVSS' map using an image-plane subtraction process. This is described in full in their paper but briefly: the brightest point source in the map was identified, fit with a 2D elliptic Gaussian sitting on top of an arbitrarily oriented plane (to account for background emission) and subtracted. The process was repeated by then subtracting the next brightest source, and so on, until a user-defined threshold was reached. This image subtraction process was performed on the SRT+NVSS map prior to convolving the image from its native \SI{45}{\arcsecond} resolution. The final `SRT+NVSS-diffuse' map, at \SI{3.5 x 3.5}{\arcminute} resolution, has a noise of \SI{3.1}{\milli \jansky \per \beam}.\footnote{Note that this is different to the value of \SI{2.5}{\milli \jansky \per \beam} given in VA18, and was calculated independently on the supplied final image. We also note that the overall mean of the image is offset from zero by \SI{-2.1}{\milli \jansky \per \beam}. When calculating detection contours, we offset multiples of our noise value by this global mean. This independent process has resulted in a small difference between the SRT+NVSS-diffuse contours published here and in VA18.}

The choice to complement existing NVSS data with the deep, single dish SRT observation arises from the assumption that nearby cosmic web emission will be large-scale, smoothly varying, and highly diffuse. Typical interferometers like the VLA lack very short and `zero spacing' baselines, and as a result are likely to be increasingly insensitive to, and eventually `resolve out', emission on these large angular scales. Single dish observations like the SRT are sensitive to these large angular scale features but typically have such low resolution that unrelated compact radio sources are blended together. In combining both together, VA18 make use of the strengths of each to get higher resolution data with excellent sensitivity to diffuse, large scale emission.

Finally, all candidate sources were identified from the SRT+NVSS-diffuse image using a threshold three times greater than the calculated map noise ($3 \sigma$). The resulting 35 sources were grouped into ten regions, labelled A through to J. Of these 35, VA18 classify five as likely to be the result of imperfect compact source subtraction, and two as known cluster halos, leaving 28 sources as candidates for large-scale, diffuse synchrotron emission.

\section{Radio observations and data processing}
In order to independently further investigate the results from VA18, these fields were observed with the MWA and ASKAP.

\begin{figure*}
    \centering
    \includegraphics[width=\linewidth]{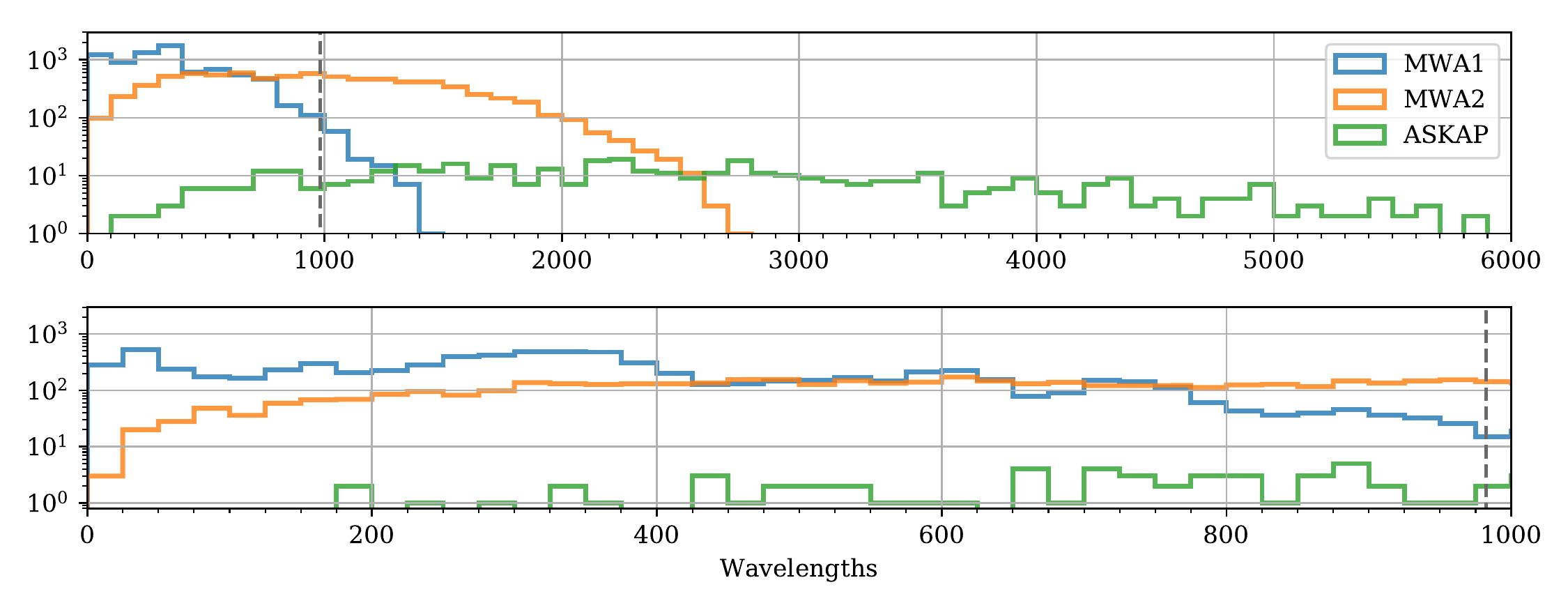}
    \caption{A comparison of baseline lengths for each of MWA Phase I (MWA1), MWA Phase 2 extended configuration (MWA2) and ASKAP. The lengths are measured in wavelengths (i.e. $\nicefrac{|b|}{\lambda}$, with $\lambda \approx 1.94$ m for the MWA and $\lambda \approx 0.34$ m for ASKAP), which allows us to compare the baseline coverage despite the different observing frequencies. All plots exclude baselines that were flagged. The dashed line indicates a baseline length that would result in a fringe pattern on the sky with angular periodicity of \SI{3.5}{\arcminute}; baselines shorter than this are sensitive to even larger spatial scales. \textit{Top:} The baselines distribution out to 6000 wavelengths, binned in intervals of 100. \textit{Bottom:} A zoom of the baselines under 1000 wavelengths, binned in intervals of 25.}
    \label{fig:baselines}
\end{figure*}

\begin{table*}
    \centering
    \begin{tabular}{lcccccc} \toprule
        Image Name & Instrument & Duration & Frequency & Briggs Weighting & Resolution & Noise \\
        & & [hours] & [\SI{}{\mega \hertz}] & & [arcsecond$^2$] & [\SI{}{\jansky \per \beam}] \\ \midrule
        MWA-1 & MWA & 2.3 & 154 & 0 & \SI{210 x 210}{} & \SI{8.4E-3}{} \\
        MWA-2 & MWA & 6.5  & 154 & 0 & \SI{79 x 62}{} & \SI{2.3E-3}{} \\
        MWA-subtracted & MWA & 6.5  & 154 & 0 & \SI{210 x 210}{} & \SI{5.4E-3}{} \\
        ASKAP-B+0.5 & ASKAP & 13  & 887 & 0.5 & \SI{21 x 17}{} & \SI{4.3E-5}{} \\
        ASKAP-B-1 & ASKAP & 13  & 887 & -1 & \SI{9.6 x 7.6}{} & \SI{5.8E-5}{} \\
        ASKAP-subtracted & ASKAP & 13  & 887 & 0.5 & \SI{210 x 210}{} & \SI{7.5E-4}{} \\
        SRT-NVSS diffuse & SRT \& VLA & 18 (SRT) & 1400 & - & \SI{210 x 210}{} & \SI{3.1E-3}{} \\ \bottomrule
    \end{tabular}
    \caption{List of images used in this work. Resolution and noise values are given for the centre of the field. Resolution values describe the major and minor axes of an elliptical Gaussian fitted to the synthesised beam. The bandwidth of all MWA images is \SI{30.72}{\mega \hertz} and the bandwidth of all ASKAP images is \SI{288}{\mega \hertz}.}
    \label{table:images}
\end{table*}

\subsection{Murchison Widefield Array}

The MWA data consists of two distinct datasets that were collected during different configurations of the array, known as `Phase I' and `Phase II', described in detail in \citet{Tingay2013} and \citet{Wayth2018}, respectively. While both configurations consisted of 128 tiles and had identical point source sensitivity, the tiles were arranged differently resulting in a different set of baselines (see \autoref{fig:baselines}). Phase I had a maximum baseline length of about \SI{2.6}{\kilo \metre} as well as a large number of short baselines, many under \SI{100}{\metre}. These short baselines gave Phase I excellent surface brightness sensitivity at the expense of poor resolution, which at \SI{154}{\mega \hertz} could be several arcminutes depending on the exact baseline weighting scheme used. Phase I is excellent at detecting faint, extended emission, however the poor resolution often necessitates additional, high resolution observations to discern whether such emission is truly extended or merely the result of blending of nearby sources (e.g.\ \cite{Hindson14,Zheng18}) . Phase II (extended configuration), on the other hand, redistributed the 128 tiles out to a maximum baseline of about \SI{5.4}{\kilo \metre} and a more sparse sampling of baselines under 500 m. Phase II has higher resolution at about \SI{65}{\arcsecond} at \SI{154}{\mega \hertz} and a better behaved synthesised beam (point spread function), but less sensitivity to diffuse emission. In this follow up, we make use of observations using both Phase I and II configurations so as to leverage their respective strengths.

The Phase I configuration data are archival observations that were collected at various times from 2013-2016 and consist of just over 2 hours of observations. The Phase II observations consist of 6 hours of observations at 154 MHz from March 2019, plus an additional 30 minutes of archival observations from the first quarter of 2018. The latter 30 minutes were observed at high elevations at which the MWA is most sensitive, so contribute a disproportionate amount of signal to the final integration. All MWA observations were made at a central frequency of \SI{154}{\mega \hertz} with a \SI{30.72}{\mega \hertz} bandwidth. The data were originally collected with a \SI{10}{\kilo \hertz} and \SI{0.5}{\second} resolution, and were averaged down to \SI{40}{\kilo \hertz} and \SI{4}{\second} prior to calibration and processing.

MWA calibration and imaging workflows operate independently on short `snapshot' observations that are typically about 2 minutes in length; this workflow is necessitated due to the complicated MWA primary beam and the stationary, non-tracking array. Snapshots are short enough in duration that we can assume a constant primary beam model and the MWA, with its more than 8,000 baselines, sufficiently samples the Fourier plane ($uv$ space) such that it is possible to image and deconvolve on time scales as short as two minutes. The downside of such a workflow is that final mosaics are only \clean{}ed down to the noise level of a single snapshot, making in-field sidelobe confusion the typically dominant source of noise, as well as prohibiting jointly imaging Phase I and Phase II observations together.

For this follow up, each snapshot was independently calibrated with an `in-field' sky model using the GLEAM extra-galactic catalogue \citep{HurleyWalker2016} and the internal MWA tool \textsc{calibrate} \citep{Offringa2016} which calculated full Jones matrix corrections across the band in 120 kHz steps. Additionally, we flagged baselines shorter than 15 wavelengths at the observing frequency, as these baselines tended to pick up significant amounts of nearby Galactic emission on scales larger than several degrees. 

After the initial sky-model calibration, snapshots were imaged using \textsc{wsclean} \citep{Offringa2014} with a shallow \clean{} and self-calibrated using the \clean{}-component model. A final snaphsot image was then produced using a Briggs 0 weighting of the baselines with a $3 \sigma$ mask and $1 \sigma$ threshold. \clean{}ing was configured to use the \textsc{wsclean} multiscale algorithm with default settings as well as joined-channel \clean{}ing with four channels and two Taylor terms (see \citealp{Offringa2017} for a description of the implementation of these algorithms). The final image was primary beam corrected and crossed-matched with the GLEAM catalogue to correct for flux. Finally, the full set of snapshots were convolved to a common beam size (using the maximum beam size of any single snapshot), regridded onto a common projection and stacked in the image domain to give the full integration.

This particular field is problematic due to the presence of a number of bright, extended sources within the large MWA field of view, specifically the Crab Nebula, the Orion Nebula and a number of large-scale supernova remnants. As a result of calibration and beam errors, these bright sources cast artifacts throughout the image and raise the noise level higher than is typical. This is particularly pronounced in the Phase I observations due to the increased power of these extended sources on the shorter baselines.

We provide two images, MWA-1 and MWA-2, using the method described here for each of the Phase I and II configurations, respectively. The properties and noise values for each of these images are provided in \autoref{table:images}. In addition, we provide a third image---`MWA-subtracted'---using the Phase II data but using a point source subtraction technique described in section
\ref{section:subtraction}.

\subsection{ASKAP}

ASKAP undertook two observations of this field as part of their early testing programme for their newly built array and data processing pipelines. The ASKAP array is situated at the Murchison Radio Observatory, alongside the MWA. The array consists of 36 tracking dishes distributed quasi-randomly so as to produce baselines ranging in length from \SI{22}{\metre} through to a maximum \SI{6.4}{\kilo \metre} (see \autoref{fig:baselines}). This large range of baselines gives ASKAP both high resolution as well as good sensitivity to extended emission, with almost a tenth of the baselines sensitive to emission on angular scales greater than \SI{3.5}{\arcminute} at \SI{887}{\mega \hertz}. Each dish is \SI{12}{\metre} in diameter, and at \SI{887}{\mega \hertz} the resulting primary beam has a full width half maximum (FWHM) of \SI{1.76}{\degree}. Additionally, each ASKAP dish is equipped with a phased array feed (PAF) allowing for 36 beams to be formed at once; depending on the configuration of these beams, this can allow for a much larger area of sky to be observed in a single pointing.

The two observations (PI: Vernstrom) occurred on 10 March 2019 and 28 June 2019 for 5 hours and 8 hours respectively, and were observed at a central frequency of 887 MHz with a bandwidth of 288 MHz. The PAF was configured in the `square6x6' configuration for the first observation and in the `closepack36' configuration for the second \citep{McConnell2019}; both allowed for the simultaneous observation of almost the entire \SI{8 x 8}{\degree} field.

Both of these observations were independently processed. The initial bandpass and calibration was completed by the automated ASKAPSoft pipelines\footnote{The ASKAPSoft pipeline does not yet have a paper describing its operation, however the current manual is available at \url{https://www.atnf.csiro.au/computing/software/askapsoft/sdp/docs/current/index.html}} using PKS B1934-638 as the primary calibrator providing both bandpass and phase calibration. Note that secondary phase calibrators are not used by ASKAP as the instrumental phases are assumed to remain stable throughout the observation. After this initial calibration, the observation was averaged to \SI{1}{\mega \hertz} channels and \SI{10}{\second} intervals. In addition, we applied two rounds of self-calibration for phase gains, and a final round of combined amplitude and phase gains using $\textsc{casa}$ \citep{CASA2007}.

Next, each of the 36 beams were imaged with \textsc{wsclean} using the following \clean{}ing configuration: $3\sigma$ mask, $1\sigma$ threshold, multiscale enabled and joined-channels configured with six channels and two Taylor terms. We were forced to exclude the six baselines under \SI{60}{\metre} in length due to large-scale fringe patterns across the field caused by these baselines; the origin of these fringes remains unclear. Each of the final 36 beam images were primary beam corrected, truncated at their half power radius, and mosaiced using their respective primary beam weights. Finally, the mosasics from each observation were summed and weighted by the mean noise across each image.

We provide two separate images, ASKAP-B+0.5 and ASKAP-B-1, imaged with Briggs weightings of 0.5 and -1, respectively. The former has good sensitivity to extended emission with a synthesised beam of about \SI{20}{\arcsecond}, whilst the latter has twice the resolution. Their combination can aid in discerning between the diffuse and compact components of regions of extended emission. Their respective noise values and other details are provided in \autoref{table:images}. Additionally, we also provide a diffuse emission map, referred hereafter as `ASKAP-subtracted', with point sources subtracted using the method described in the next section.

\subsection{Source subtraction} \label{section:subtraction}

From each of the MWA Phase II and ASKAP observations we created additional, lower resolution images with point sources subtracted so as to emphasise diffuse emission. Rather than attempt to fit and subtract point sources from the final, deconvolved images as was done to produce the SRT+NVSS-diffuse image, we took advantage of the \clean{} deconvolution process itself. Recall that \clean{} runs in a loop whereby it finds the brightest peak in the dirty image, models a point source at this position with some fraction of the measured peak value (the `gain' parameter, typically 0.1), and subtracts this from the image (during `minor cycles') and the visibilities (during `major' cycles). This loop continues, each time searching for a peak in the residual image and subtracting it out, until some stopping condition is met, typically when the brightest peak remaining falls under some threshold. An output of this process is a final residual image, with the \clean{} components fully subtracted. This residual image will be devoid of any bright sources, however large-scale, faint emission will typically still be present hidden in amongst the noise, and it is this image that we use to construct our diffuse maps.

We used \textsc{wsclean} to perform the imaging and deconvolution with stopping conditions controlled by the mask and threshold options. The first of these options constructs a mask such that we only search for peaks within a masked region that is some factor of the noise, and the second determines that we stop \clean{}ing when there are no more peaks within the masked region greater than this factor of the noise.

For the ASKAP-subtracted image, we set the threshold value to $1.5 \sigma$, which is fairly typical, however we set the mask value to $8 \sigma$, which is higher than usual. The result of this is that all bright regions of the map (greater than $8 \sigma$) are \clean{}ed all the way down to $1.5 \sigma$, whilst any regions with faint emission beneath this $8 \sigma$ threshold are left in the final residual map. This first round of \clean{}ing was run with \textsc{wsclean} multiscale disabled. Next, we continued to \clean{} the residual map, but with \textsc{wsclean}'s multiscale \clean{} algorithm enabled and with a deeper mask of $3 \sigma$. The \clean{} components found in this second round of deconvolution were not subtracted, and were either very faint point sources or large scale extended emission. Finally, this image was convolved up to a resolution of \SI{3.5 x 3.5}{\arcminute} so as to emphasise any diffuse emission present whilst suppressing any remaining faint point sources. The final image has a noise of \SI{0.75}{\milli \jansky \per \beam} and identical resolution to the SRT+NVSS-diffuse image.

We used a similar process for the MWA-subtracted image. However, since we image and deconvolve each snapshot independently we use different values for each of the mask and threshold parameters. Typical two minute snapshots have a noise of about \SI{12}{\milli \jansky \per \beam}, whilst the final MWA-2 image has a noise of \SI{2.3}{\milli \jansky \per \beam}. To obtain the same \clean{}ing thresholds as in ASKAP would require us to \clean{} to a threshold under the noise of the individual snapshots, which is both unstable and unphysical. Instead, we set each of the mask and threshold to their lowest, stable values of 3 and 1, respectively, meaning the residual images contain faint emission up to approximately \SI{35}{\milli \jansky \per \beam}. Since we are already \clean{}ing down to the limits, the residual images for each snapshot are not further \clean{}ed using multiscale. Finally, as with the ASKAP-subtracted image, we convolved each snapshot to a resolution of \SI{3.5 x 3.5}{\arcminute} and stacked the images. The final MMA-subtracted image has a noise of \SI{5.4}{\milli \jansky \per \beam}. 

\section{Surface Brightness Sensitivity}
\label{sec:surfacebrightness}

\begin{figure*}
    \centering
    \begin{subfigure}{0.32\linewidth}
        \includegraphics[width=1\linewidth,trim={1.4cm 0 0.3cm 0}]{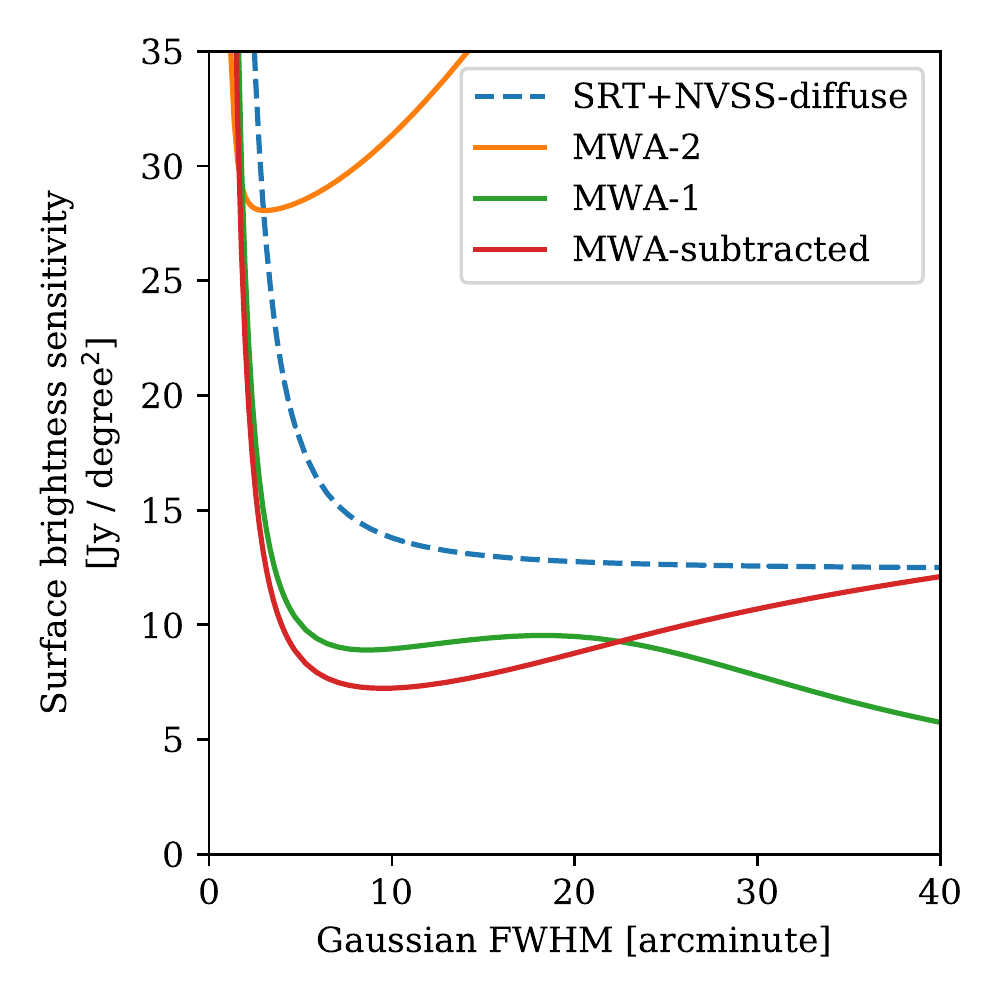}
        \caption{MWA comparison 154 MHz}
        \label{fig:sensitivity:a}
    \end{subfigure}
    \begin{subfigure}{0.32\linewidth}
        \includegraphics[width=1\linewidth,trim={1.4cm 0 0.3cm 0},clip]{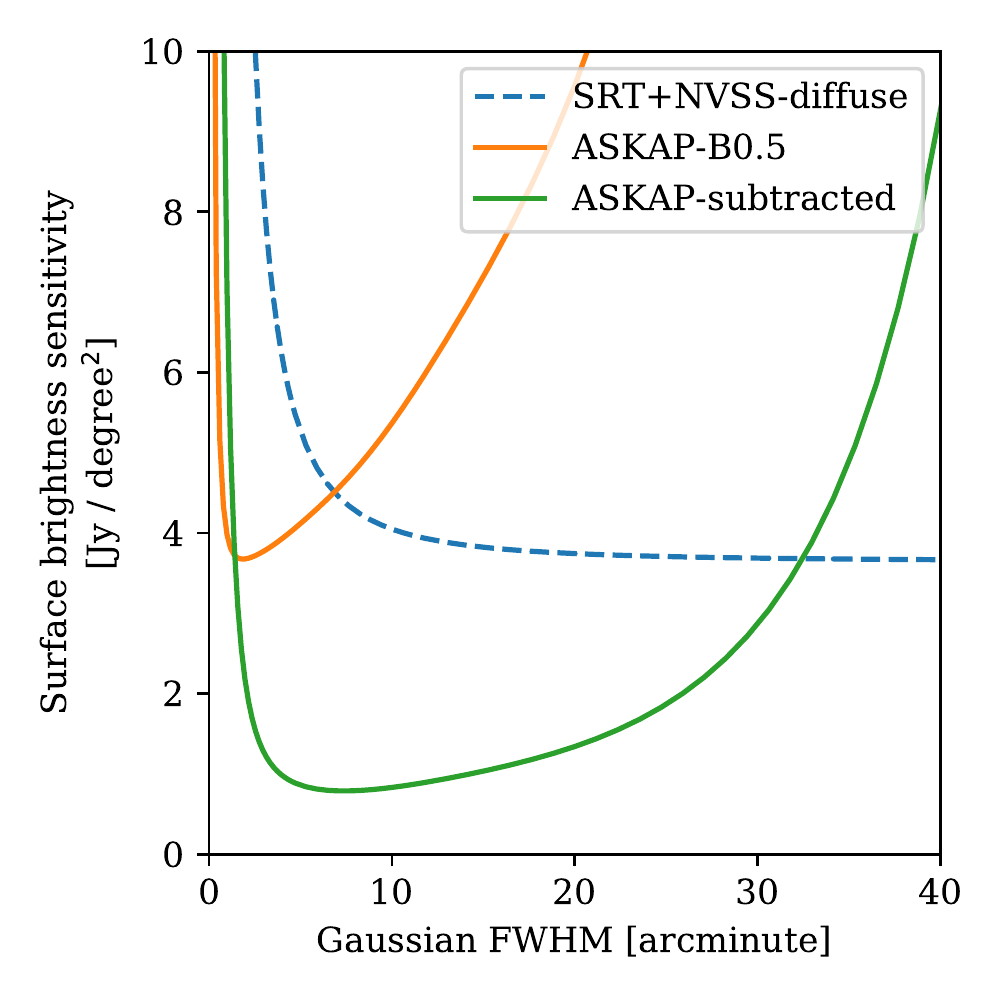}
        \caption{ASKAP comparison 887 MHz}
        \label{fig:sensitivity:b}
    \end{subfigure}
    \begin{subfigure}{0.32\linewidth}
        \includegraphics[width=1\linewidth,trim={1.4cm 0 0.3cm 0},clip]{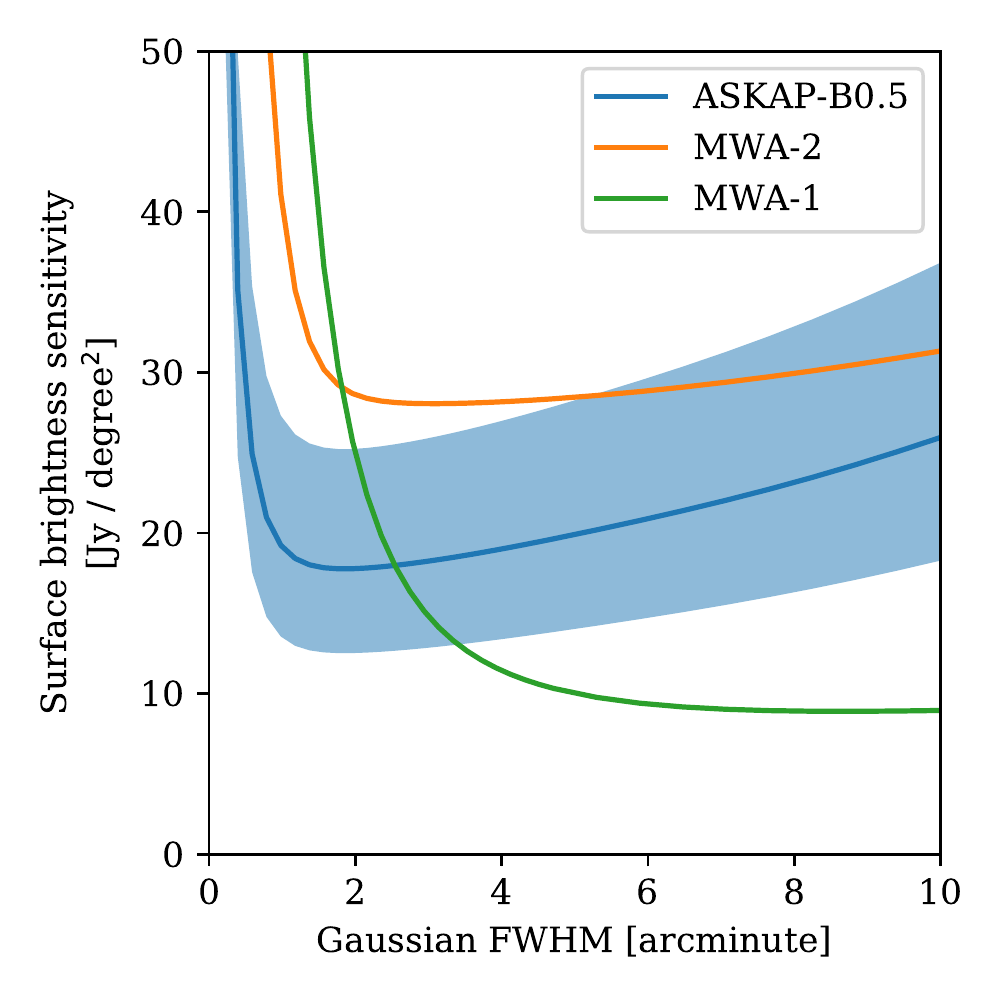}
        \caption{MWA \& ASKAP 154 MHz}
        \label{fig:sensitivity:c}
    \end{subfigure}
    \caption{Surface brightness sensitivity values: \textbf{(a)} 154 MHz (MWA-1, MWA-2, and MWA-subtracted); \textbf{(b)} 887 MHz (ASKAP-B+0.5, ASKAP-subtracted). The SRT+NVSS-diffuse values (dashed blue line) are frequency adjusted from 1.4 GHz, and represent the minimum surface brightness required to corroborate candidate sources in VA18 assuming a spectral index of -0.7 or steeper. \textbf{(c)} Direct comparison at 154 MHz of MWA and ASKAP surface brightness sensitivity, where ASKAP has been frequency adjusted from 887 MHz assuming a spectral index range $-0.7 < \alpha < -1.1$, with the solid line at the midpoint $\alpha = -0.9$.}
    \label{fig:sensitivity}
\end{figure*}

Surface brightness sensitivity, $\sigma_{\text{SB}}$, measures an interferometer's response to extended emission; specifically it is the minimum surface brightness that is detectable above the noise. As we are searching for large extended emission---which we assume to be smoothly varying---surface brightness is a more useful measure than the more typically quoted point source sensitivity. In this section we measure and compare the surface brightness sensitivity of each of MWA Phase I, Phase II and ASKAP.

An interferometer's sensitivity to extended emission is dependent on the same factors that contribute to point source sensitivity (such as system temperature, effective collecting area, number of antennae and baselines, etc.) but, crucially, also depends on the geometry of the array. In particular, as the angular scale of emission increases, in visibility space the power spectrum of the source shifts towards the zeroth spacing and therefore short baselines are essential to sample this region.

Surface brightness sensitivity varies based on angular scale of the emission. For sources with an angular scale smaller than the synthesised beam, sensitivity scales approximately with the area of the source, until becoming most sensitive when the scale of the source matches the scale of the synthesised beam. On the other hand, extended emission above a threshold angular scale will have its power spectrum so condensed around the zeroth spacing that few baselines will properly sample its power and the sensitivity to sources above this scale will drop as we `resolve out' the source.

We attempt to estimate our surface brightness sensitivity in the following way. We simulate two-dimensional, circular Gaussian sources with constant peak brightness, $P$ [Jy degree$^{-2}$], and varying FWHM values into the visibilities of the MWA Phase I, Phase II and ASKAP measurement sets. We then produce dirty images of each and measure the peak flux response $S_{\text{peak}}$ [Jy beam$^{-1}$] at the center of each Gaussian in the resulting image. We estimate the surface brightness sensitivity as:
\begin{equation}
    \sigma_{\text{SB}} = n \sigma_{\text{RMS}} \frac{P}{S_{\text{Peak}}}
    \label{eqn:surfacebrightness}
\end{equation}
where $\sigma_{\text{RMS}}$ [Jy beam$^{-1}$] is the measured noise of our final images as detailed in \autoref{table:images}, and $n$ is the factor above the noise required for a detection (which was $3 \sigma$ in all cases). The fraction $\nicefrac{P}{S_{\text{Peak}}}$ measures the response to the simulated surface emission and is solely a function of the shape of the synthesised beam (i.e\@. the PSF); this is constant irrespective of the actual value of the simulated surface emission. Given this constant fraction, \autoref{eqn:surfacebrightness} allows us to calculate just how bright the simulated surface brightness would need to be for the response to rise above the threshold for detection (i.e\@. $n\sigma_{\text{RMS}}$).

In this sensitivity estimation we use the dirty image as opposed to the deconvolved image as this better simulates how very faint sources are processed. At the limits of surface brightness sensitivity, emission in our images is buried amongst the image noise, and \clean{}ing thresholds will result in such emission being at most only partially deconvolved. Moreover, deconvolving a source makes it brighter, and so by using the dirty images we are properly modelling the worst case.

To compare these values with the SRT+NVSS-diffuse image, we use their stated beam size of \SI{3.5 x 3.5}{\arcminute} and simply convolve our Gaussian sky models with a Gaussian beam of this size. From the resulting images, we measure the peak flux response. This process assumes perfect and complete $uv$ coverage with no interfering sidelobes, and so is a lower limit (i.e\@. best case) for the surface brightness sensitivity of the SRT+NVSS-diffuse image. 

There is one further complication. We would like to answer the question: if emission is detectable in the SRT+NVSS-diffuse image at 1.4 GHz, what level of sensitivity is required at 154 MHz and 887 MHz to be able to detect the same emission? To make this comparison, we need to make assumptions about the spectrum of such emission. Shock emission, such as relic, halo or filamentary accretion shocks typically have spectral indices of approximately -1 or steeper, while -0.7 is more typical of AGN emission. We choose here to use the more conservative value of -0.7. We can then scale the surface brightness sensitivity limits of the SRT+NVSS-diffuse image by this factor for each of the MWA and ASKAP observing frequencies:
\begin{equation}
    \sigma_\text{min} = \left(\frac{\nu}{1.4 \text{ GHz}}\right)^{-0.7} \sigma_\text{SB}
\end{equation}
This frequency-adjusted limit thus represents the minimum sensitivity required to corroborate detection of a source at the limit of the SRT+NVSS sensitivity for any sources with a spectral index of -0.7 or steeper.

Using this method, \autoref{fig:sensitivity} compares the surface brightness sensitivity of the MWA and ASKAP with the frequency-adjusted surface brightness of the SRT+NVSS-diffuse image. In \autoref{fig:sensitivity:a}, we compare the surface brightness sensitivity of the \SI{154}{\mega \hertz} images of the MWA with the SRT+NVSS-diffuse image. We can see that the MWA-2 image surpasses the surface brightness sensitivity of the SRT+NVSS-diffuse image only out to angular scales of approximately \SI{3}{\arcminute}. Emission on angular scales larger than this, however, is increasingly resolved out. It is interesting to note that this reduction in sensitivity occurs on angular scales much smaller than we would expect just from calculating the fringe patterns of the shortest baselines of the MWA phase 2; this discrepancy arises from the weighted addition of each baseline's respective fringe pattern that ultimately forms the shape of the synthesised beam. On the other hand, both MWA-1 and MWA-subtracted have a greater surface brightness sensitivity than the frequency-adjusted SRT+NVSS-diffuse image on all angular scales out to at least \SI{40}{\arcminute}. MWA-1 achieves this by its dense sampling of the inner region of the $uv$-plane, whilst MWA-subtracted achieves this sensitivity as a result of the extra convolution step that decreased the resolution to \SI{3.5 x 3.5}{\arcminute}.

In \autoref{fig:sensitivity:b}, we compare the surface brightness sensitivity of ASKAP observing at \SI{887}{\mega \hertz}. The ASKAP-B0.5 image has greater surface brightness sensitivity than SRT+NVSS-diffuse out to angular scales of approximately \SI{7}{\arcminute}. The ASKAP-subtracted image, on the other hand, is able to exceed the frequency-adjusted limit required to corroborate synchrotron emission out to angular scales of approximately \SI{32}{\arcminute}, which is, again, solely a result of the extra convolution step used in the point source subtraction process. We can conclude that both images have the required sensitivity to detect the kind of large scale emission reported by VA18.


We can also directly compare the surface brightness sensitivity of MWA and ASKAP by frequency adjusting the sensitivity values of ASKAP from \SI{887}{\mega \hertz} down to \SI{154}{\mega \hertz}. As can been seen in \autoref{fig:sensitivity:c}, we use a range of spectral indices, ranging from -0.7 to the steeper -1.1 with a solid line indicating an intermediate spectral index of -0.9. We find that the ASKAP-B0.5 image is significantly more sensitive than MWA-2 on all angular scales out to approximately \SI{5}{\arcminute}, beyond which the MWA-2 image is more sensitive to those sources with the very steepest spectral indices. ASKAP is more sensitive than MWA-1 on angular scales smaller than approximately \SI{2.5}{\arcminute}; for larger angular scales, the prevalence of short baselines in the MWA phase I array result in MWA-1 having superior surface brightness sensitivity. Nonetheless, this suggests a surprising result: ASKAP is ideally suited to the detection of synchrotron emission on scales both small and large, even for sources with moderately steep spectral indices.

\section{Results}

\begin{table*}
    \centering
    \begin{tabular}{cccccccp{4cm}} \toprule
        Source & RA (J2000) & Dec (J2000) & SRT significance & MWA & ASKAP & HII & Notes \\
        & h:m:s & d:m:s & $\sigma$ & detection & detection & region & \\\midrule
        A1 & 04:59:08.81 & +08:48:52 & 6 & Yes & Yes & No & Radio halo in A523\\
        A2 & 04:57:43.81 & +08:47:03 & 3 & No & No & No & \\
        A3 & 04:56:23.85 & +09:27:59 & 3 & No & No & No & \\
        B1 & 04:49:29.06 & +08:30:16 & 3 & No & No & Yes & \\
        B2 & 04:53:19.21 & +07:48:11 & 3 & No & No & No &\\
        B3 & 04:51:39.15 & +07:15:01 & 3 & No & No & No & Double-lobed radio galaxy immediately South of source\\
        \multirow{2}{*}{C1*} & 05:15:39.81 & +06:51:47 & 5 & No & No & Partly & Excluding Northern zoom \\
        & 05:15:31.00 & +06:49:40 & 5 & Yes & Yes & No & Northern zoom only \\
        C2 & 05:12:24.80 & +07:25:01 & 3 & No & No & No & \\
        C3 & 05:10:39.64 & +07:06:07 & 4 & No & No & No & \\
        C4 & 05:12:34.29 & +06:49:01 & 3 & No & No & No & \\
        C5 & 05:11:21.76 & +06:49:35 & 3 & No & No & No &\\
        C6* & 05:12:26.81 & +06:20:31 & 4 & No & Yes* & Partly & North West contour only, but does not overlap\\
        C7 & 05:07:44.04 & +06:26:13 & 4 & No & Yes & Yes & \\
        C8 & 05:06:57.73 & +06:21:59 & 3 & No & No & Yes & \\
        C9* & 05:05:57.34 & +06:14:45 & 3 & Yes & No & No & \\
        C10 & 05:06:19.45 & +06:04:59 & 3 & No & Yes & Yes & \\
        D1 & 05:05:00.00 & +06:44:00 & 3 & No & No & No & \\
        D2 & 05:01:52.93 & +06:06:57 & 4 & No & No & No & \\
        D3 & 05:00:19.57 & +05:44:24 & 3 & No & No & No & \\
        E1 & 04:57:26.67 & +06:52:01 & 5 & No & Yes & No & \\
        E2* & 04:55:05.24 & +06:17:21 & 4 & Yes & No & No & \\
        E3 & 04:57:10.28 & +06:04:15 & 3 & No & No & No & \\
        F1 & 05:11:24.89 & +03:46:42 & 3 & Yes & No & No & SRT and MWA contours only partially overlap \\
        G1 & 05:02:21.28 & +05:26:12 & 3 & No & No & No & \\
        G2 & 04:55:03.01 & +05:33:20 & 3 & No & Yes & Yes & \\
        G3 & 05:00:28.92 & +05:03:38 & 3 & No & No & No & \\
        G4 & 04:59:12.63 & +05:01:05 & 3 & No* & No & No & SRT contours sit immediately North of large extended emission system in MWA \\
        G5 & 04:57:59.36 & +04:58:01 & 3 & No & No & No & \\
        G6* & 04:58:34.65 & +04:42:47 & 4 & Yes & No & No & \\
        H1 & 04:49:56.16 & +04:48:46 & 3 & No & No & No & \\
        H2 & 04:49:28.39 & +04:31:12 & 3 & No & No & No &\\
        I1 & 04:54:06.90 & +02:33:02 & 5 & Yes & Yes & No & Radio halo in A520 \\
        I2 & 04:55:06.23 & +02:33:02 & 3 & No & N/a* & No & Source beyond ASKAP primary beam \\
        I3 & 04:55:06.23 & +02:30:33 & 3 & No & N/a* & No & Source beyond ASKAP primary beam \\
        J1 & 04:48:37.81 & +03:00:55 & 3 & No & No & No & \\\bottomrule
    \end{tabular}
    \caption{Diffuse large-scale emission regions identified by VA18. An asterisk by the name indicates that VA18 considered it possible that the region was contaminated by residuals from compact source subtraction.}
    \label{table:results}
\end{table*}

In \autoref{table:results} we present each of the 35 sources reported by VA18, the maximum significance of their detection in the SRT+NVSS-diffuse image, and whether either MWA (in any of MWA-1, MWA-2 or MWA-subtracted) or ASKAP are able to detect emission in the same region to $3\sigma$ significance. In the Appendix we provide images of every VA18 region. In \autoref{fig:askaps} we show each of the 35 sources as imaged in ASKAP-B+0.5, with contours from the SRT+NVSS-diffuse (blue) and ASKAP-subtracted (red). In \autoref{fig:mwa-2} we present each of the regions as imaged in MWA-2, with contours again from SRT+NVSS-diffuse (blue) and MWA-subtracted (red). Finally, in \autoref{fig:mwa-1}, we present the full \SI{8 x 8}{\degree} region as imaged in MWA-1, with contours from SRT+NVSS-diffuse. This latter image is scaled such that saturated black represents a $5 \sigma$ detection.

\section{Discussion}

\begin{figure*}
    \centering
    \begin{subfigure}{0.49\linewidth}
        \includegraphics[width=1\linewidth,trim={0 0.2cm 0 0.2},clip]{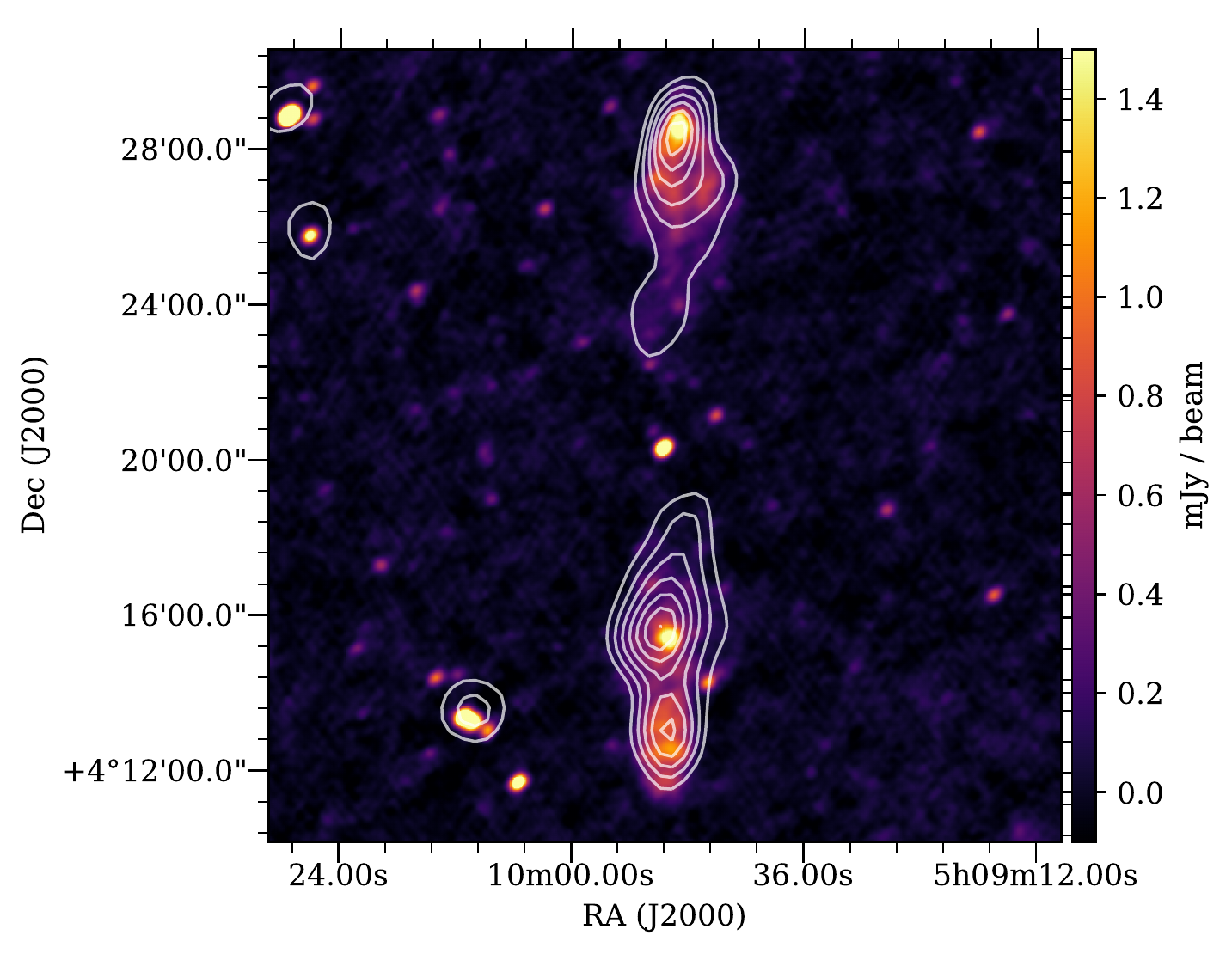}
        \caption{RA \ra{5;09;50} Dec \ang{4;20;19}}
        \label{fig:radiogals:a}
    \end{subfigure}
    \begin{subfigure}{0.49\linewidth}
        \includegraphics[width=1\linewidth,trim={0 0.2cm 0 0.2},clip]{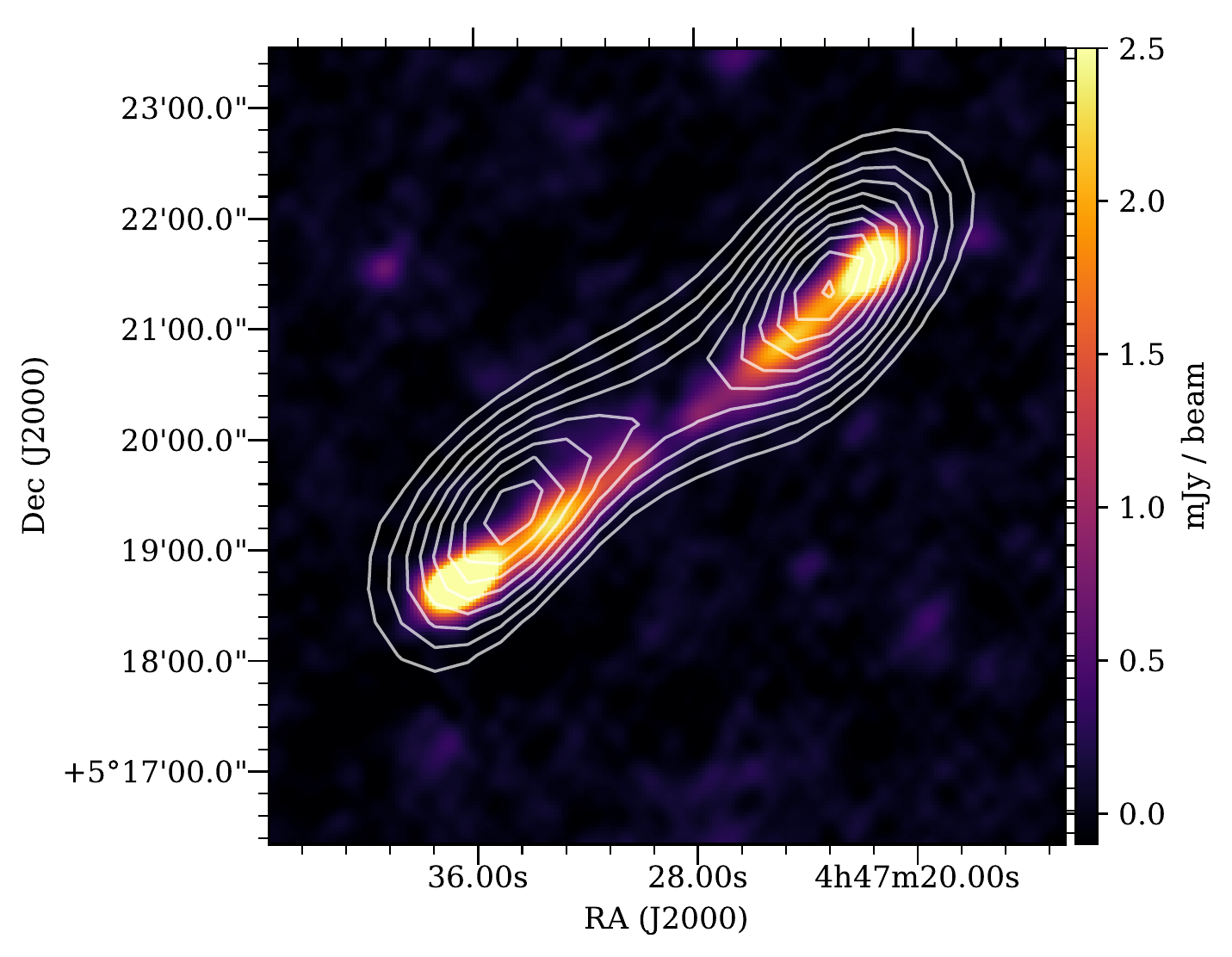}
        \caption{RA \ra{4;47;23.9} Dec \ang{5;18;50}}
        \label{fig:radiogals:b}
    \end{subfigure}
    \caption{Images from ASKAP-B+0.5 at 887 MHz of two radio galaxies in the field mentioned by VA18. The white contours are MWA-2 at 154 MHz, starting at $3\sigma$ and increasing in increments of $+2\sigma$.}
    \label{fig:radiogals}
\end{figure*}

The known, large-scale synchrotron sources in this field are detected in all our images with strong statistical significance, and this provides a initial validation of the angular sensitivity our observations. For example, the radio halo in Abell 523 (source A1) is detected in each of ASKAP-subtracted and MWA-subtracted well above the noise (statistical significance of $20 \sigma$ and $11 \sigma$, respectively), as is the radio halo in Abell 520 (source I1; statistical significance of $8 \sigma$ and $8 \sigma$, respectively). Both are also visible in MWA-2 and MWA-1, though in the latter the more compact emission is blended in with the diffuse components. In addition, the large extended lobes of the radio galaxy that VA18 report in region F are visible in all images, as we show in \autoref{fig:radiogals:a}. The core, on the other hand, is only visible in the higher frequency ASKAP image; this is typical of galactic core emission which is dominated by free-free mechanisms and thus tends to have have a flatter spectrum at low radio frequencies.\footnote{We also identify an optical candidate for the core of this radio galaxy, which is clearly visible both in Digital Sky Survey \citep{Blanton2017} and Panoramic Survey Telescope and Rapid Response System (Pan-STARRS; \citealp{Chambers2016}) optical surveys and has previously been catalogued in the infrared as WISEA J050950.55+042021.0. The calculations in VA18 that inferred a minimum size of the radio galaxy from the magnitude limit of the DSS survey are therefore invalid.} Similarly, the lobes of the smaller radio galaxy located at RA \ra{4;47;24} Dec \ang{5;18;50} are also clearly detected in all images as shown in \autoref{fig:radiogals:b}.

Despite demonstrating that we can detect the known synchrotron sources in this field, 23 of the 35 candidate sources are undetected in any of our direct observations as well as our `subtracted' treatments. If we assume that these sources are both real and have spectra that are well approximated by a power law at this frequency range ($S \propto \nu^\alpha$), then we can calculate a lower limit value for the spectral index of these sources from ASKAP-subtracted map as $\alpha > 2.5$. The MWA-subtracted map places a less stringent constraint of $\alpha > -0.37$. Such a steep positive spectral index is atypical for synchrotron sources, with the exception of sources that exhibit a turnover due to synchrotron self-absorption or free-free absorption mechanisms. Both these mechanisms, however, are unusual to observe in this frequency range for large, diffuse systems.

We turn now to discuss the sources for which we make a potentially corroborating detection, or are otherwise noteworthy.

\subsection{Source B1}

\begin{figure}
    \centering
    \includegraphics[width=1\linewidth]{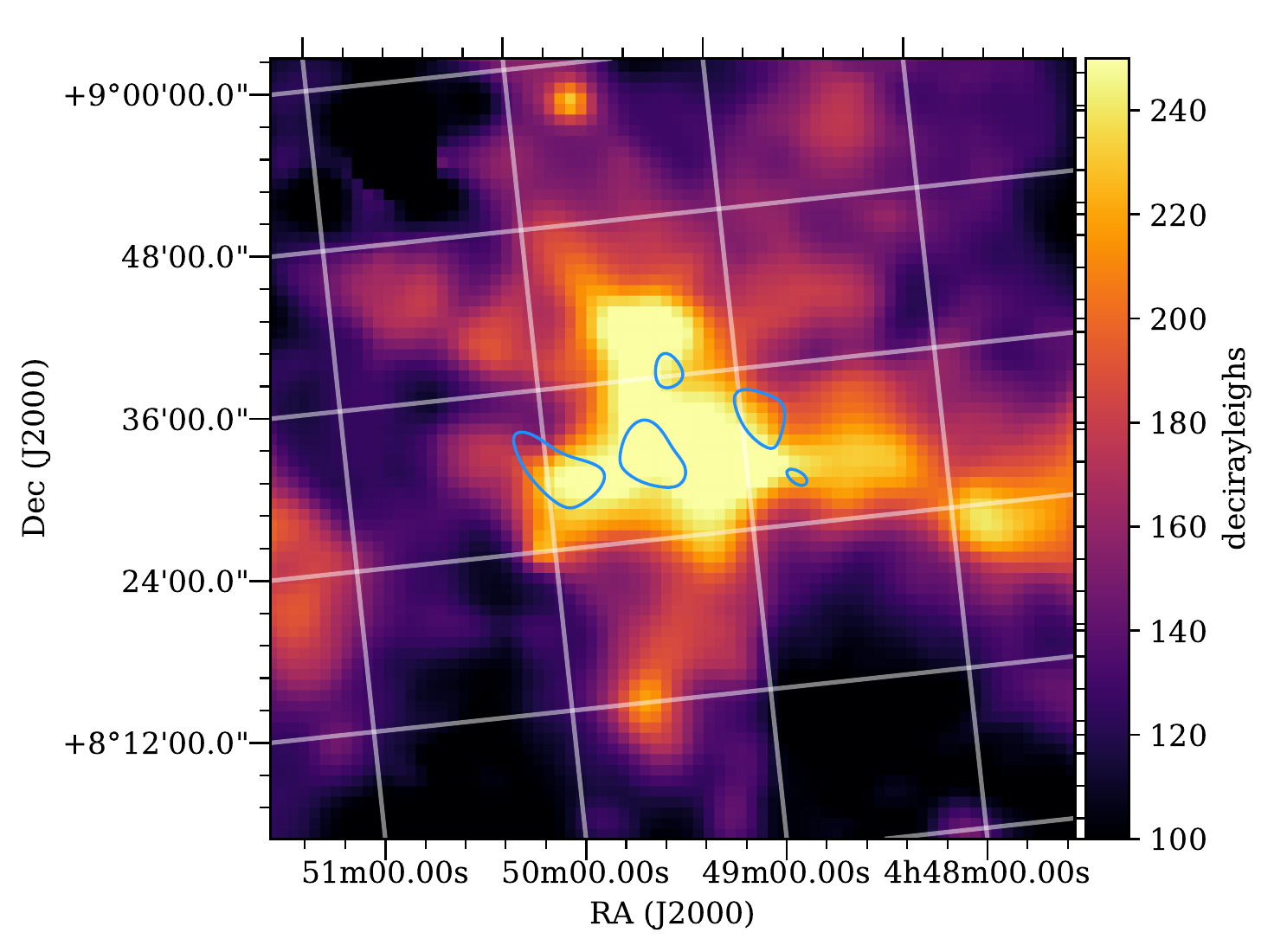}
    \caption{An H-alpha map of region B1 from SHASSA showing the coincident H-alpha emission. SRT+NVSS-diffuse contours (blue) indicate $3\sigma$, $4\sigma$, $5\sigma$, etc.}
    \label{fig:regionB1}
\end{figure}

Source B1 appears in the SRT+NVSS-diffuse map as a $3 \sigma$ detection at 04:49:29.06 +08:30:16, for which we find no radio emission in either ASKAP-subtracted or MWA-subtracted. However, in \autoref{fig:regionB1} we present the associated Southern H-alpha Sky Survey Atlas (SHASSA; \citealp{Gaustad2001}) image showing this is a region of strong H-alpha emission, and indicating that this is a Galactic HII region. We propose that source B1 is likely a faint detection of associated thermal free-free emission produced by this Galactic HII region, and that the non-detection by both ASKAP and MWA is due to the typically inverted, blackbody spectrum of such sources, placing its surface brightness below the detection levels of our lower-frequency observations.

\subsection{Sources C1, C6, C7, C8, C10}

\begin{figure*}
    \centering
    \includegraphics[width=1\linewidth]{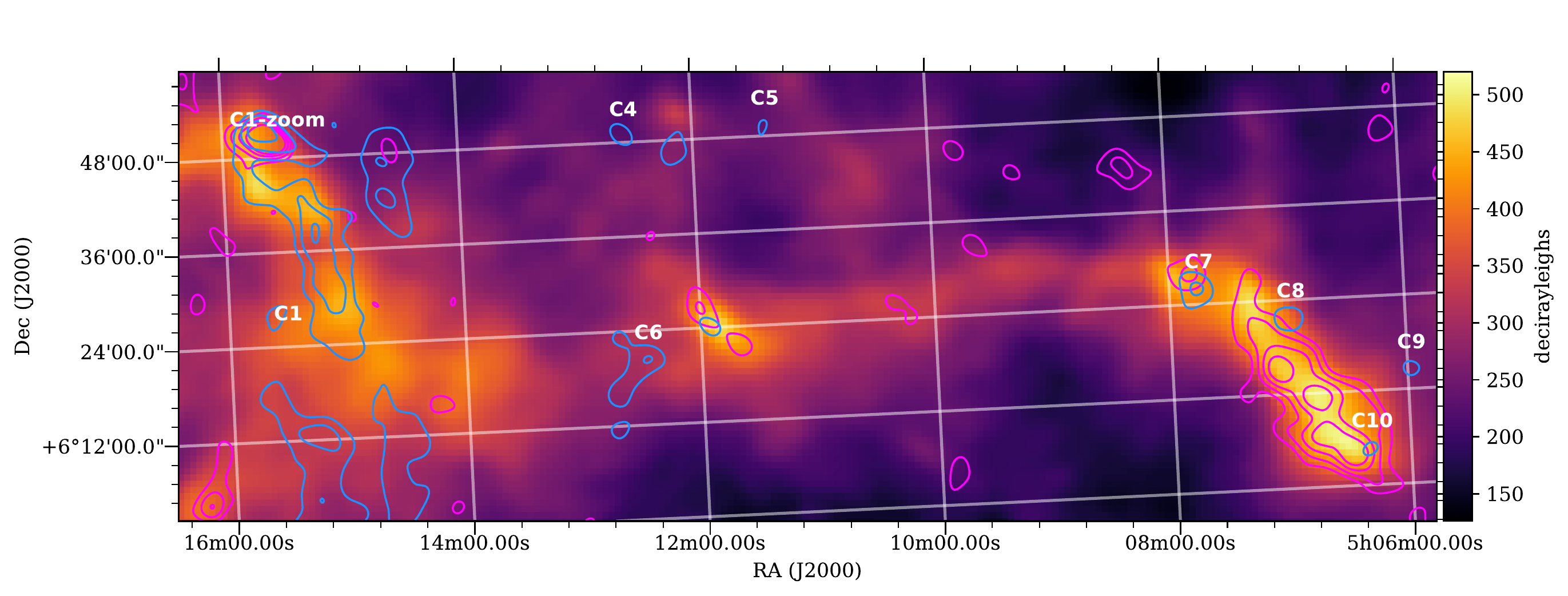}
    \caption{An H-alpha map of region C from SHASSA. SRT+NVSS-diffuse contours (blue) indicate $3\sigma$, $4\sigma$, $5\sigma$, etc. ASKAP diffuse contours (magenta) indicate $2 \sigma$, $3 \sigma$, $4 \sigma$ etc.}
    \label{fig:regionC}
\end{figure*}

VA18 report a very large-scale detection in the vicinity of Abell 539, spanning multiple large-scale islands of emission (C1) as well as numerous small regions of diffuse emission to the West (C2-C10).

In \autoref{fig:regionC} we show the SHASSA image for a large section of region C, overlaid with the SRT+NVSS-diffuse contours (blue) and the ASKAP-subtracted contours (magenta). From the contours, we can observe that the ASKAP-subtracted map shows a clearly visible ridge of flux extending approximately 40 arcminutes in a North-Easterly orientation, approximately joining the regions C7, C8 and C10. This ridge has a peak flux of \SI{6.3}{\milli \jansky \per \beam}, whilst it is undetectable in the lower frequency MWA images suggesting a shallow or inverted spectral index. From the background SHASSA map, we observe that this ridge of emission traces a similarly bright region of Galactic H-alpha emission which extends West from the Galactic equator through C1 and C6, and peaks along the ridge adjoining C7, C8 and C10.

The coincident emission of H-alpha and radio strongly suggests that the ridge we are observing is a Galactic HII region, and that we are detecting the thermal free-free component of this region in the radio. Moreover, the lack of radio emission in the MWA observations is consistent with the inverted spectrum of thermal free-free emission.

\autoref{fig:regionC} includes, in addition to the bright ridge of emission on the right, regions C1 and C6. We include these regions to suggest the possibility that the Western component of C6 as well as the North-East island of C1 (with the exception of the Northern `C1-zoom') may also be a detection of the extended Galactic HII region. Indeed, despite C1 lying beyond the half-power point of the ASKAP primary beam, we still detect a radio component coincident with a peak in the H-alpha map. This strongly suggests that the North-East island of C1, which lies closest to the centre of Abell 539, is not extra-galactic in origin.

\subsection{Source C1 `zoom'}

\begin{figure}
    \centering
    \includegraphics[width=1 \linewidth,trim={0 0 1.7cm 0},clip]{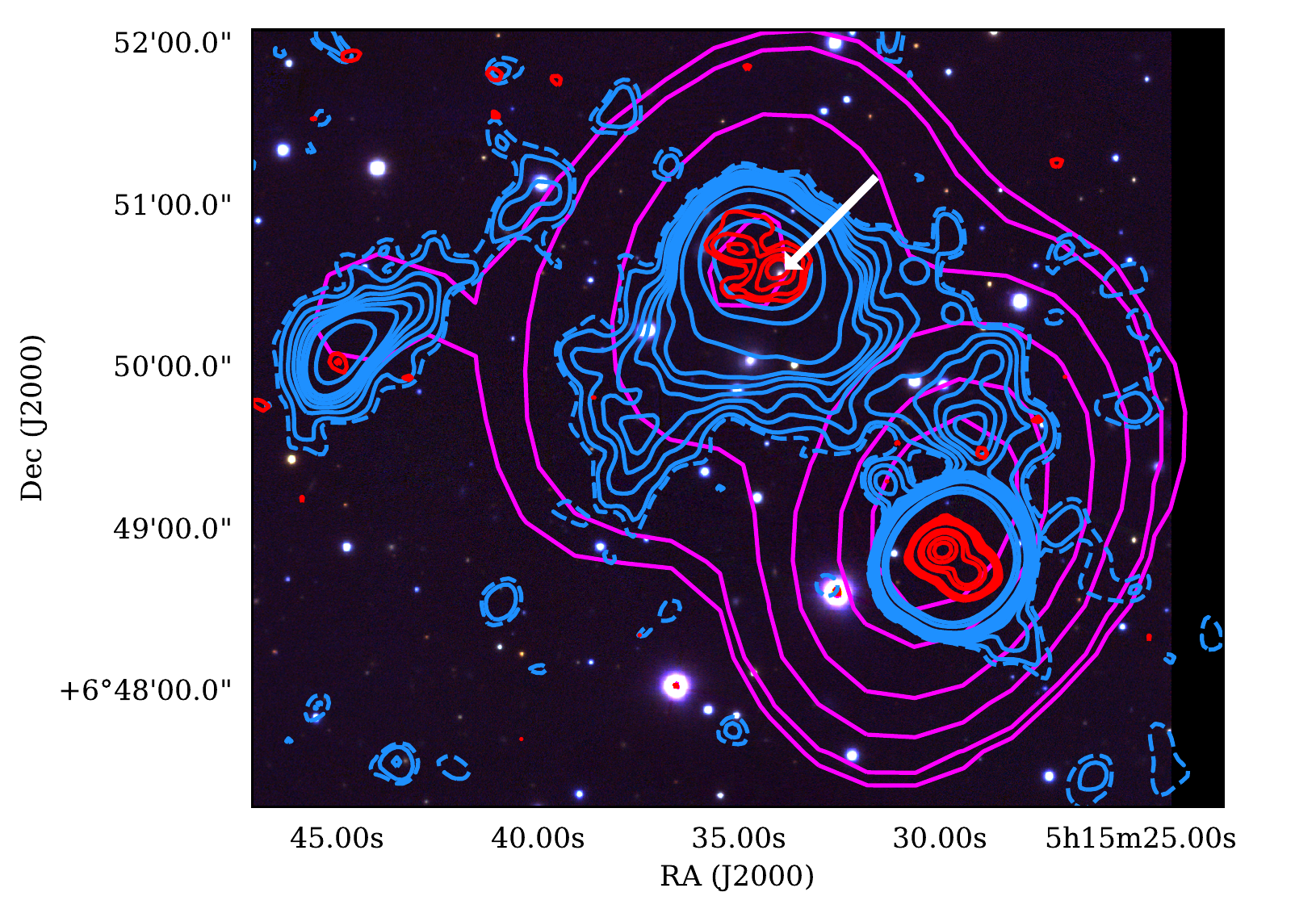}
    \caption{The Pan-STARRS three-colour (bands Y, I, G) image of `C1-zoom', showing the presumed optical host indicated by the white arrow. The contours are: ASKAP-B+0.5 (blue) at $1.5 \sigma$ (dashed) and then 2, 3, 4, 5, 6, 7, 10, 20, 30$\sigma$; ASKAP-B-1 (red) at 3, 4, 6, 8, 30, 50, 100, 150$\sigma$; MWA-2 (magenta) at 3, 5, 15, 35, 80, 120$\sigma$.}
    \label{fig:c1zoom}
\end{figure}

The C1 Northern zoom, centred at 05:15:31 +06:49:40 and located at the very periphery of Abell 539, contains significant diffuse emission that is detected in the MWA (Phase I \& II) and ASKAP images. In \autoref{fig:c1zoom} we show the three-colour optical image from the Panoramic Survey Telescope and Rapid Response System (Pan-STARRS; \citealp{Chambers2016}) overlaid with contours from ASKAP-B-1, ASKAP-B+0.5 and MWA-2. 

The C1 Northern zoom contains a number of bright points of emission. The brightest, located at 5:15:29.52 +6:48:46.23 (lower right), is resolved into two conjoined points in ASKAP-B-1 with no optical association in Pan-STARRS, whilst in ASKAP-B+0.5 it has a faint extension along the same axis; we propose this source is a pair of radio lobes of a distant, background galaxy and unrelated to the extended emission in this region.

The second brightest source of emission in the C1 North zoom is centered at 5:15:33.93 +6:50:33.3 and is surrounded by diffuse radio emission. It is clearly extended in the ASKAP-B+0.5 image with a largest angular scale of approximately \SI{180}{\arcsecond}. A central hotspot is visible in the ASKAP-B-1 image and in addition, two satellite patches of extended emission appear in ASKAP-B-1 to the South East and South West. The source is also visible in all MWA images, and using the MWA-2 and ASKAP-B+0.5 images we can calculate a spectral index for the total integrated flux as -0.97. In the associated Pan-STARRS image we observe a candidate host galaxy 2MASX J05153393+0650333 indicated by the arrow sitting near the peak of the emission, for which there is unfortunately no currently available redshift information, as neither of the satellite regions have any optical candidate. Given the existence of a host galaxy and the hotpots, it seems likely that this is diffuse radio-galaxy emission. The presence of a bright core suggests this is a Fanaroff \& Riley class I (FRI) radio galaxy, however there are clearly weakly emitting lobes which would suggest the presence of some environmental pressure. The overall morphology of the source is certainly atypical of normal radio jet structure, however it is suggestive of a head-tail galaxy. Whilst additional observations may aid in understanding its complex morphology, we feel confident to classify this diffuse emission as originating from 2MASX J05153393+0650333.

In addition, a secondary diffuse radio source is visible in the top left of the image. This appears to be an FRII radio galaxy, with the left lobe significantly brighter than the right, possibly due to relativistic beaming. The left-most lobe is visible in lowest MWA-2 contour, that is, a $3 \sigma$ detection at 154 MHz. There is no obvious optical candidate visible in Pan-STARRS, suggesting that this is in the background of the 2MASX J05153393+0650333 system.






\subsection{Source C9}

Source C9 is detected at $3 \sigma$ significance in MWA-subtracted. The ASKAP-B+0.5 image shows five point sources in a small angular area, and MWA-2 detects and resolves at least 3 of these. However, the brightest of these sources in MWA-2 is just \SI{13}{\milli \jansky \per \beam}, meaning that none of these sources will have been subtracted from individual snapshots; any flux present in the MWA-subtracted image is likely unsubtracted point source emission. In agreement with VA18, source C9 is most likely the result of residual point sources.

\subsection{Source E1}

There is a trace of a detection at the central peak of E1 in ASKAP-subtracted (peak $3.1 \sigma$), whilst there is nothing in MWA, either in MWA-1, 2 or subtracted. In the case of MWA-1, this is a region with no nearby sources that might produce a false positive result due to blending, and given its superb surface brightness sensitivity, the absence of a lower frequency detection strongly suggests against this region as being synchrotron in origin. SHASSA, however, does not indicate any associated peak in H-alpha emission in this region. Given the low statistical significance of the ASKAP detection, and that the region above the $3 \sigma$ threshold has a maximum angular extent of just \SI{0.8}{\arcminute} (compared to a beam size of \SI{3.5 x 3.5}{\arcminute}), we would be inclined to suggest that this is noise in our own image if it were not so clearly aligned with the SRT+NVSS-diffuse peak contour. We measure a peak brightness of \SI{2.7}{\milli \jansky \per \beam} at 887 MHz, and \SI{14.5}{\milli \jansky \per \beam} in the SRT+NVSS-diffuse image at 1.4 GHz, giving a steep positive spectral index of +3.7. Whilst we conclude this is unlikely to be synchrotron, we leave open the possibility that this is emission by some other mechanism with an inverted spectrum.

\subsection{Source E2}

The MWA-subtracted image detects a small area of diffuse emission at E2, whilst nothing is detected in ASKAP-subtracted. The ASKAP-B+0.5 image resolves 5 bright radio sources in this small region. At least two of these are very slightly extended in the ASKAP image: the source located at 4:55:07.7 +6:16:31.6 is a star-forming spiral galaxy with a bright compact core visible in Pan-STARRS but whose spiral arms are also weakly visible in radio; the source located at 4:54:58.0 +6:17:22.5 is extended in ASKAP-B+0.5 with an extension towards the North and a bright core or hotspot visible in the ASKAP-B-1 image but no obvious optical counterpart.

As VA18 suggest, the source E2 is most likely due to a blending of numerous radio sources and not due to diffuse radio emission.

\subsection{Source F1}

VA18 report a small region of $3 \sigma$ significance located at 05:11:24.89 +03:46:42. The MWA-subtracted image finds a small region of extended emission offset North of this, which encompasses four distinct radio sources in ASKAP-B+0.5. This extended emission signal is almost certainly just the result of blended emission from these point sources, and does not corroborate the F1 candidate region.

\subsection{Source G2}

\begin{figure}
    \centering
    \includegraphics[width=1\linewidth]{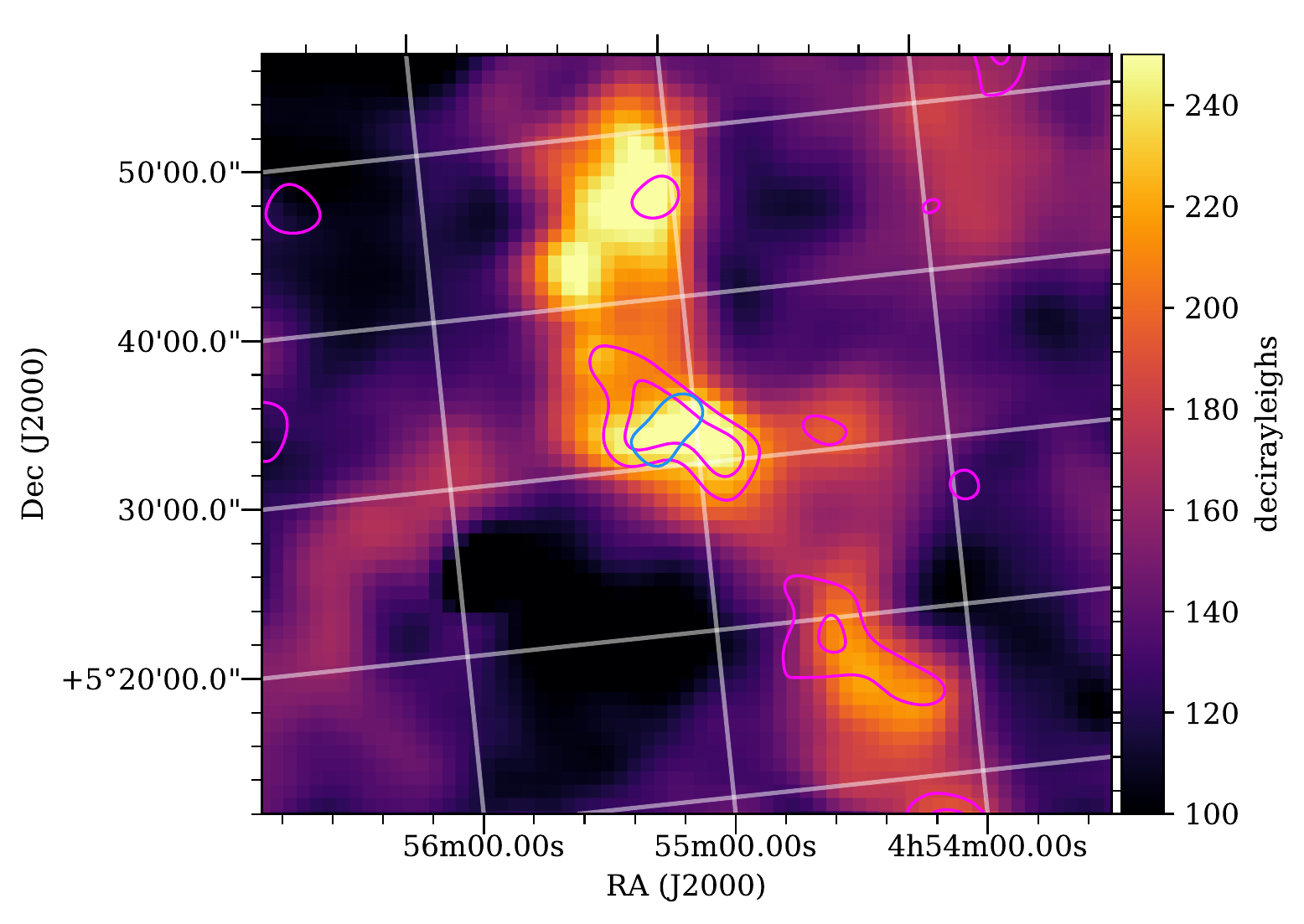}
    \caption{An H-alpha map of region G2 from SHASSA showing the coincident H-alpha emission. SRT+NVSS-diffuse contours (blue) indicate $3\sigma$, $4\sigma$, $5\sigma$, etc. ASKAP-diffuse contours (magenta) indicate $2 \sigma$, $3 \sigma$, $4 \sigma$, etc.}
    \label{fig:regionG2}
\end{figure}

In the SRT+NVSS-diffuse image, source G2 is a small $3 \sigma$ detection. MWA-subtracted makes no detection in this region, however ASKAP-subtracted makes a similarly weak $3 \sigma$ detection in the same region. In \autoref{fig:regionG2} we show the H-alpha emission in this region from SHASSA with contour overlays from SRT+NVSS-diffuse (blue) and ASKAP-diffuse (magenta). Once again we find a correlation between a peak in the H-alpha emission and the detected diffuse radio emission, suggesting that the radio is free-free emission from a Galactic HII region. Indeed, the ASKAP-diffuse $2 \sigma$ contours appear to trace two additional H-alpha peaks both North and South of G2.





\section{Conclusion}

We are unable to corroborate the candidate synchrotron sources of VA18. Careful examination of each of the 35 sources suggests five classes: known halo systems (A1, I1); radio galaxies (C1-zoom); HII emission (B1, North-East C1, North-West C6, C7, C8, C10, G2); blended compact sources (C9, F2, E1); and finally one non-synchrotron but otherwise unknown source (E1). The remaining sources are not detected in our observations.

The non-detections strongly suggest against these sources being synchrotron in origin. Synchrotron sources in general exhibit negative spectral indices, and models suggest the shocked emission from the cosmic web proper to have a spectral index $\alpha \lessapprox -1$. These properties ensure that synchrotron sources are brightest at lower radio frequencies, and given the surface brightness sensitivity of the MWA and ASKAP images, any large scale synchrotron emission should surely be visible at these lower frequencies. As we have noted, the ASKAP non-detection puts a stringent condition on the candidate sources as having a steep, positive spectral index of $\alpha > 2.4$, and this can only be explained if these are regions exhibiting a turnover due to synchrotron self-absorption or free-free absorption.

We suggest three explanations for these non-detections. Firstly, these may be real emission that have a positive spectral index and which renders them undetectable at lower frequencies, for example thermal free-free emission. However, given the extreme spectral steepness of such a population, we consider this an unlikely scenario. Secondly, given the low $3 \sigma$ threshold used to identify the candidate sources, some fraction may simply be noise. This may be especially applicable to those regions that were small in angular extent, typically much smaller than the \SI{3.5}{\arcminute} resolution of the SRT+NVSS-diffuse image. Finally, given the significant image processing employed by VA18, which included combining systematics from both SRT and NVSS, as well as a complex and imperfect point source subtraction process, some fraction of these sources may be the result of spurious image artifacts. VA18 acknowledge this possibility but, as they detailed in Appendix C, their own simulations excluded gain fluctuations from within their pipeline as being significant, and Galactic foreground simulations suggested that less than 20\% of the candidate sources could be attributed to this foreground.

Whilst this is a disappointing result, we wish to raise the possibility that large scale, extended emission may be the wrong parameter space for searching for the synchrotron cosmic web. There has been an assumption to date that the synchrotron cosmic web would match the spatial scales of the underlying filaments, which is evident both in the work of VA18 as well as others (see e.g\@. \citealp{Brown2017, Vernstrom2017}). However, the mechanism for synchrotron emission is primarily by way of accretion shocks, which are by definition regions of discontinuity. Such mechanisms may be more likely to produce sharp and smaller scale emission features as opposed to the broad, smooth and extended features that have been assumed to date. Indeed, such compact features can already be observed in simulations \citep{ArayaMelo2012,Vazza2015,Vazza2019}, suggesting that we may have in fact been looking in the wrong place. Future work in this area will be required to properly understand the characteristic spatial scales of this radio emission and constrain the parameter space as we continue to search for evidence of the synchrotron cosmic web.

\section{Acknowledgements}

This scientific work makes use of the Murchison Radio-astronomy Observatory, operated by CSIRO. We acknowledge the Wajarri Yamatji people as the traditional owners of the Observatory site. Support for the operation of the MWA is provided by the Australian Government (NCRIS), under a contract to Curtin University administered by Astronomy Australia Limited. We acknowledge the Pawsey Supercomputing Centre which is supported by the Western Australian and Australian Governments.

The Australian SKA Pathfinder is part of the Australia Telescope National Facility which is managed by CSIRO. Operation of ASKAP is funded by the Australian Government with support from the National Collaborative Research Infrastructure Strategy. ASKAP uses the resources of the Pawsey Supercomputing Centre. Establishment of ASKAP, the Murchison Radio-astronomy Observatory and the Pawsey Supercomputing Centre are initiatives of the Australian Government, with support from the Government of Western Australia and the Science and Industry Endowment Fund. We acknowledge the Wajarri Yamatji people as the traditional owners of the Observatory site.

This work was supported by resources provided by the Pawsey Supercomputing Centre with funding from the Australian Government and the Government of Western Australia.

\bibliography{refs}

\appendix
\setcounter{figure}{0}
\renewcommand{\thefigure}{A\arabic{figure}}
\renewcommand{\thesubfigure}{\roman{subfigure}}
\begin{figure*}
    \centering
    \caption{ASKAP-B+0.5 image with SRT+NVSS-diffuse contours (blue) and ASKAP-subtracted contours (red). Contours start at $3 \sigma$ of their respective map noise and increase in increments of $1 \sigma$. All images scaled linearly from -50 to \SI{1000}{\micro \jansky \per \beam}.}
    \label{fig:askaps}
    \begin{subfigure}{0.32\linewidth}
        \caption{A1}
        \includegraphics[width=1\linewidth,trim={0 -0.2cm 1cm 0.3cm},clip]{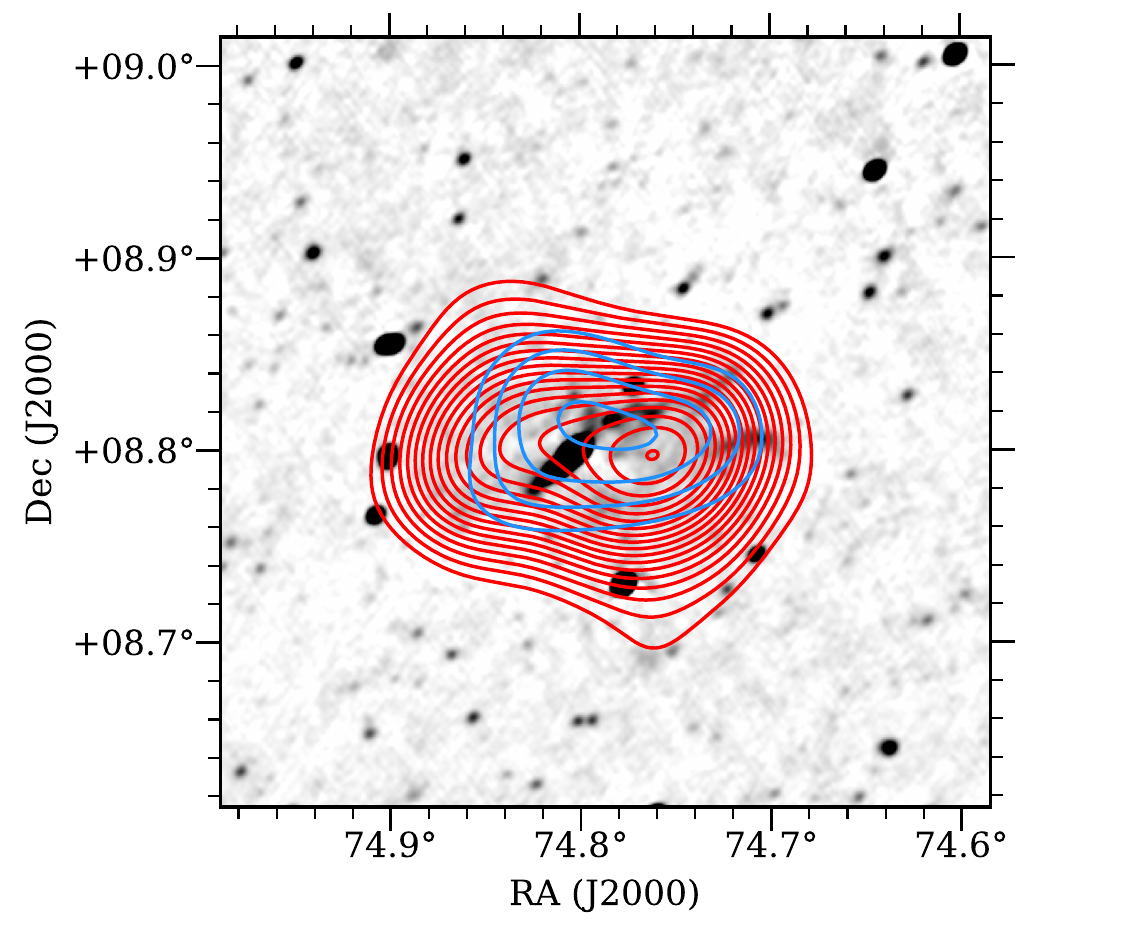}
    \end{subfigure}
    \begin{subfigure}{0.32\linewidth}
        \caption{A2}
        \includegraphics[width=1\linewidth,trim={0 -0.2cm 1cm 0.3cm},clip]{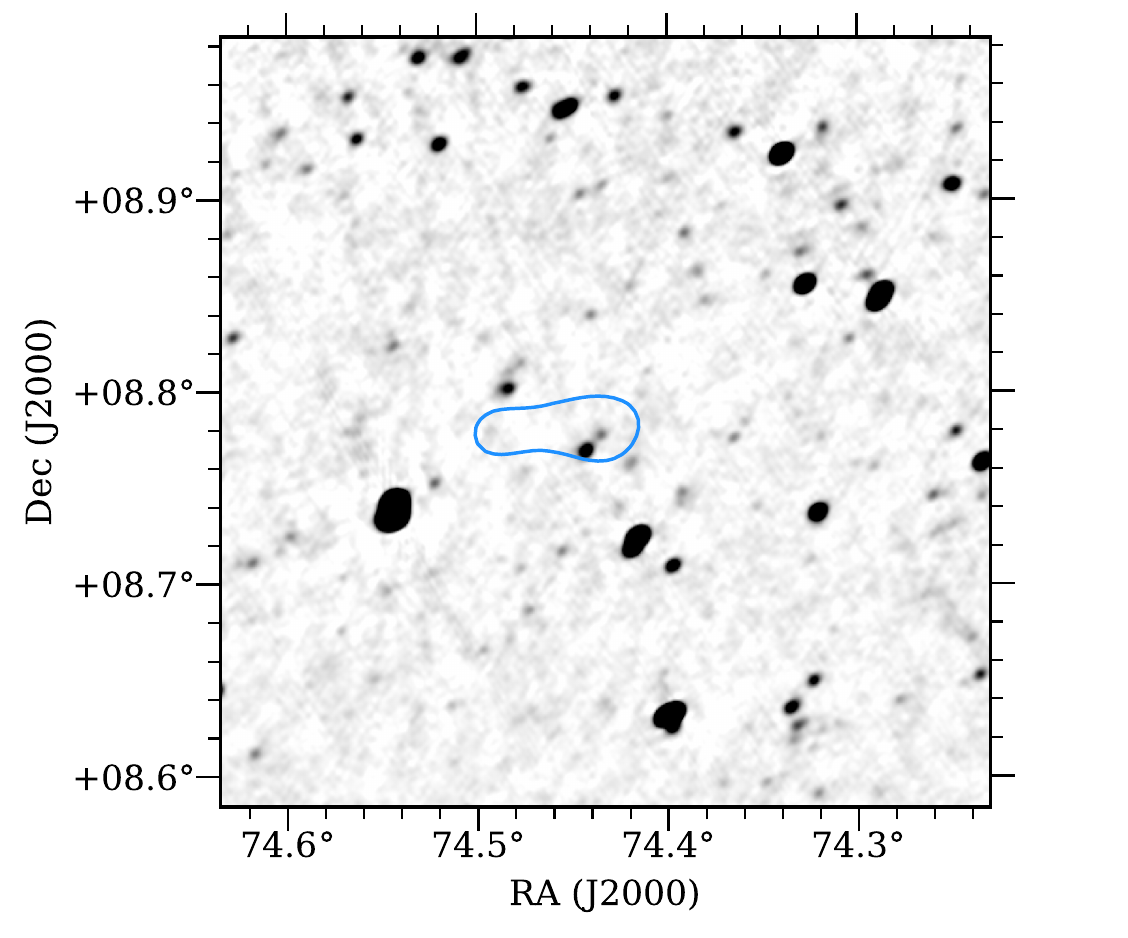}
    \end{subfigure}
    \begin{subfigure}{0.32\linewidth}
        \caption{A3}
        \includegraphics[width=1\linewidth,trim={0 -0.2cm 1cm 0.3cm},clip]{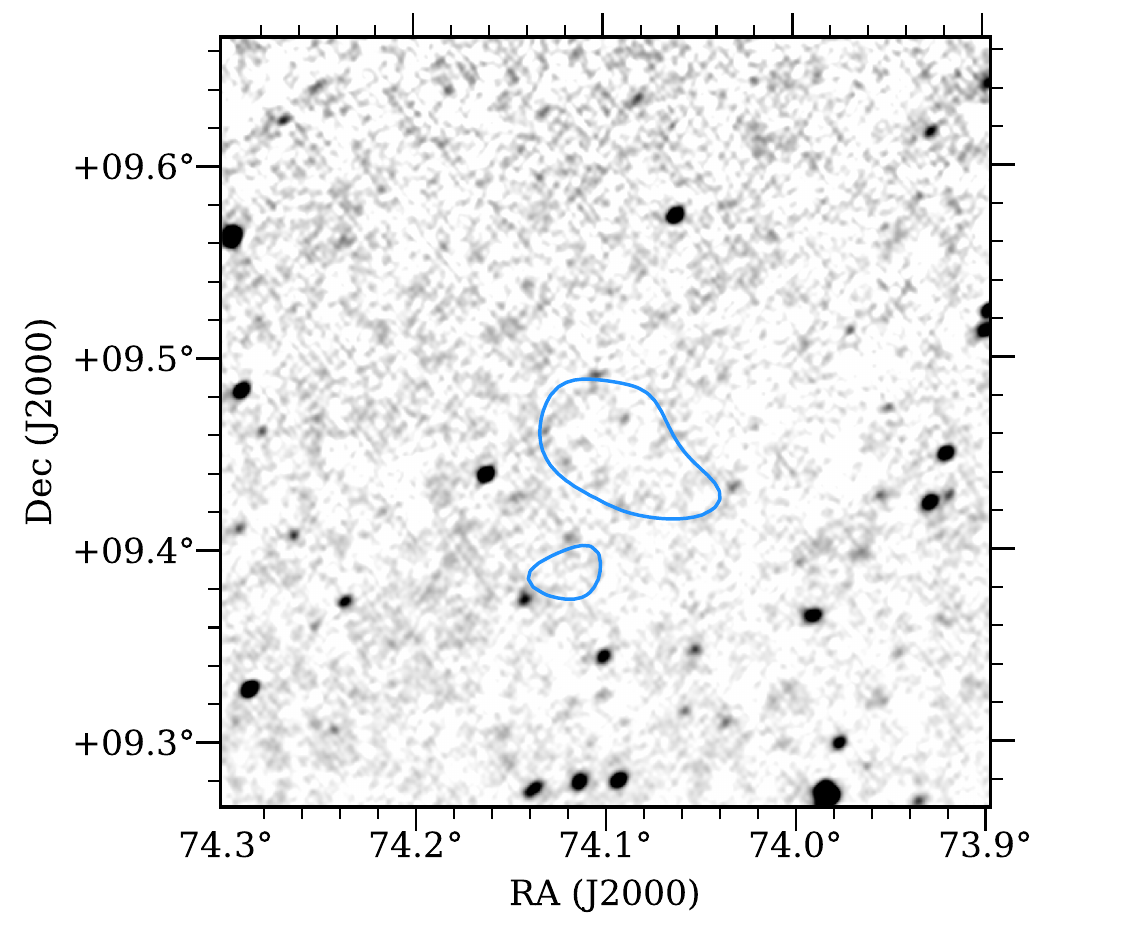}
    \end{subfigure}
    \begin{subfigure}{0.32\linewidth}
        \caption{B1}
        \includegraphics[width=1\linewidth,trim={0 -0.2cm 1cm 0.3cm},clip]{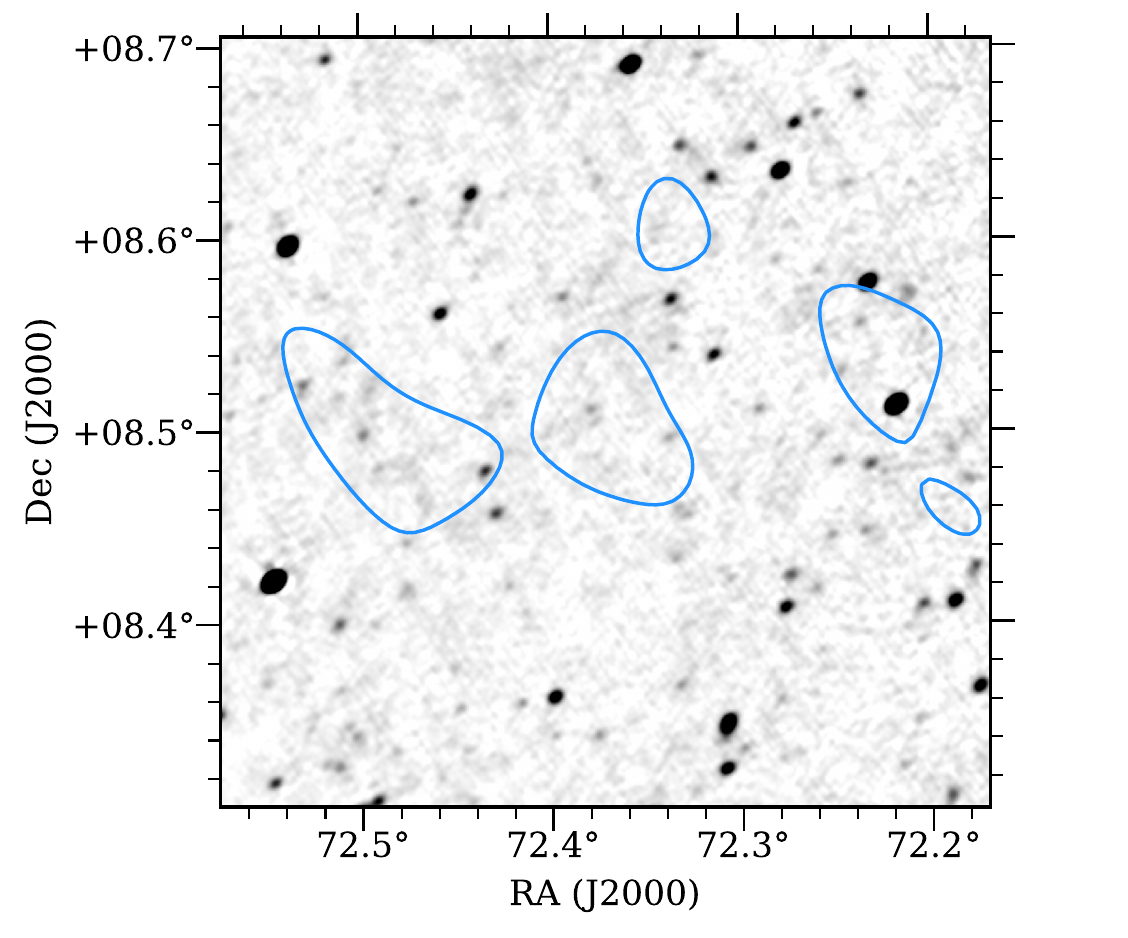}
    \end{subfigure}
    \begin{subfigure}{0.32\linewidth}
        \caption{B2}
        \includegraphics[width=1\linewidth,trim={0 -0.2cm 1cm 0.3cm},clip]{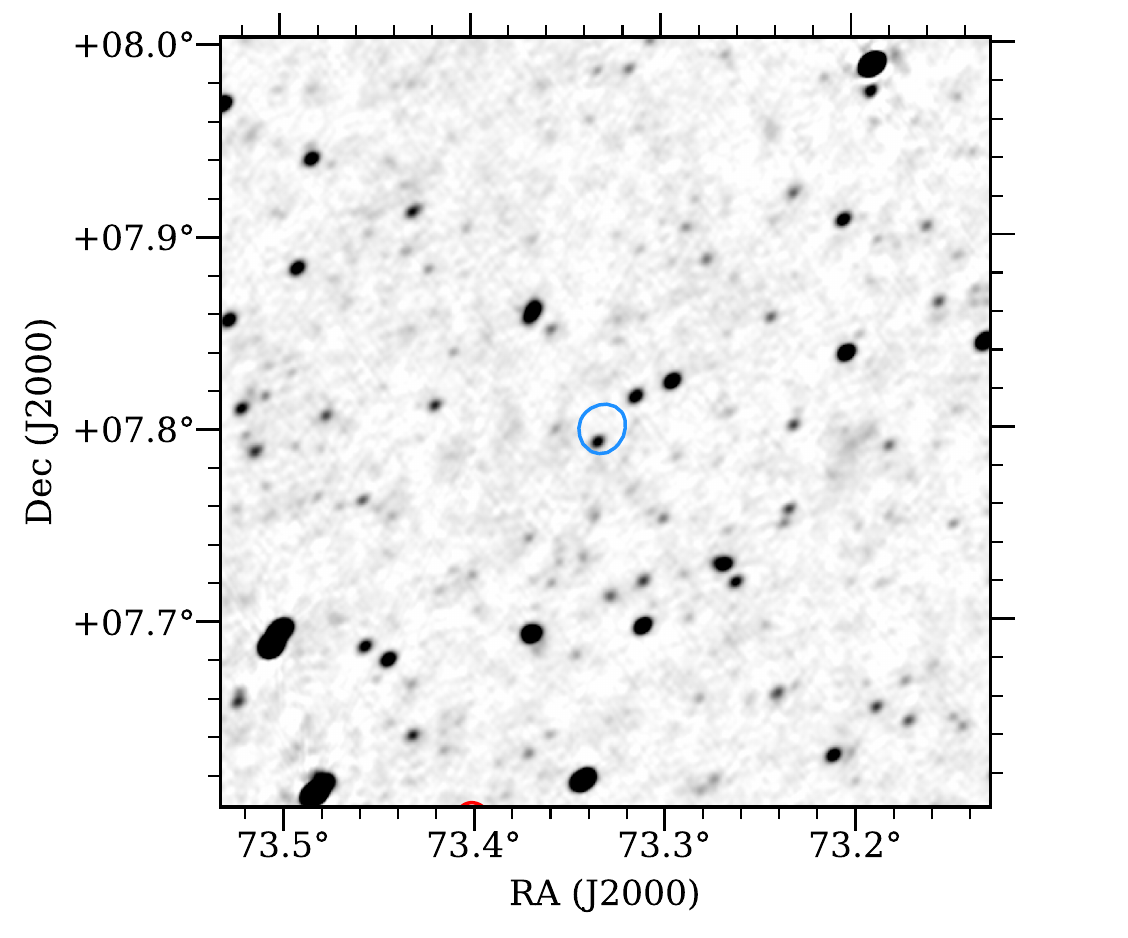}
    \end{subfigure}
    \begin{subfigure}{0.32\linewidth}
        \caption{B3}
        \includegraphics[width=1\linewidth,trim={0 -0.2cm 1cm 0.3cm},clip]{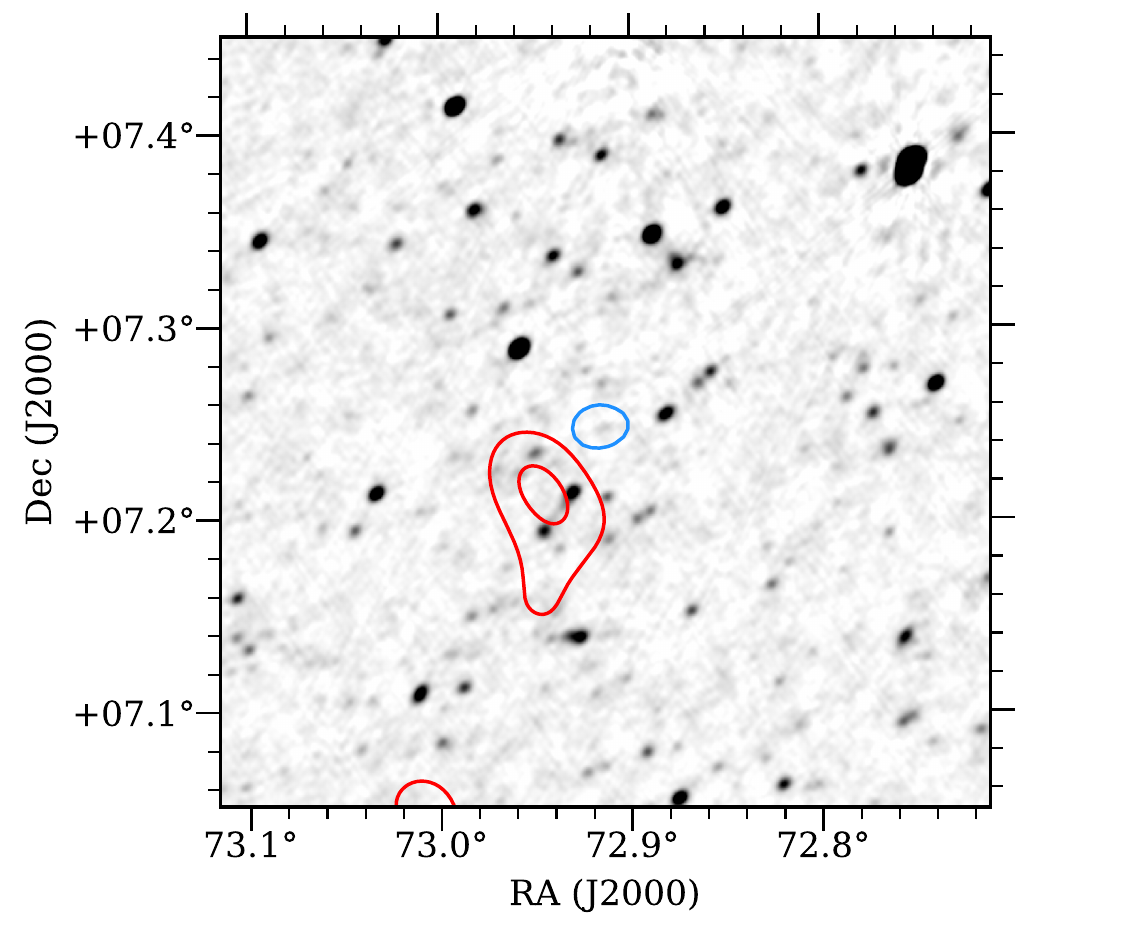}
    \end{subfigure}
    \begin{subfigure}{0.32\linewidth}
        \caption{C1}
        \includegraphics[width=1\linewidth,trim={0 -0.2cm 1cm 0.3cm},clip]{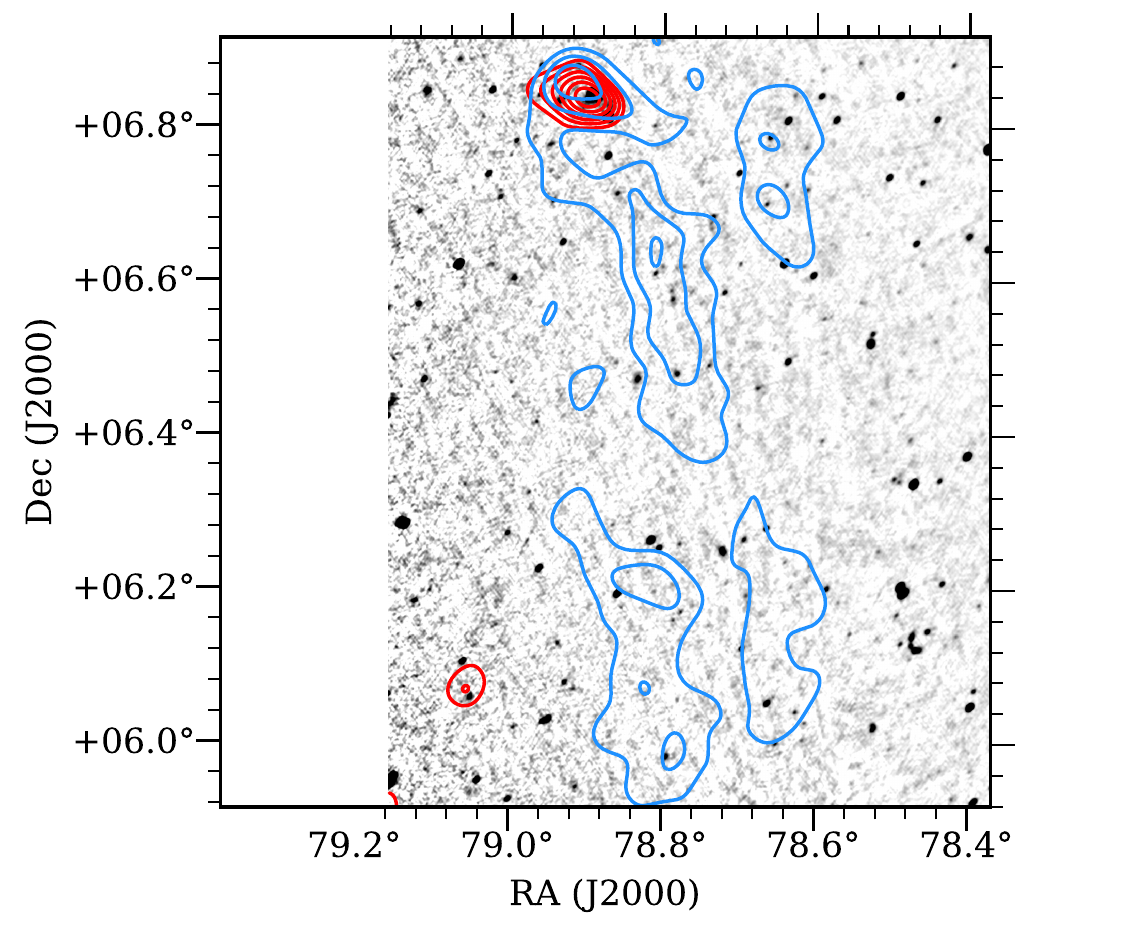}
    \end{subfigure}
    \begin{subfigure}{0.32\linewidth}
        \caption{C1 Zoom}
        \includegraphics[width=1\linewidth,trim={0 -0.2cm 1cm 0.3cm},clip]{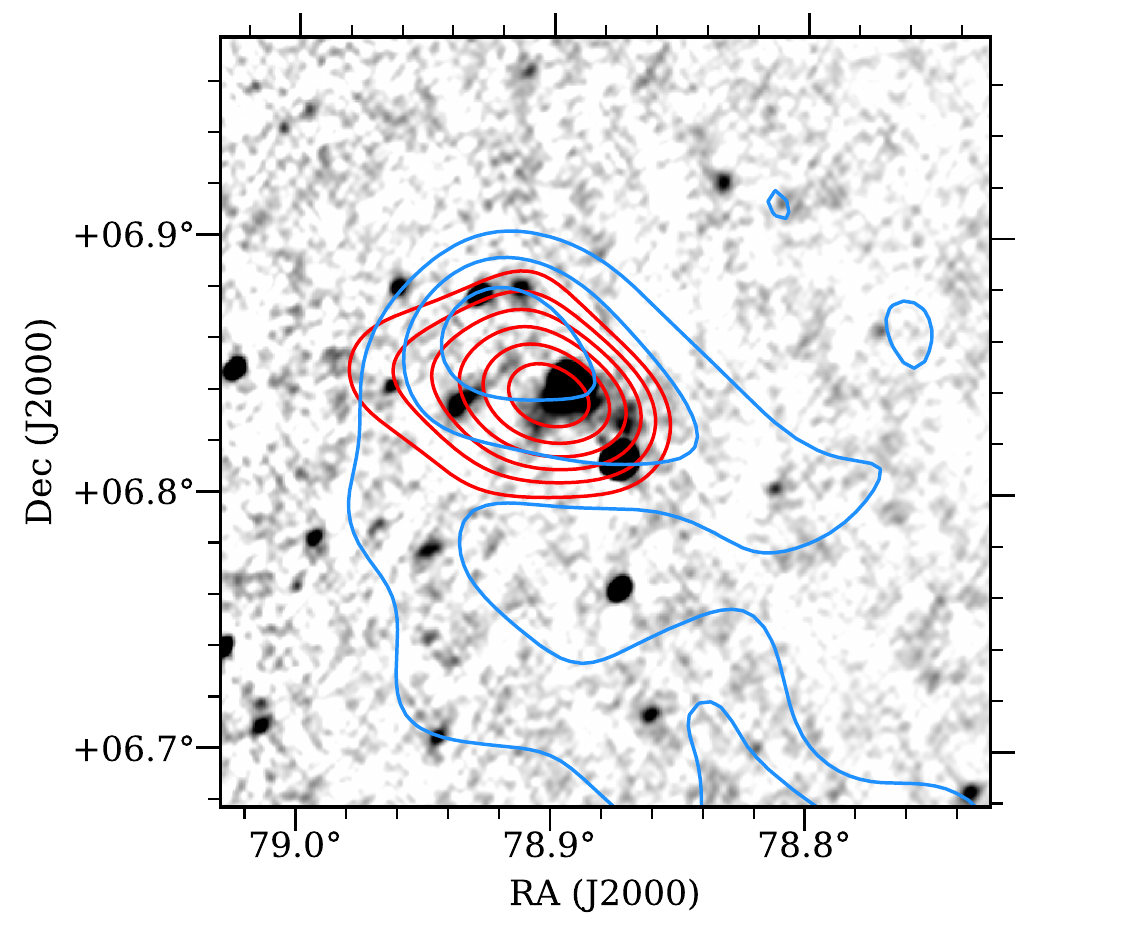}
    \end{subfigure}
    \begin{subfigure}{0.32\linewidth}
        \caption{C2}
        \includegraphics[width=1\linewidth,trim={0 -0.2cm 1cm 0.3cm},clip]{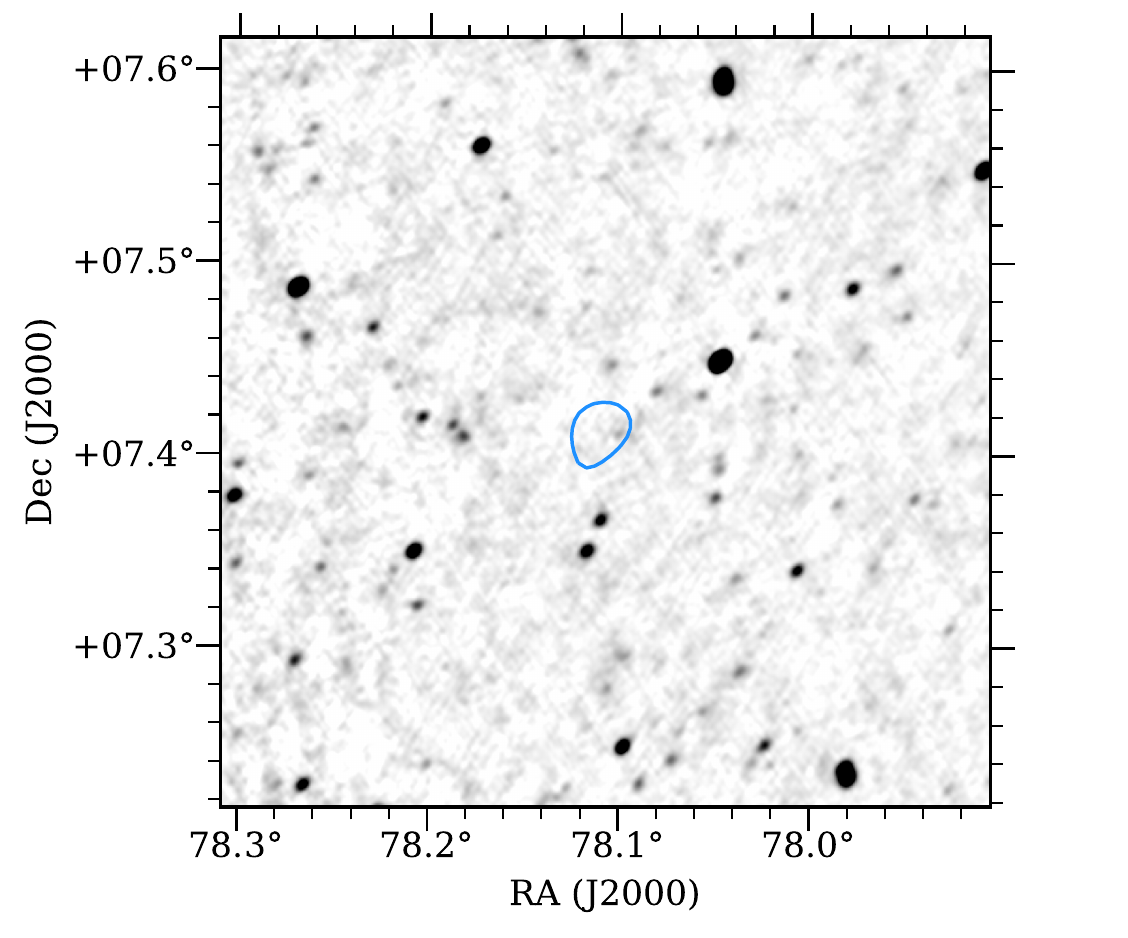}
    \end{subfigure}
    \begin{subfigure}{0.32\linewidth}
        \caption{C3}
        \includegraphics[width=1\linewidth,trim={0 -0.2cm 1cm 0.3cm},clip]{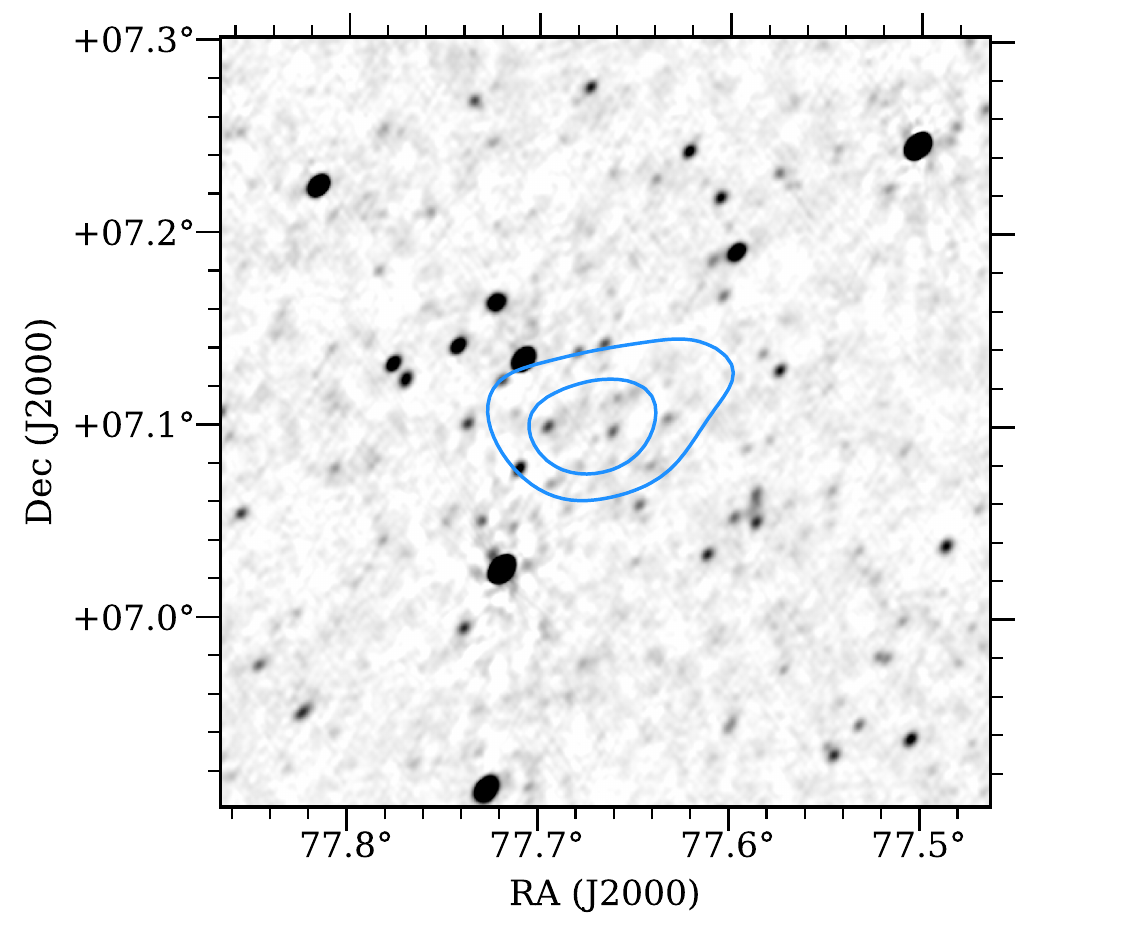}
    \end{subfigure}
    \begin{subfigure}{0.32\linewidth}
        \caption{C4}
        \includegraphics[width=1\linewidth,trim={0 -0.2cm 1cm 0.3cm},clip]{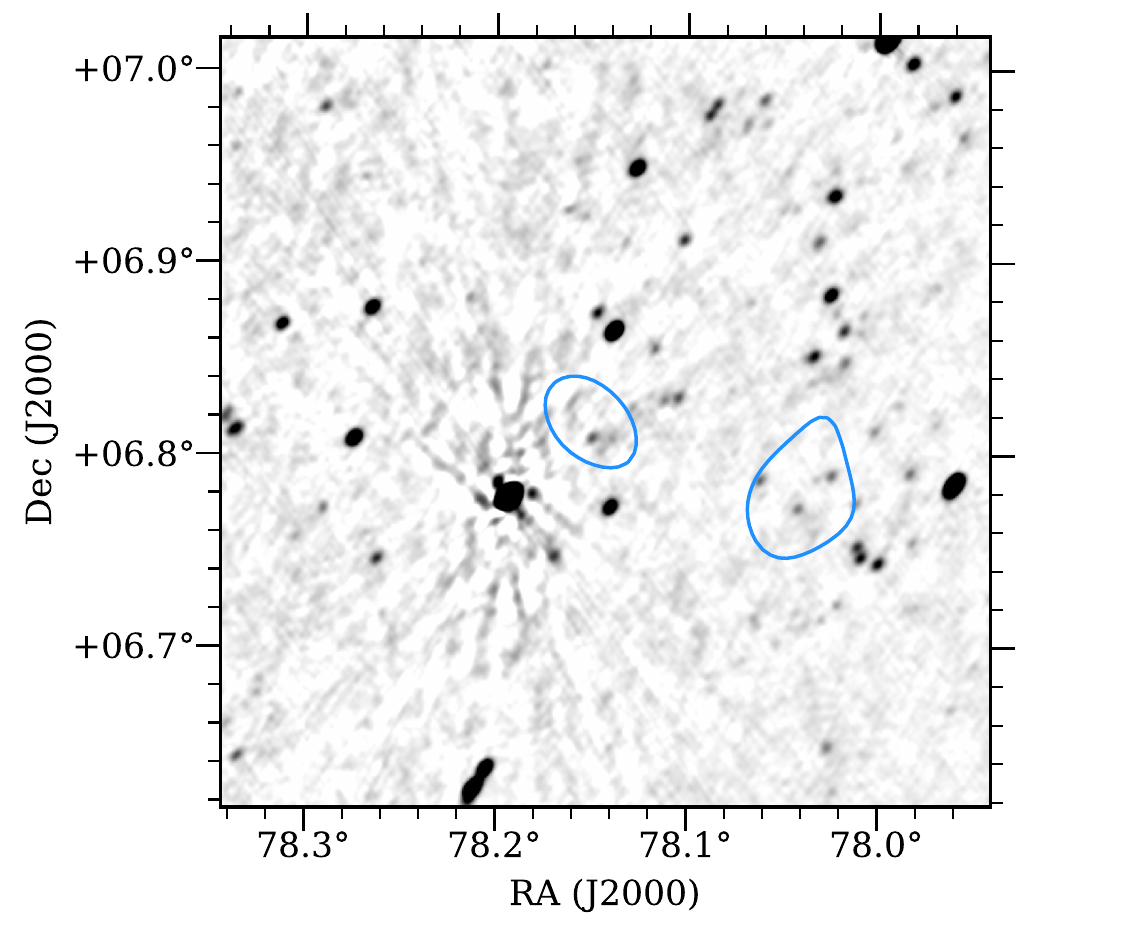}
    \end{subfigure}
    \begin{subfigure}{0.32\linewidth}
        \caption{C5}
        \includegraphics[width=1\linewidth,trim={0 -0.2cm 1cm 0.3cm},clip]{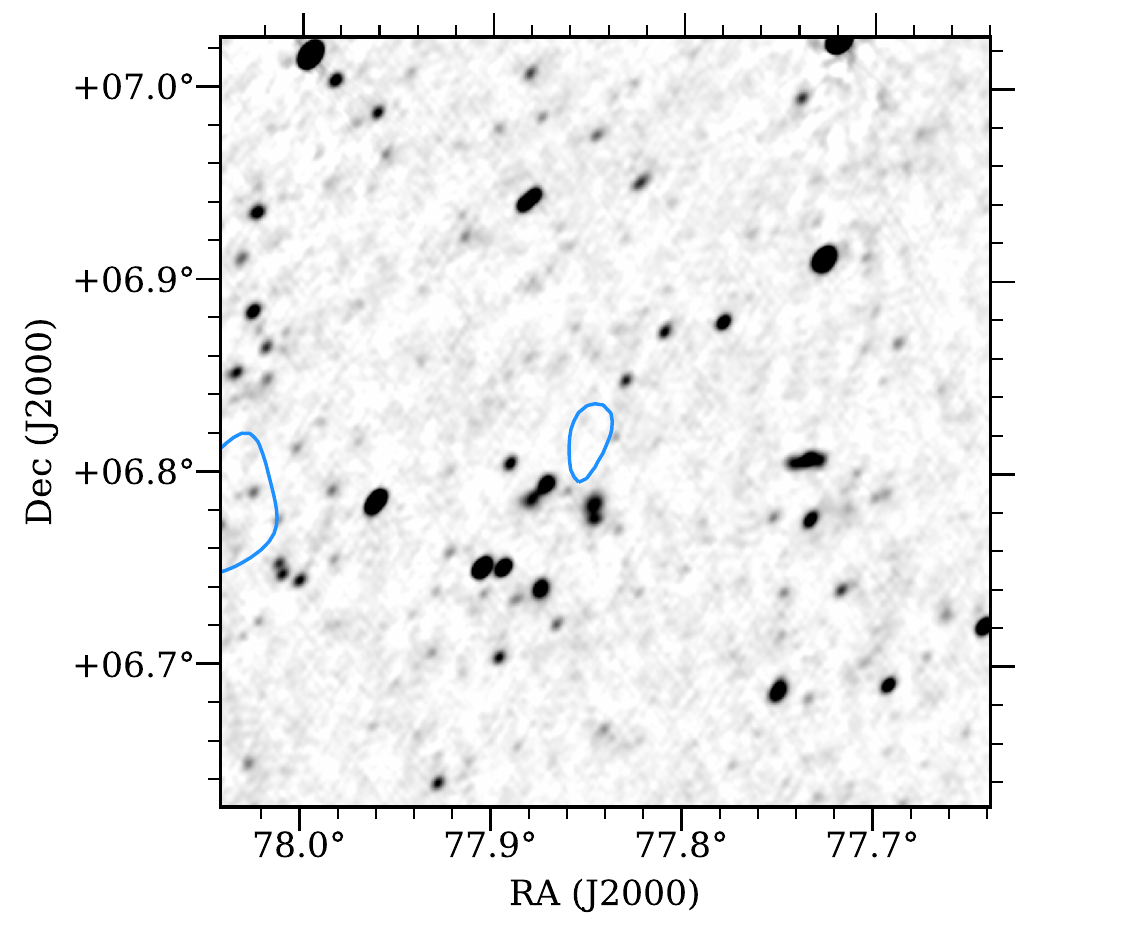}
    \end{subfigure}
\end{figure*}
\begin{figure*}\ContinuedFloat
    \centering
    \begin{subfigure}{0.32\linewidth}
        \caption{C6}
        \includegraphics[width=1\linewidth,trim={0 -0.2cm 1cm 0.3cm},clip]{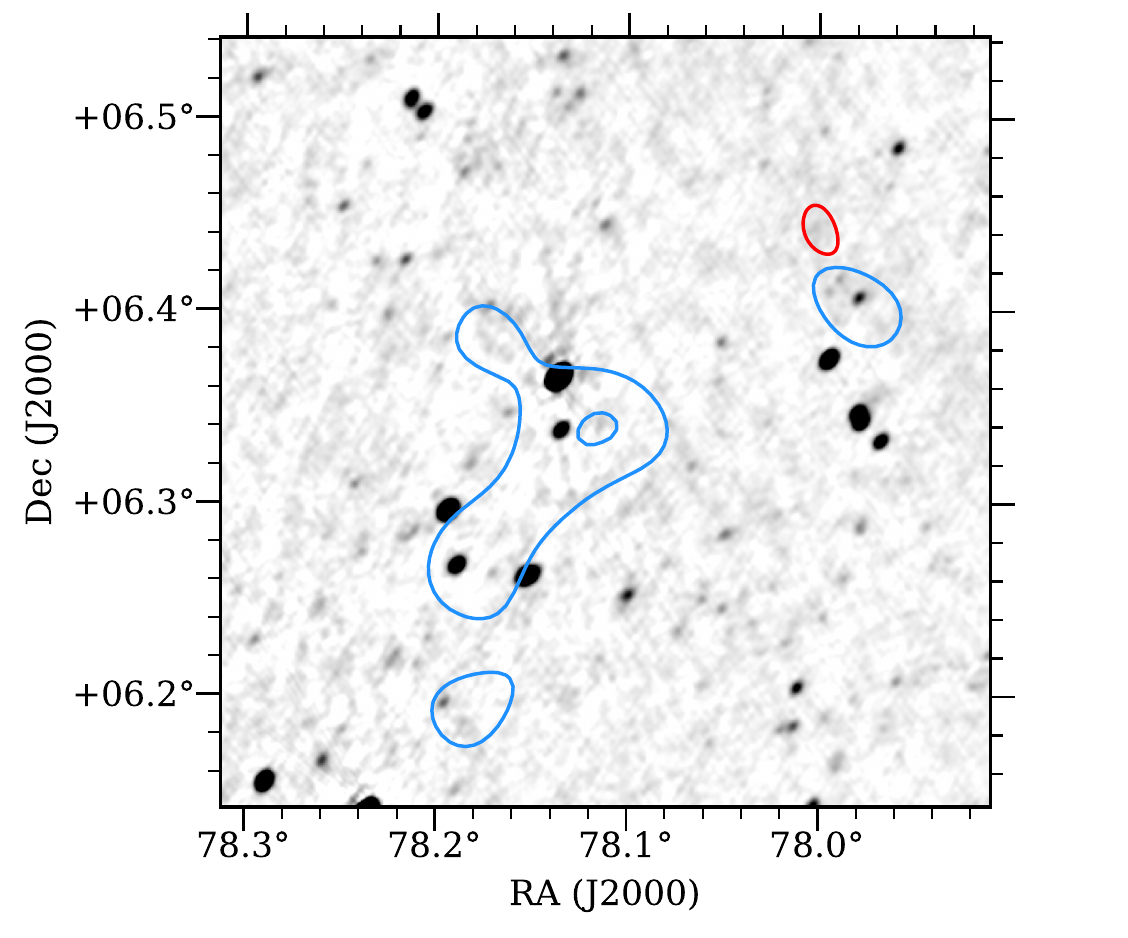}
    \end{subfigure}
    \begin{subfigure}{0.32\linewidth}
        \caption{C7}
        \includegraphics[width=1\linewidth,trim={0 -0.2cm 1cm 0.3cm},clip]{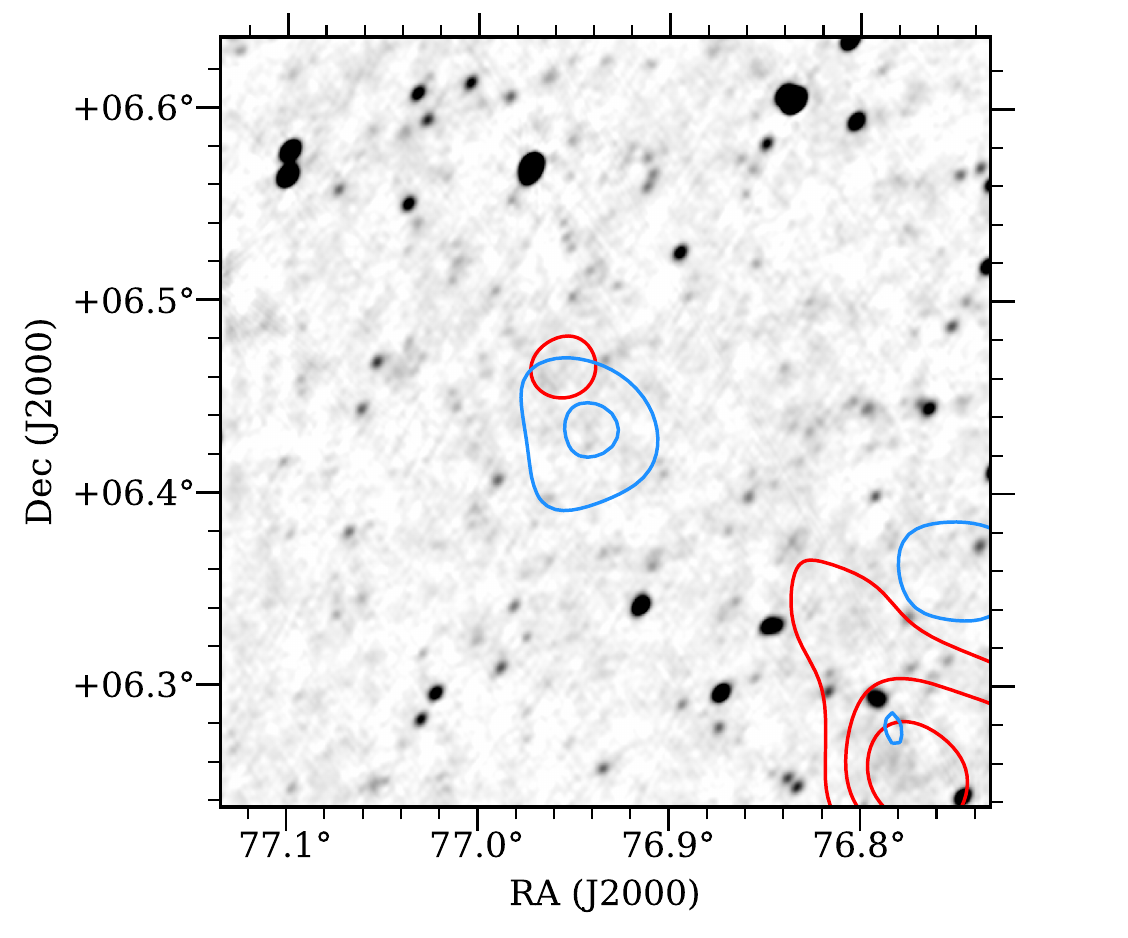}
    \end{subfigure}
    \begin{subfigure}{0.32\linewidth}
        \caption{C8}
        \includegraphics[width=1\linewidth,trim={0 -0.2cm 1cm 0.3cm},clip]{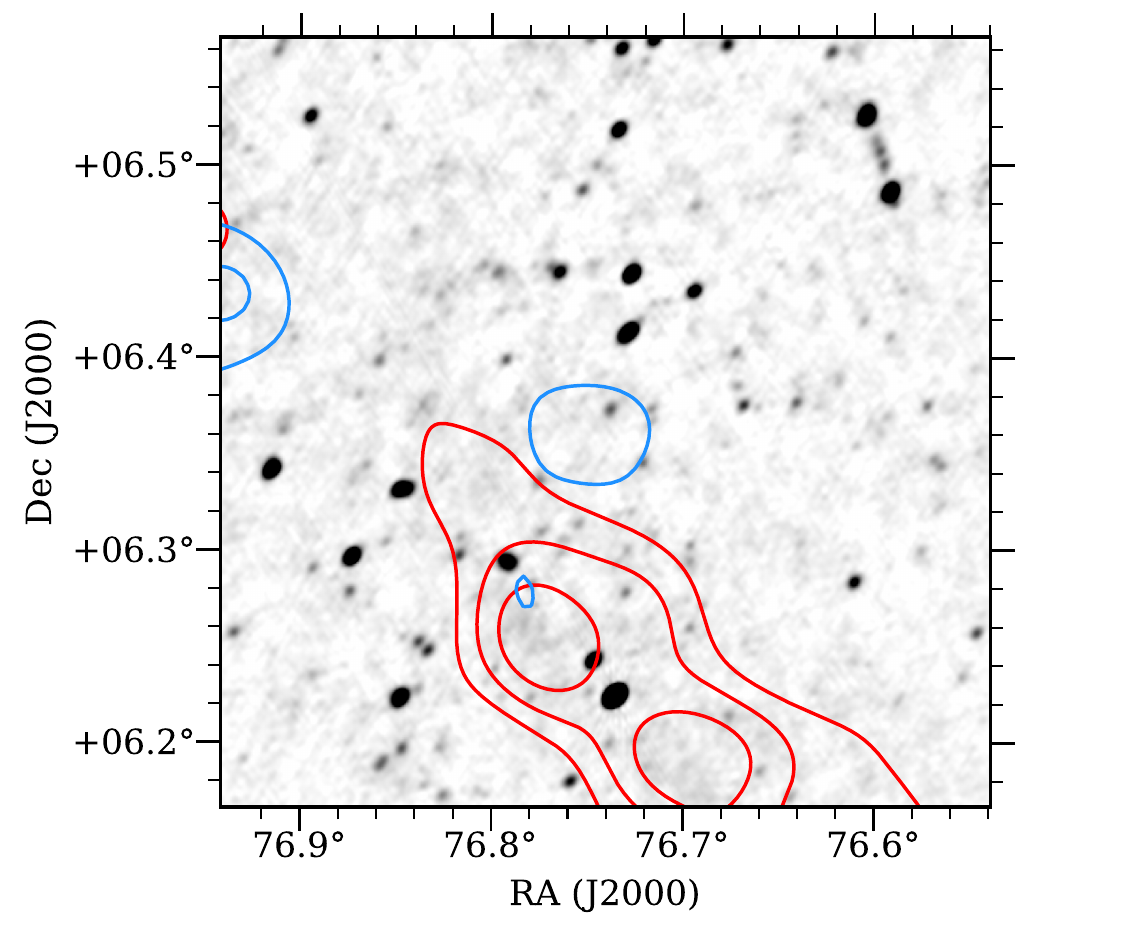}
    \end{subfigure}
    \begin{subfigure}{0.32\linewidth}
        \caption{C9}
        \includegraphics[width=1\linewidth,trim={0 -0.2cm 1cm 0.3cm},clip]{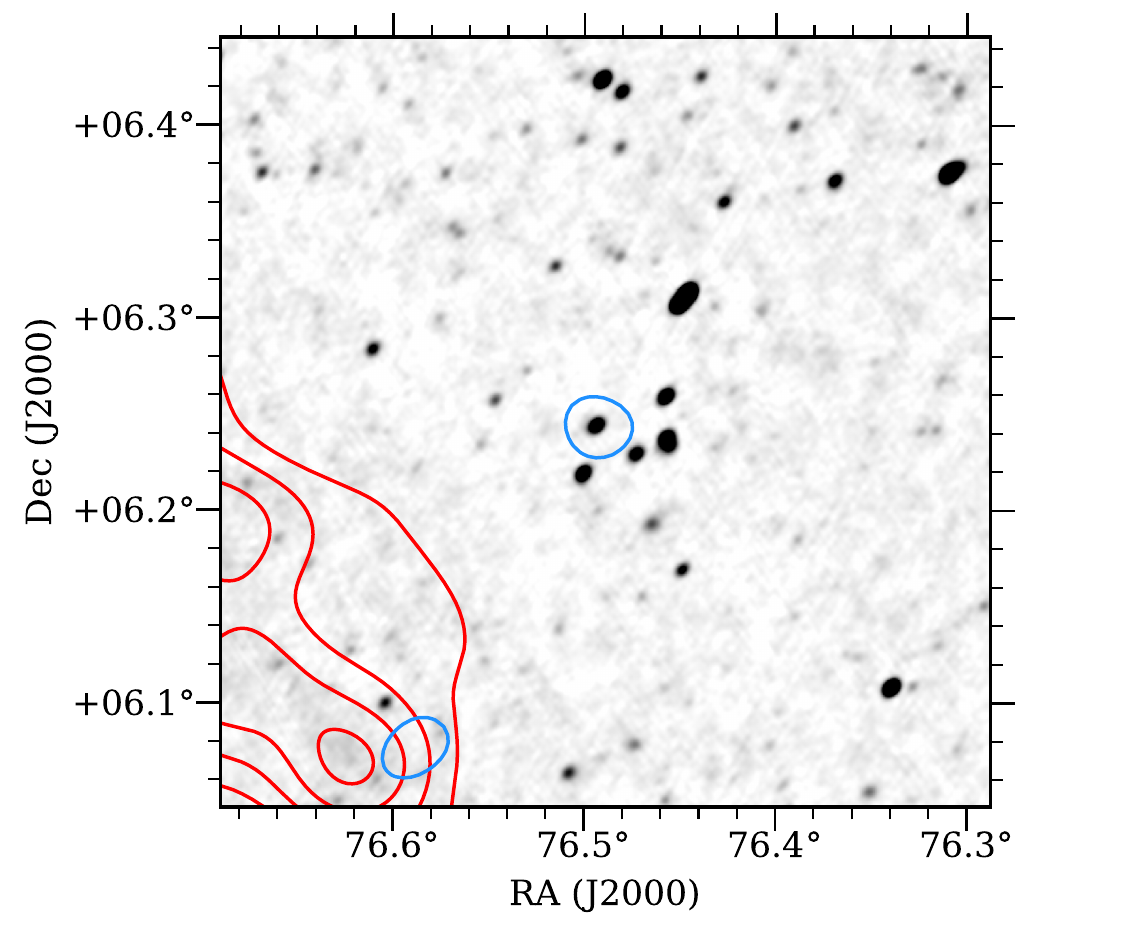}
    \end{subfigure}
    \begin{subfigure}{0.32\linewidth}
        \caption{C10}
        \includegraphics[width=1\linewidth,trim={0 -0.2cm 1cm 0.3cm},clip]{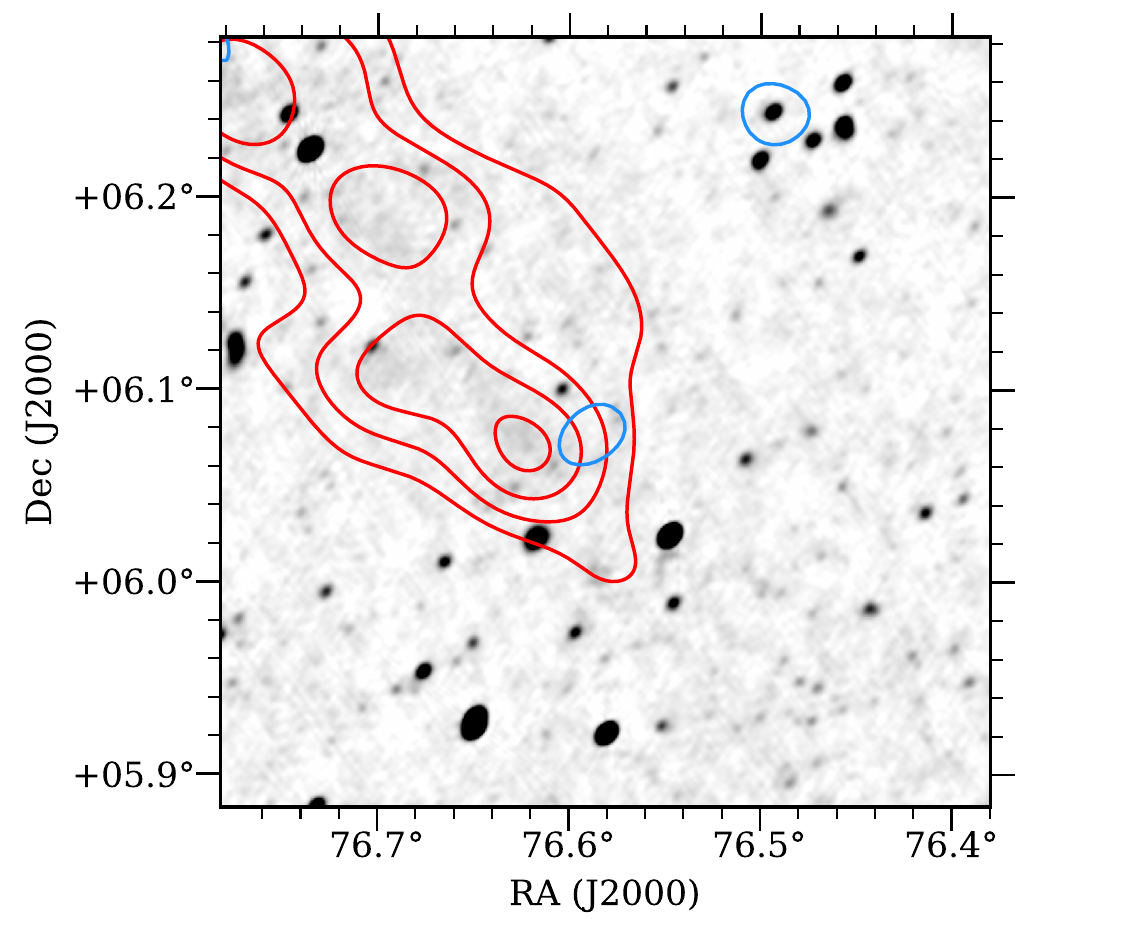}
    \end{subfigure}
    \begin{subfigure}{0.32\linewidth}
        \caption{D1}
        \includegraphics[width=1\linewidth,trim={0 -0.2cm 1cm 0.3cm},clip]{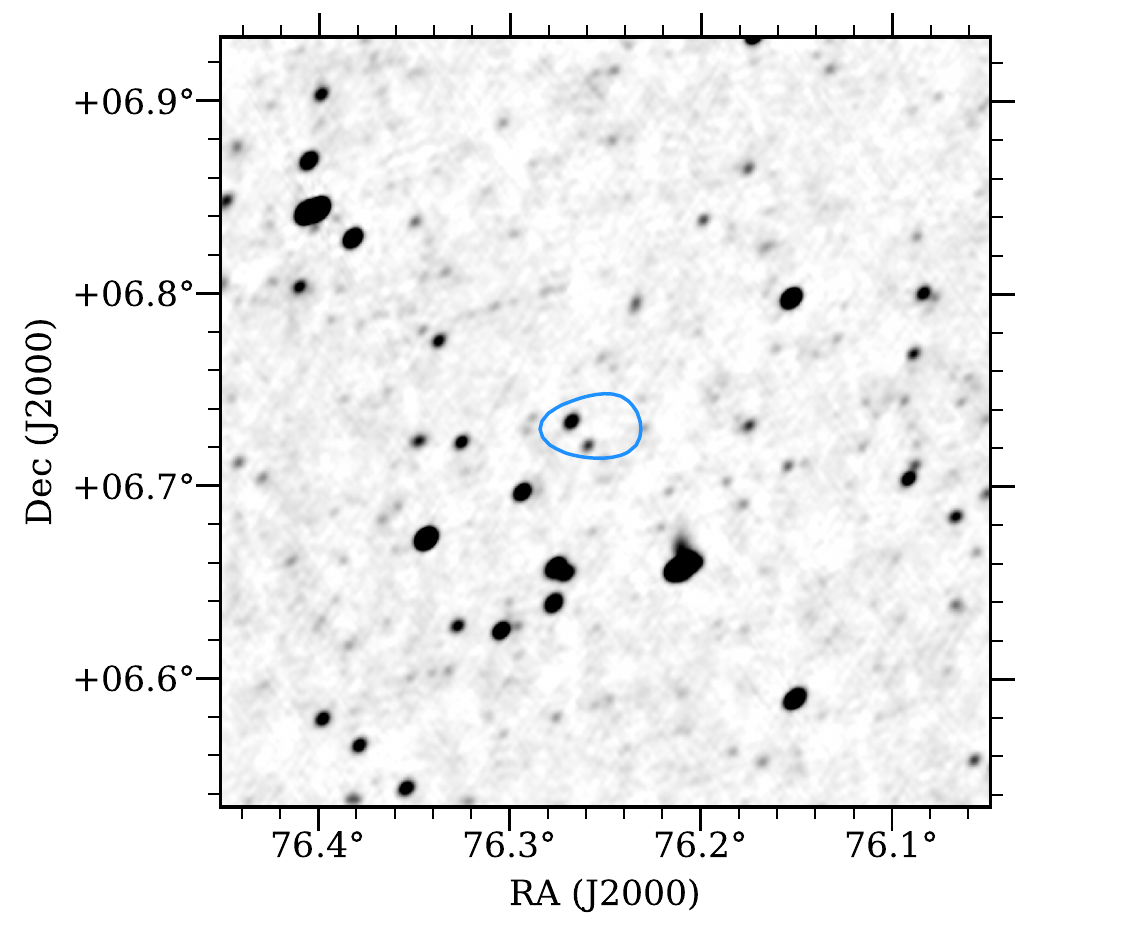}
    \end{subfigure}
    \begin{subfigure}{0.32\linewidth}
        \caption{D2}
        \includegraphics[width=1\linewidth,trim={0 -0.2cm 1cm 0.3cm},clip]{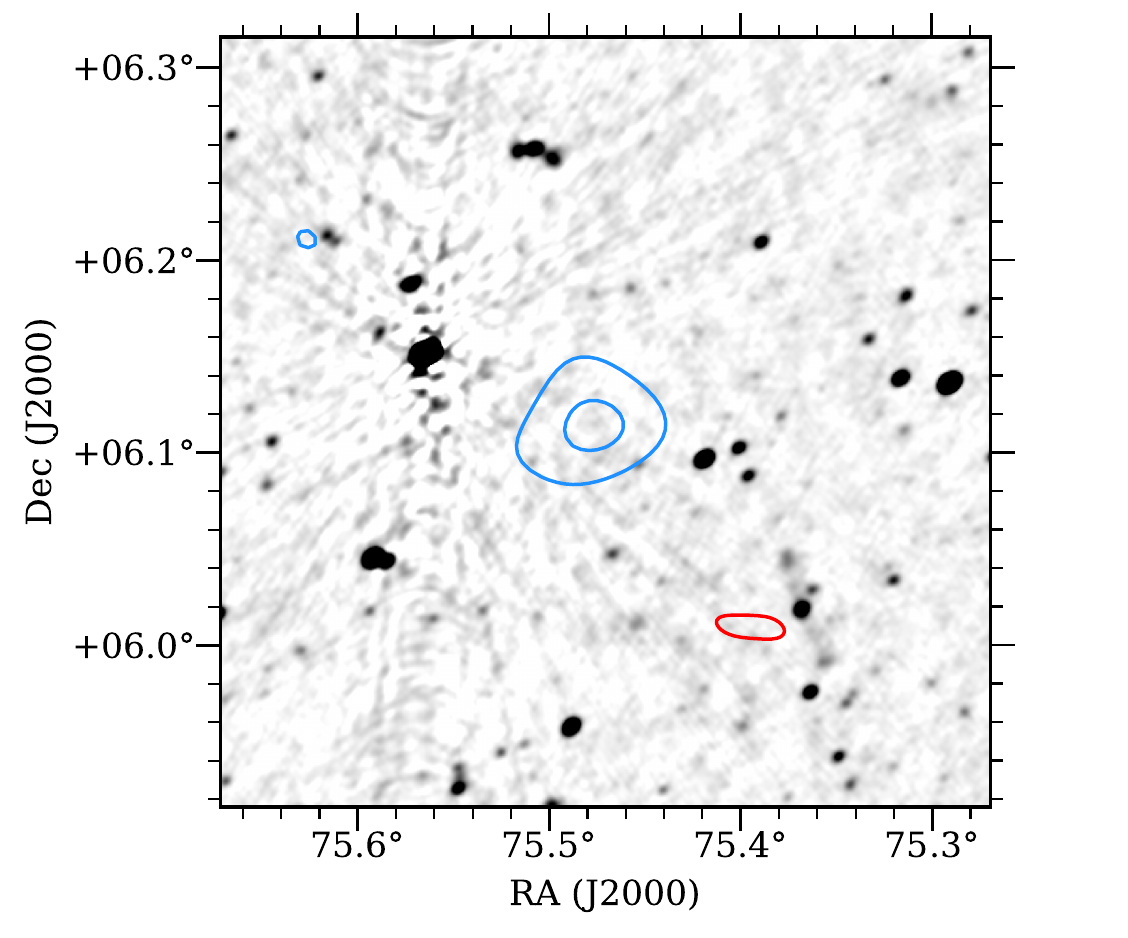}
    \end{subfigure}
    \begin{subfigure}{0.32\linewidth}
        \caption{D3}
        \includegraphics[width=1\linewidth,trim={0 -0.2cm 1cm 0.3cm},clip]{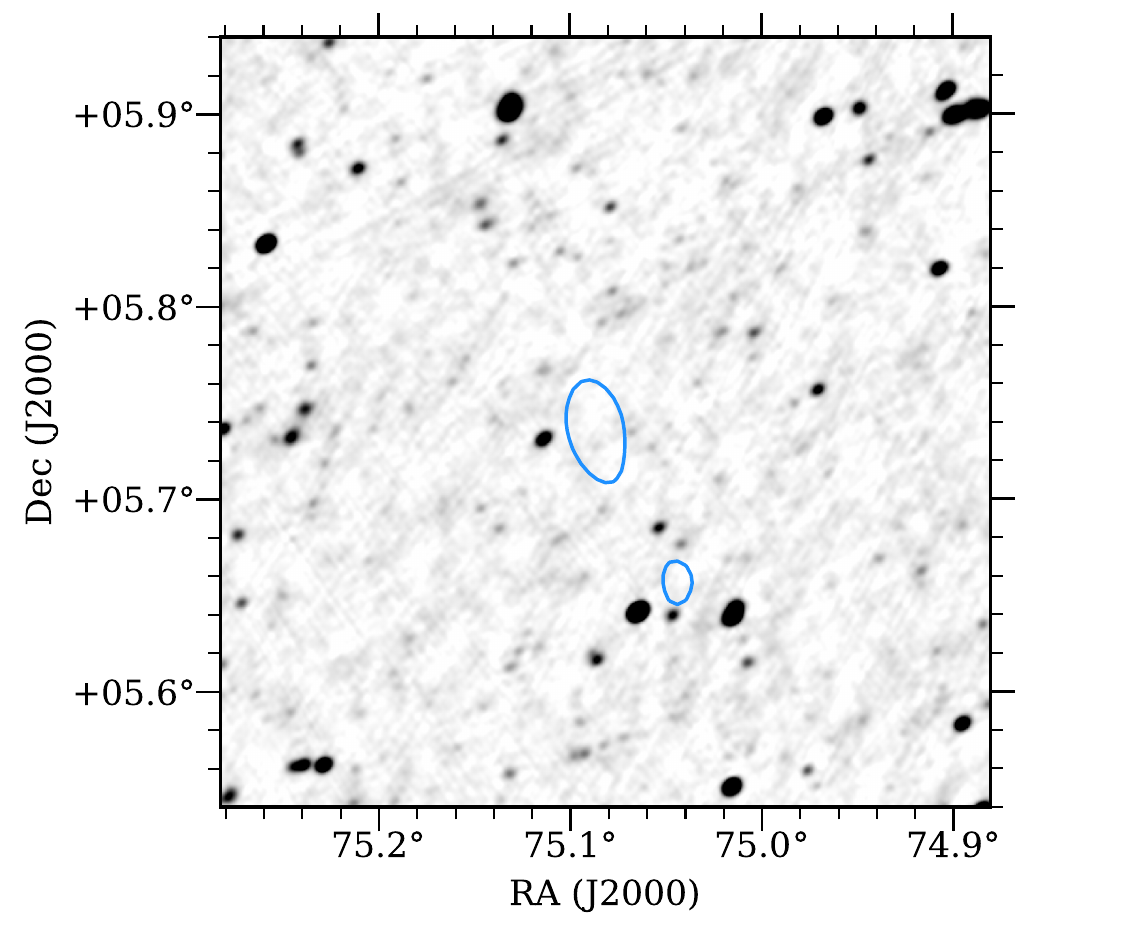}
    \end{subfigure}
    \begin{subfigure}{0.32\linewidth}
        \caption{E1}
        \includegraphics[width=1\linewidth,trim={0 -0.2cm 1cm 0.3cm},clip]{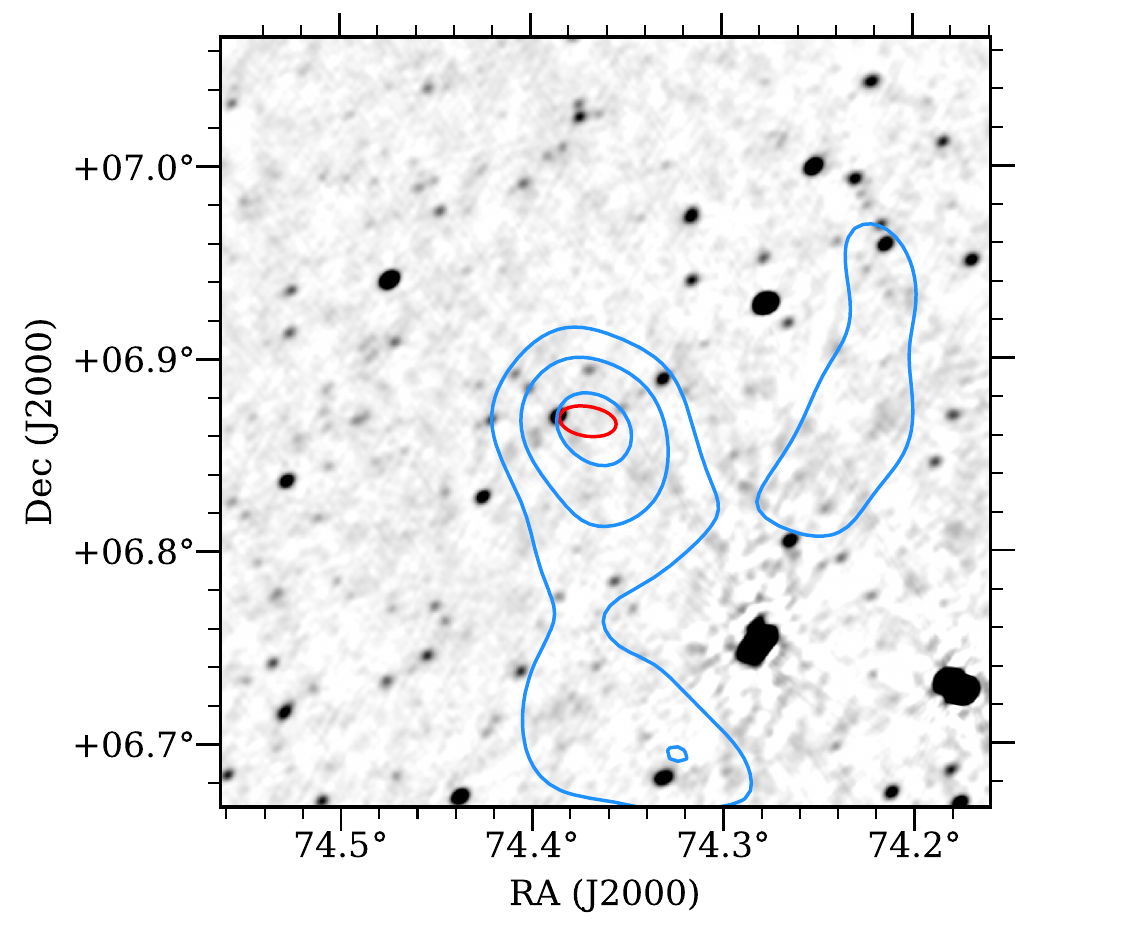}
    \end{subfigure}
    \begin{subfigure}{0.32\linewidth}
        \caption{E2}
        \includegraphics[width=1\linewidth,trim={0 -0.2cm 1cm 0.3cm},clip]{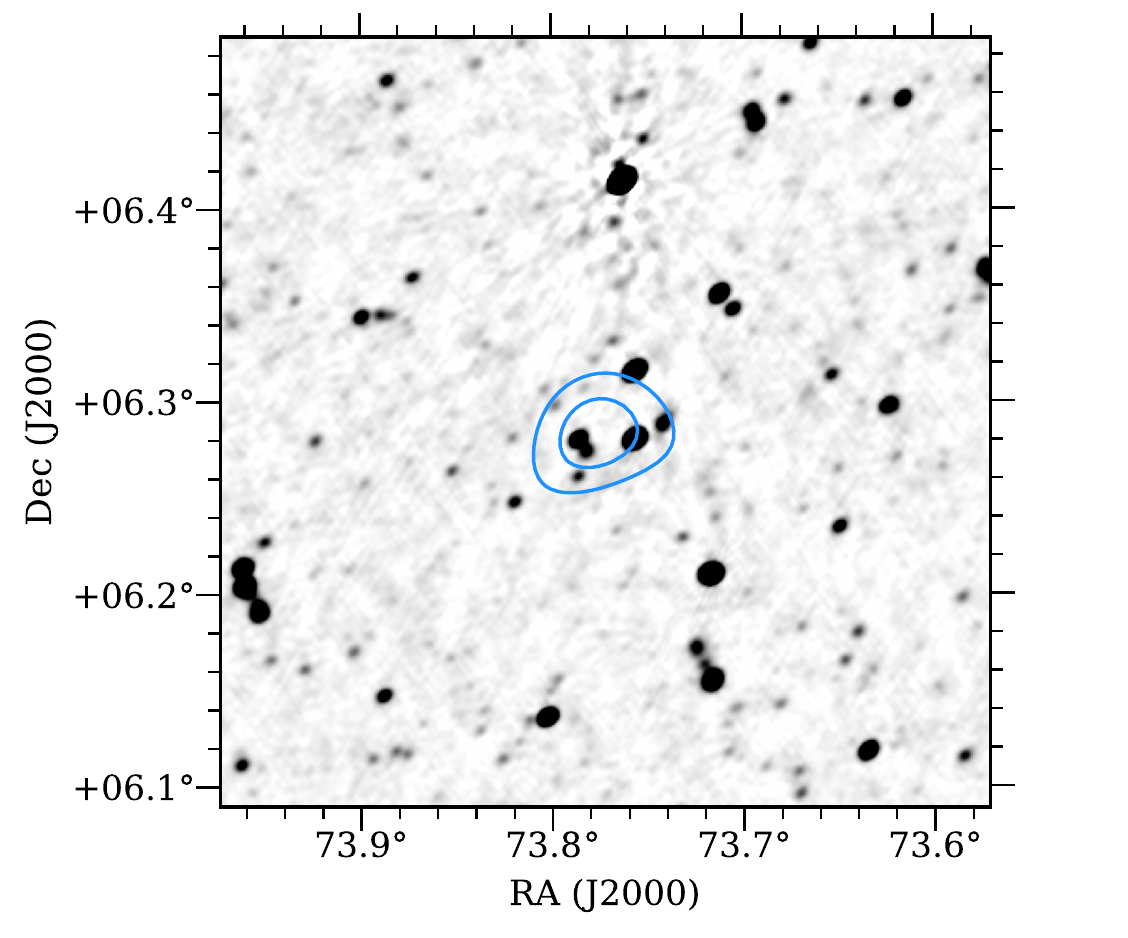}
    \end{subfigure}
    \begin{subfigure}{0.32\linewidth}
        \caption{E3}
        \includegraphics[width=1\linewidth,trim={0 -0.2cm 1cm 0.3cm},clip]{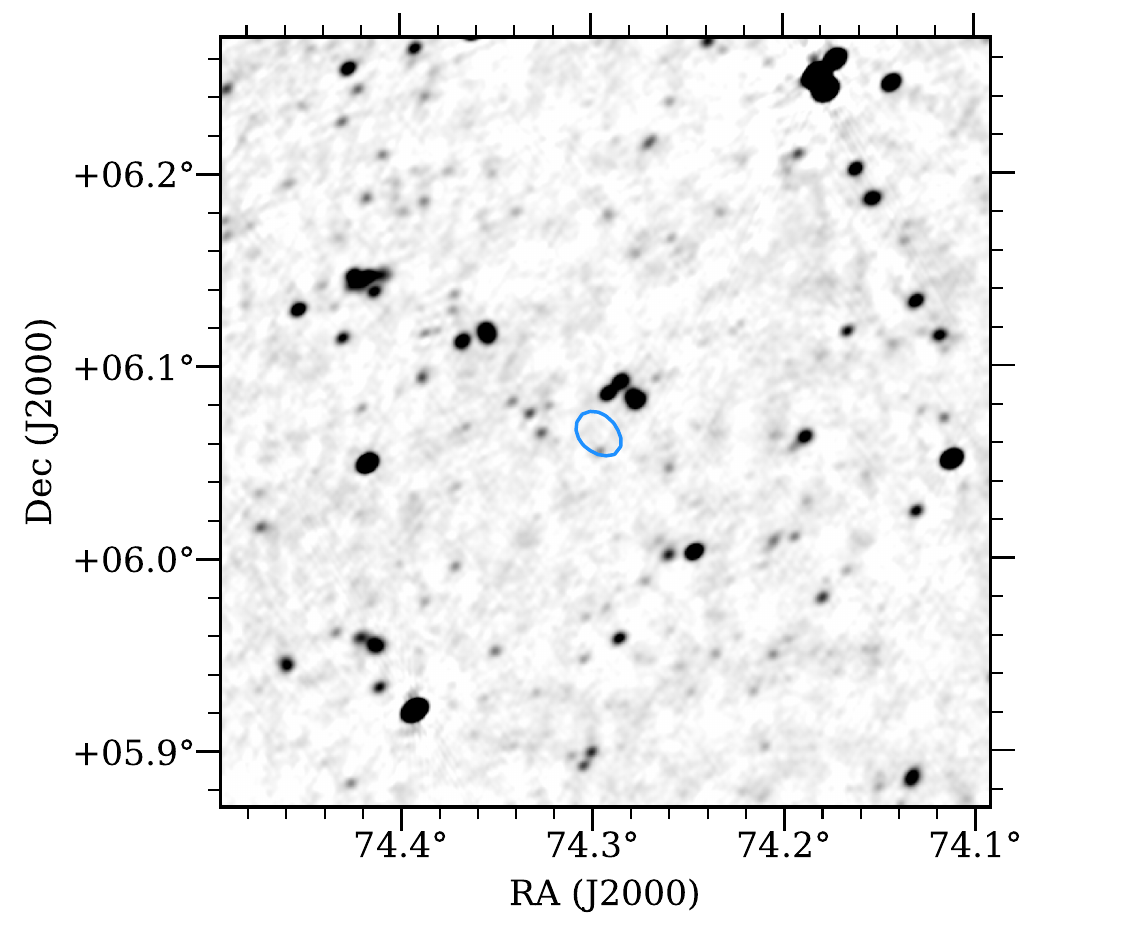}
    \end{subfigure}
    \begin{subfigure}{0.32\linewidth}
        \caption{F1}
        \includegraphics[width=1\linewidth,trim={0 -0.2cm 1cm 0.3cm},clip]{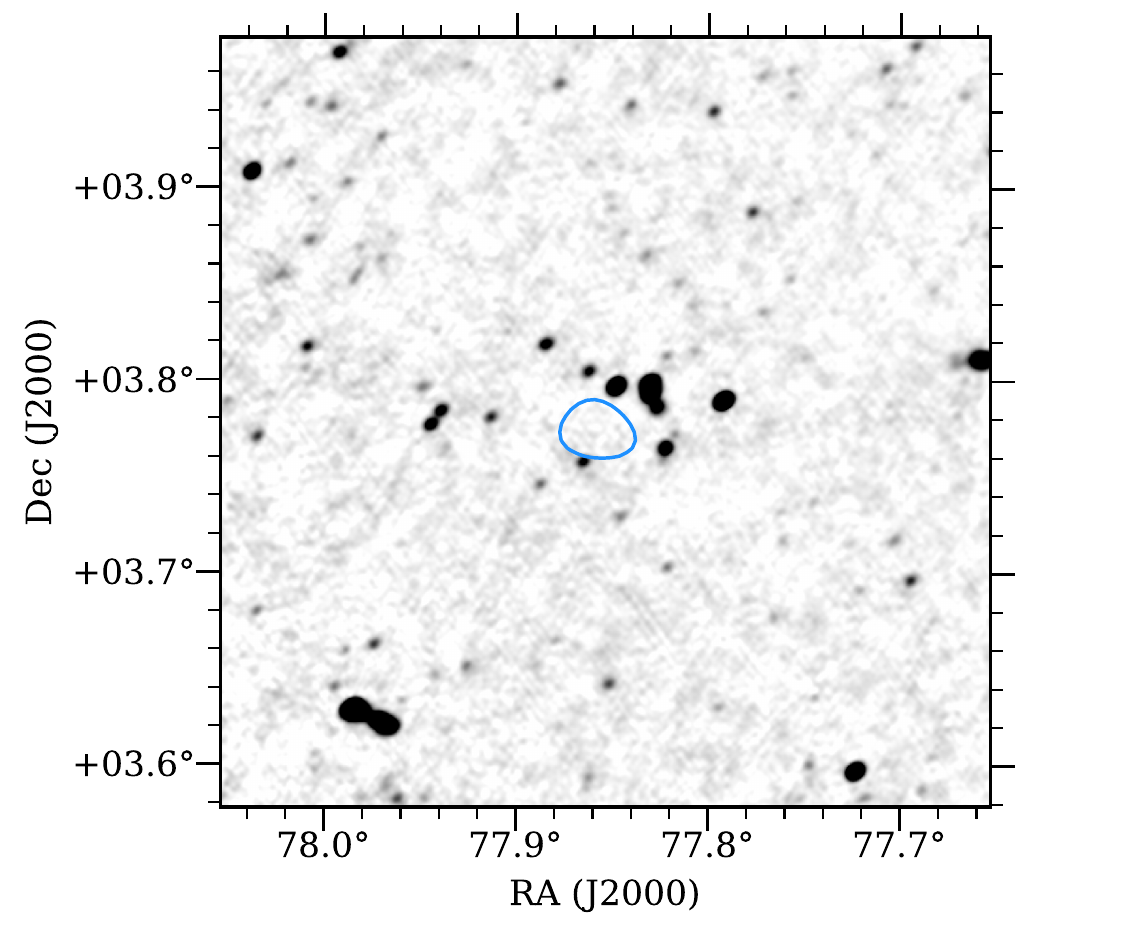}
    \end{subfigure}
\end{figure*}
\begin{figure*}\ContinuedFloat
    \centering
    \begin{subfigure}{0.32\linewidth}
        \caption{G1}
        \includegraphics[width=1\linewidth,trim={0 -0.2cm 1cm 0.3cm},clip]{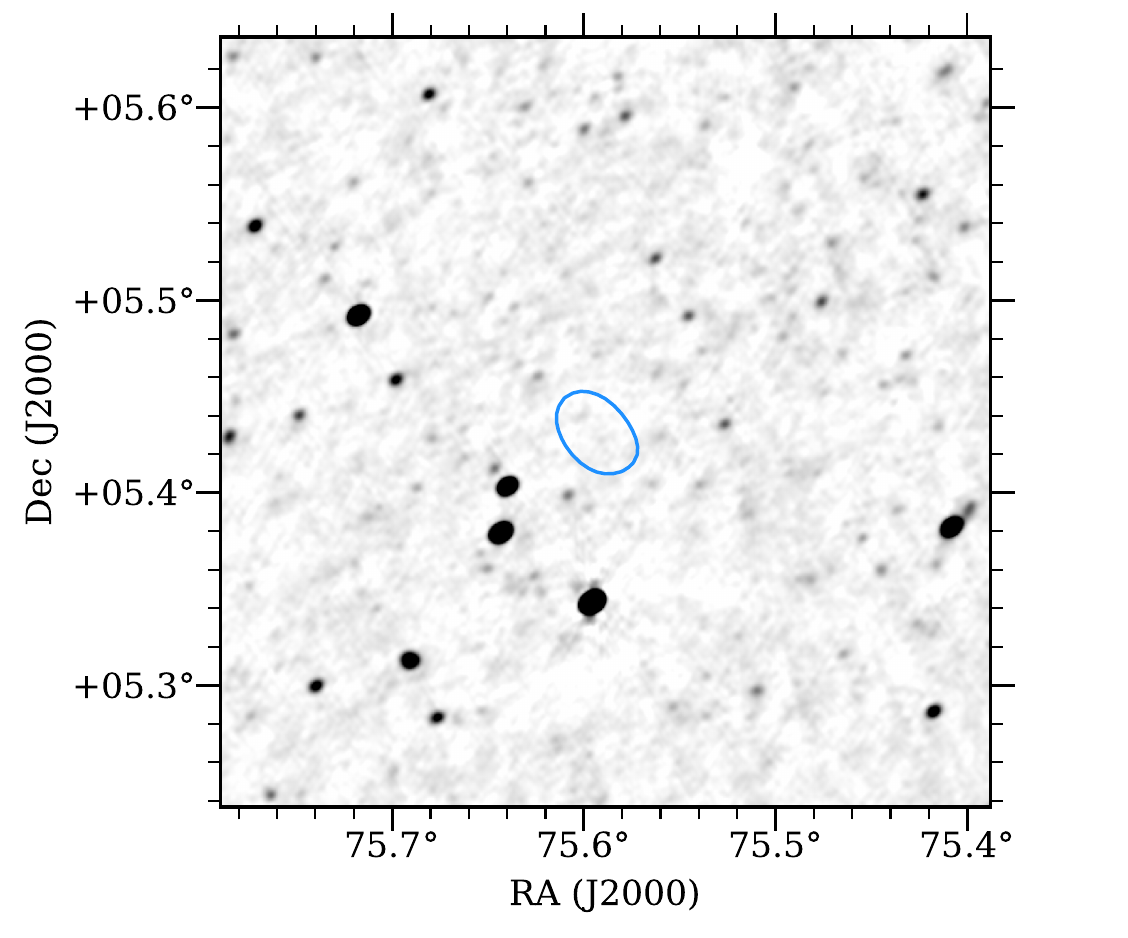}
    \end{subfigure}
    \begin{subfigure}{0.32\linewidth}
        \caption{G2}
        \includegraphics[width=1\linewidth,trim={0 -0.2cm 1cm 0.3cm},clip]{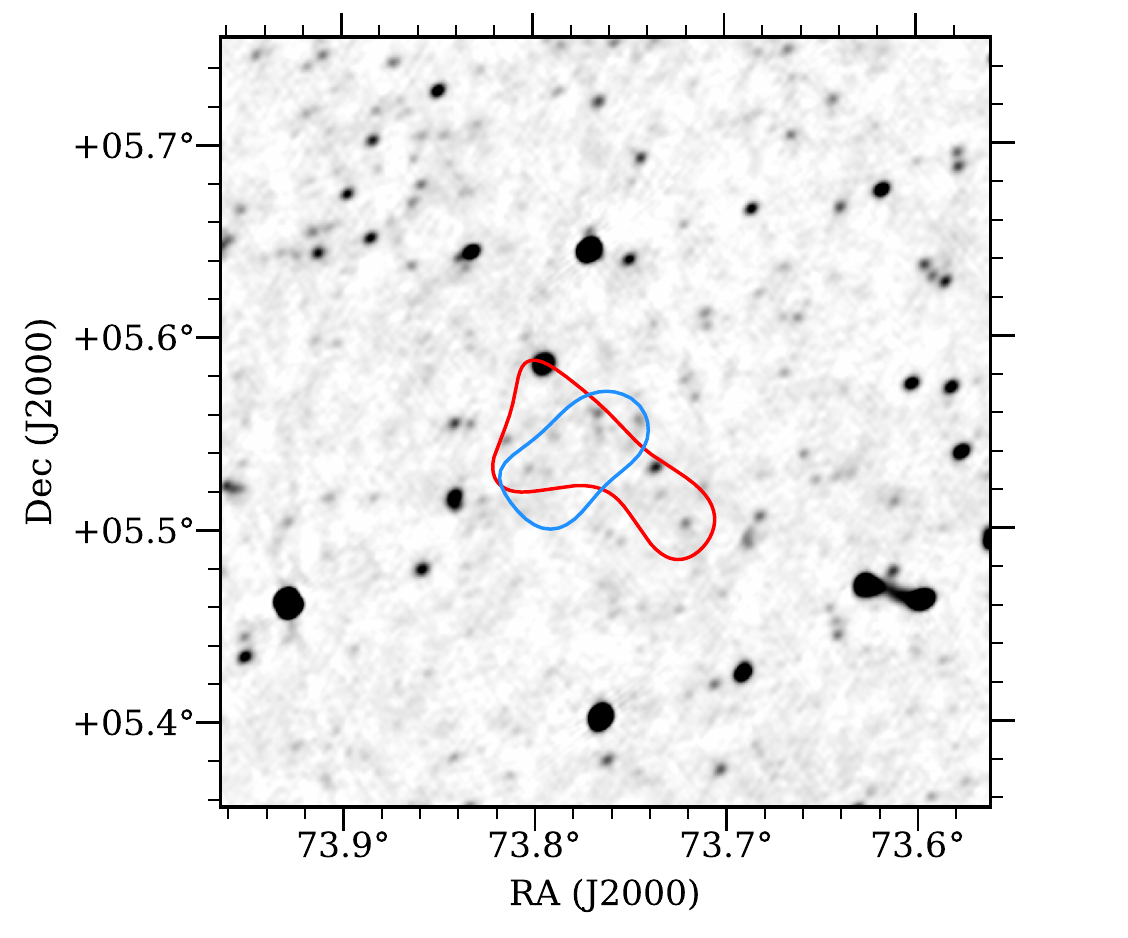}
    \end{subfigure}
    \begin{subfigure}{0.32\linewidth}
        \caption{G3}
        \includegraphics[width=1\linewidth,trim={0 -0.2cm 1cm 0.3cm},clip]{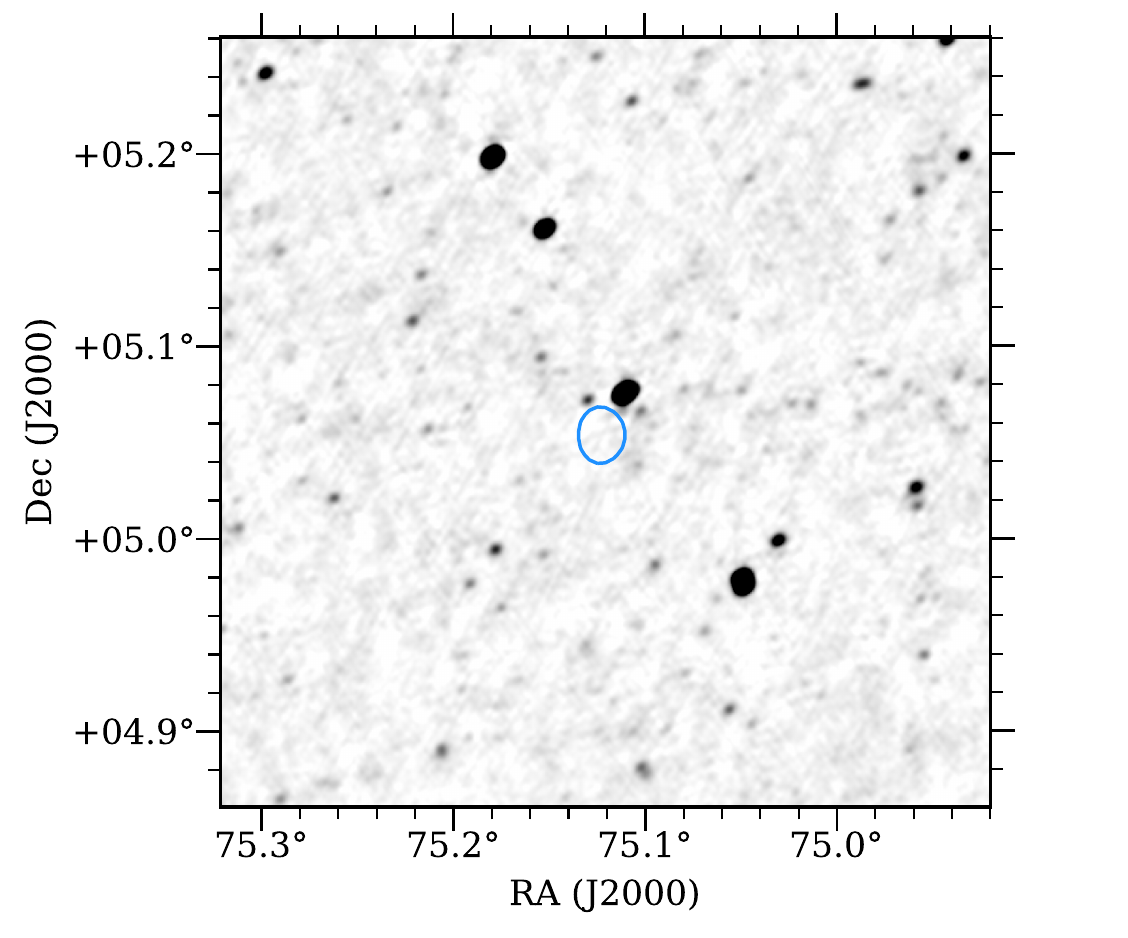}
    \end{subfigure}
    \begin{subfigure}{0.32\linewidth}
        \caption{G4}
        \includegraphics[width=1\linewidth,trim={0 -0.2cm 1cm 0.3cm},clip]{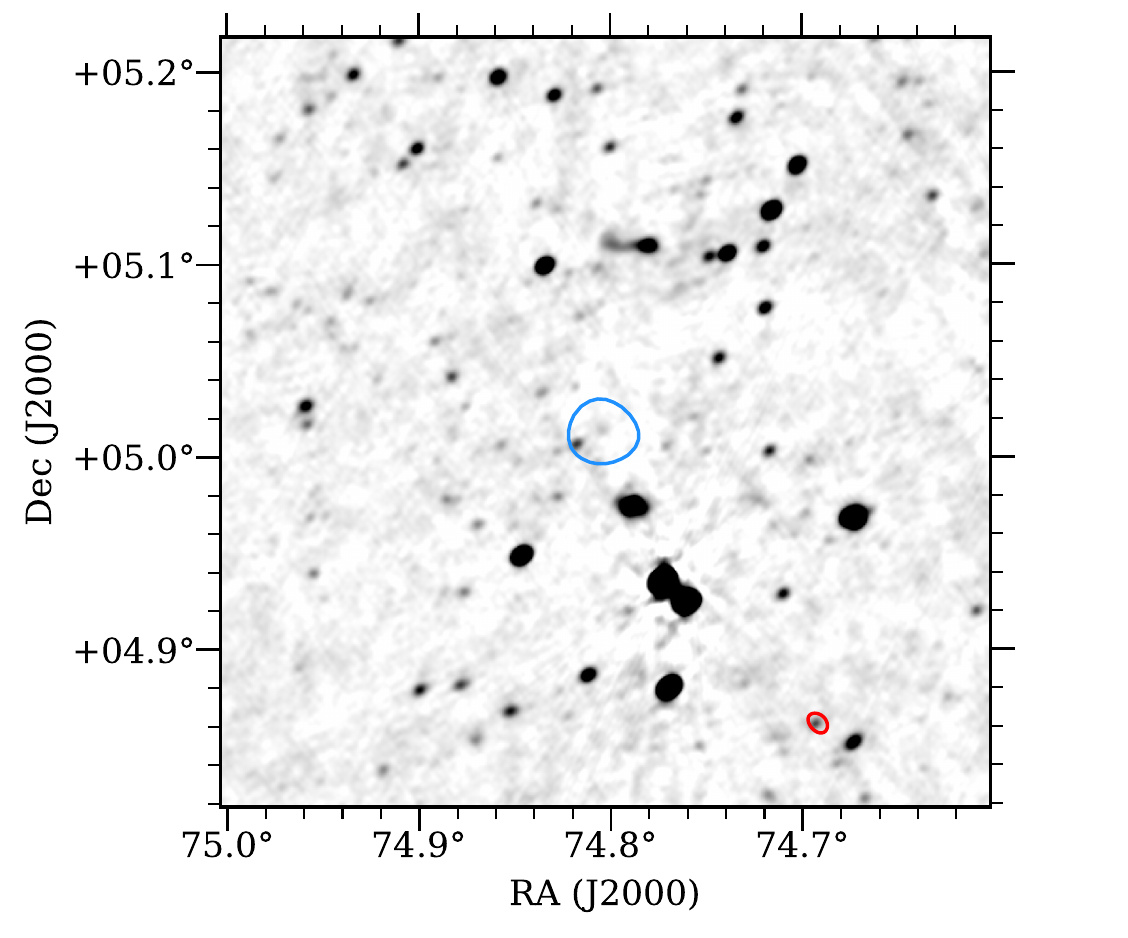}
    \end{subfigure}
    \begin{subfigure}{0.32\linewidth}
        \caption{G5}
        \includegraphics[width=1\linewidth,trim={0 -0.2cm 1cm 0.3cm},clip]{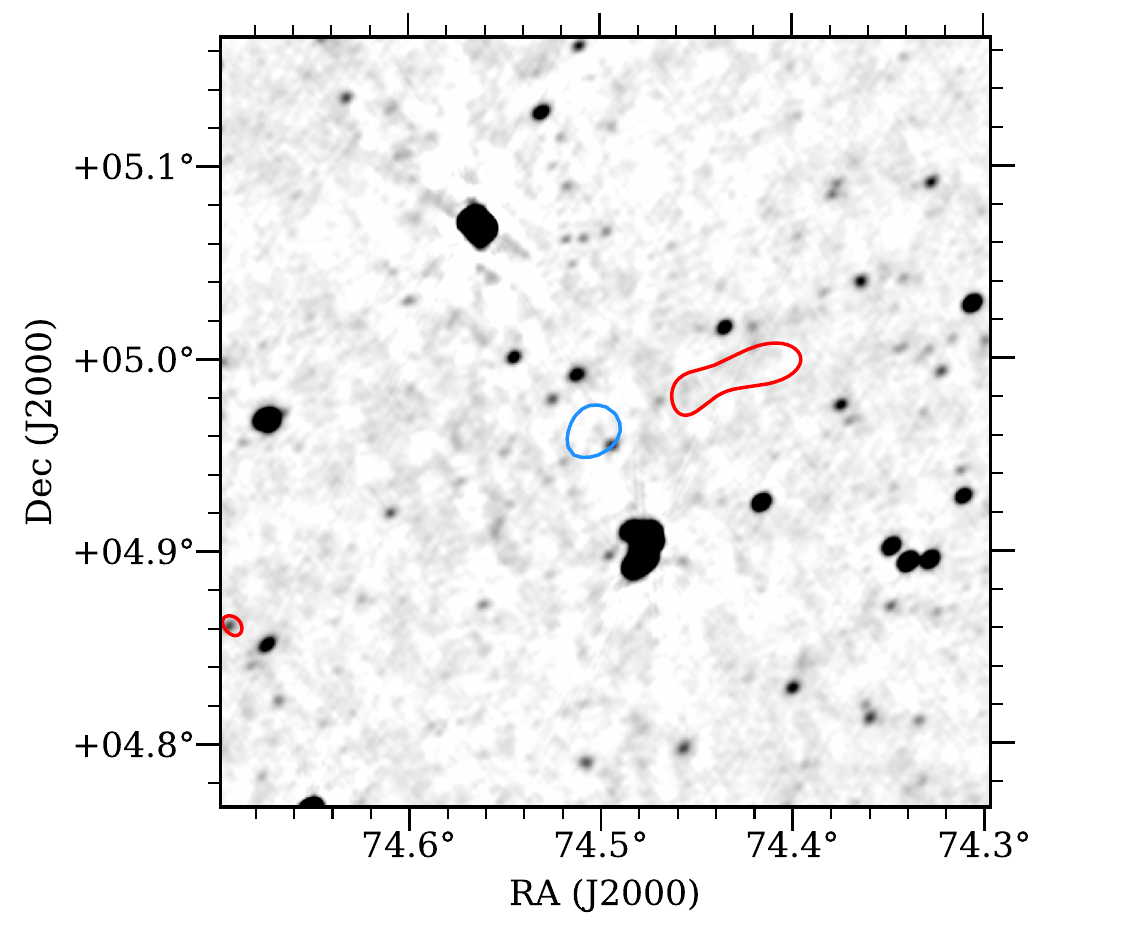}
    \end{subfigure}
    \begin{subfigure}{0.32\linewidth}
        \caption{G6}
        \includegraphics[width=1\linewidth,trim={0 -0.2cm 1cm 0.3cm},clip]{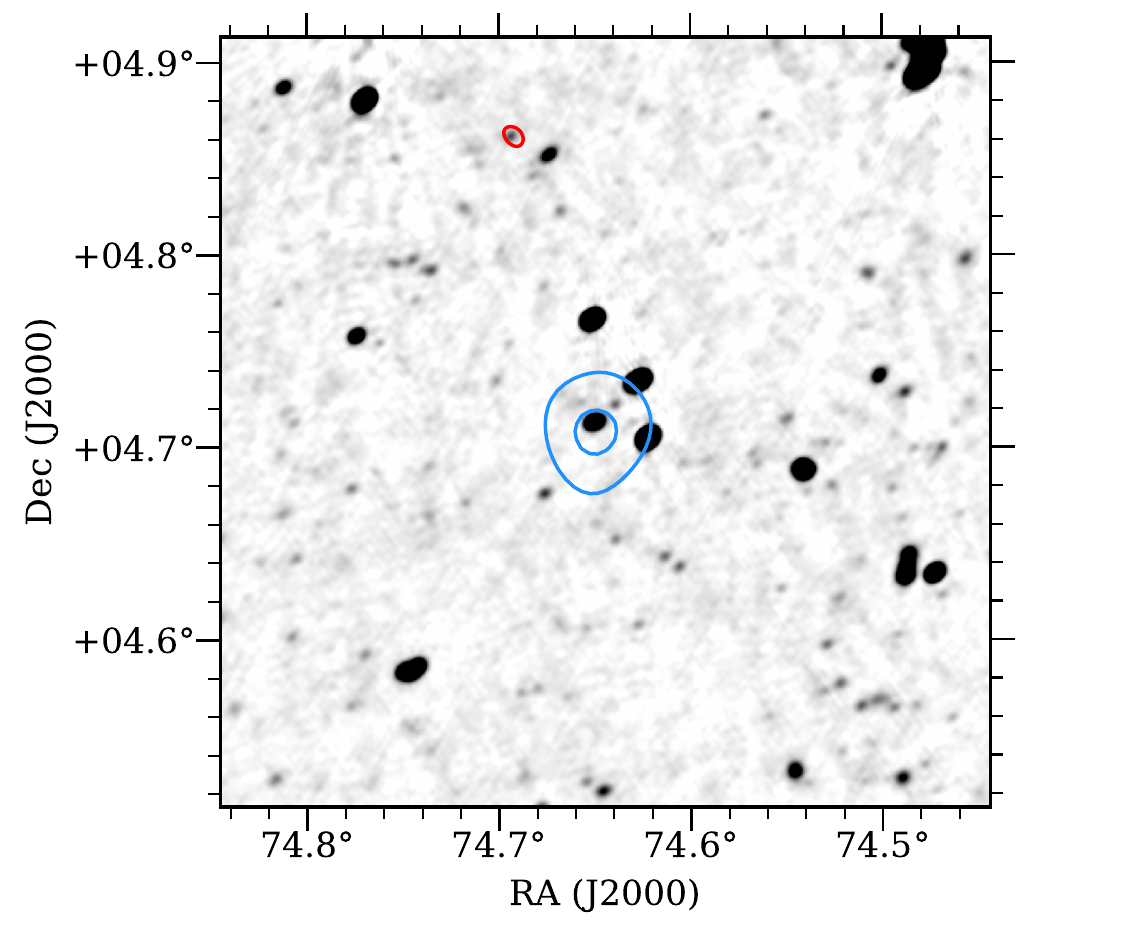}
    \end{subfigure}
    \begin{subfigure}{0.32\linewidth}
        \caption{H1}
        \includegraphics[width=1\linewidth,trim={0 -0.2cm 1cm 0.3cm},clip]{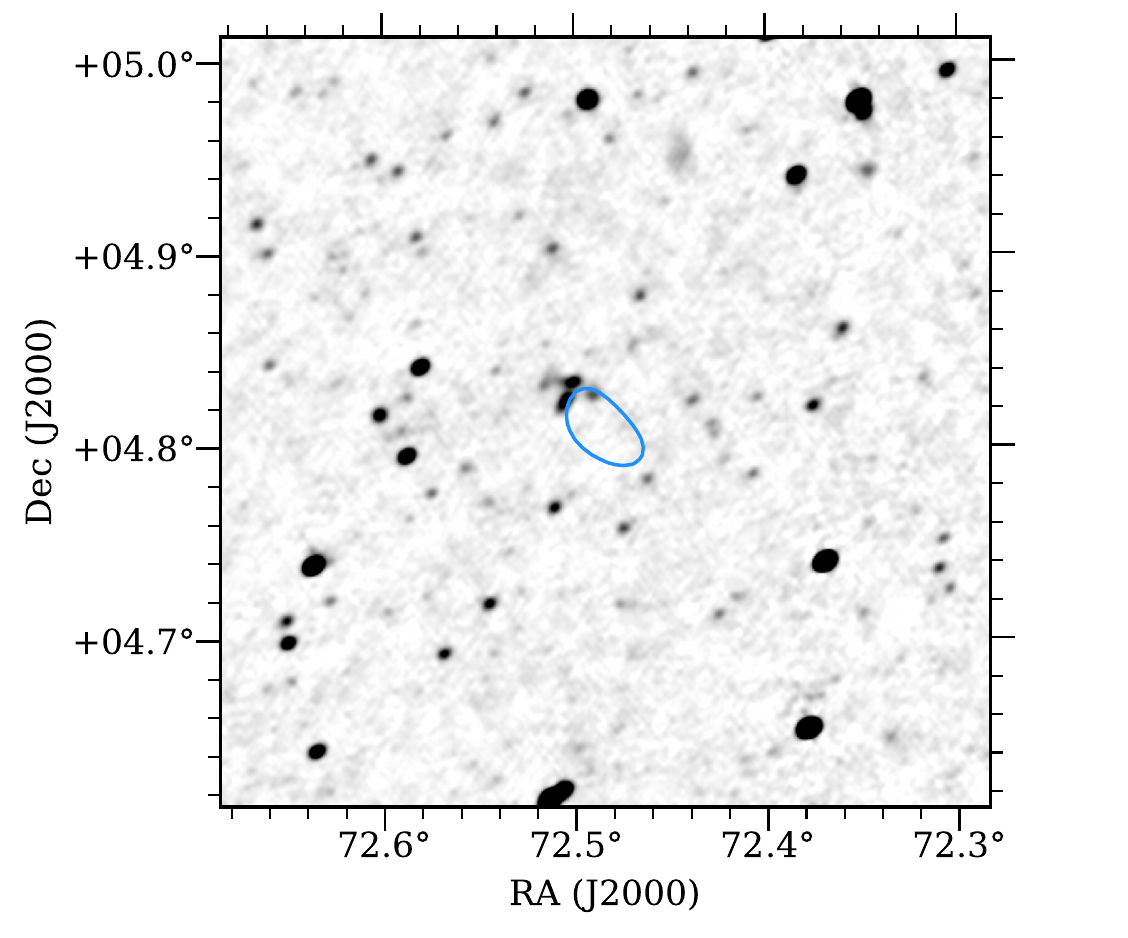}
    \end{subfigure}
    \begin{subfigure}{0.32\linewidth}
        \caption{H2}
        \includegraphics[width=1\linewidth,trim={0 -0.2cm 1cm 0.3cm},clip]{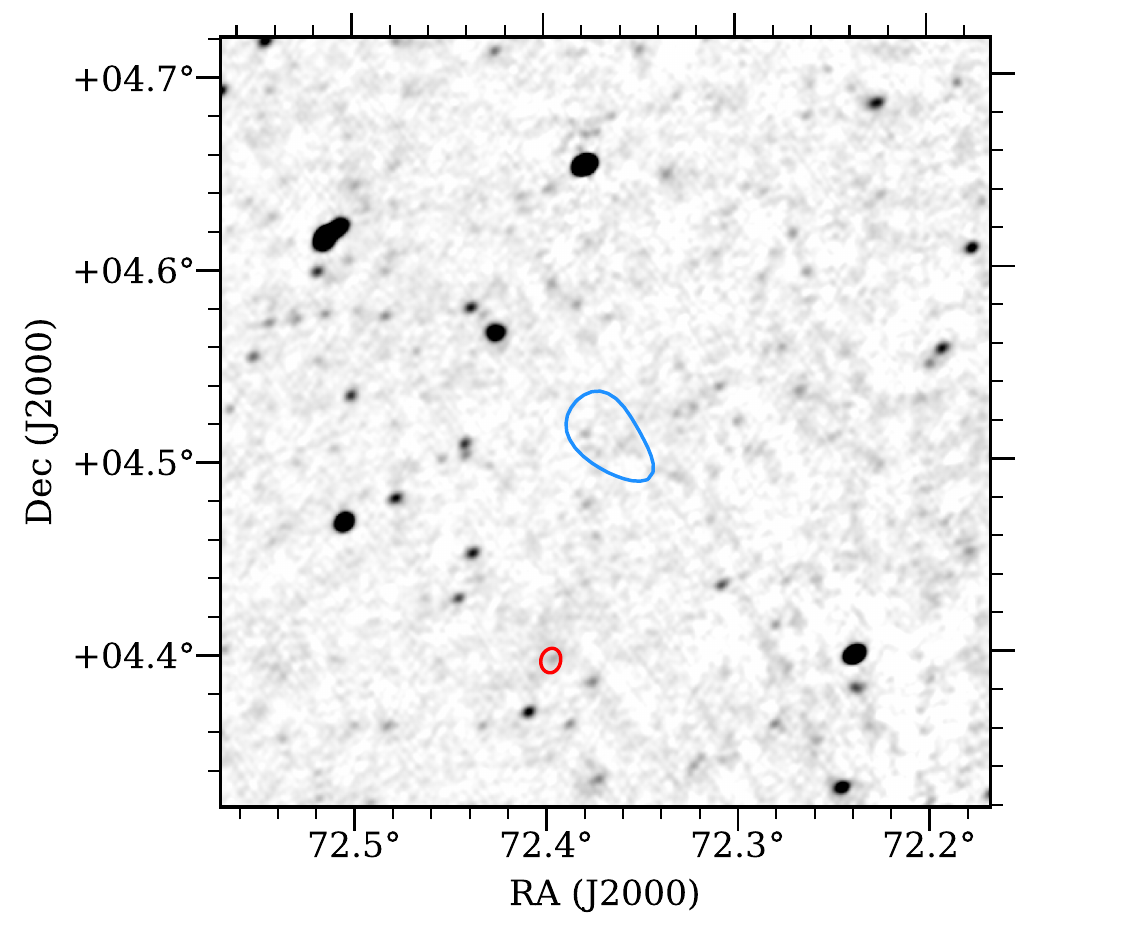}
    \end{subfigure}
    \begin{subfigure}{0.32\linewidth}
        \caption{I1}
        \includegraphics[width=1\linewidth,trim={0 -0.2cm 1cm 0.3cm},clip]{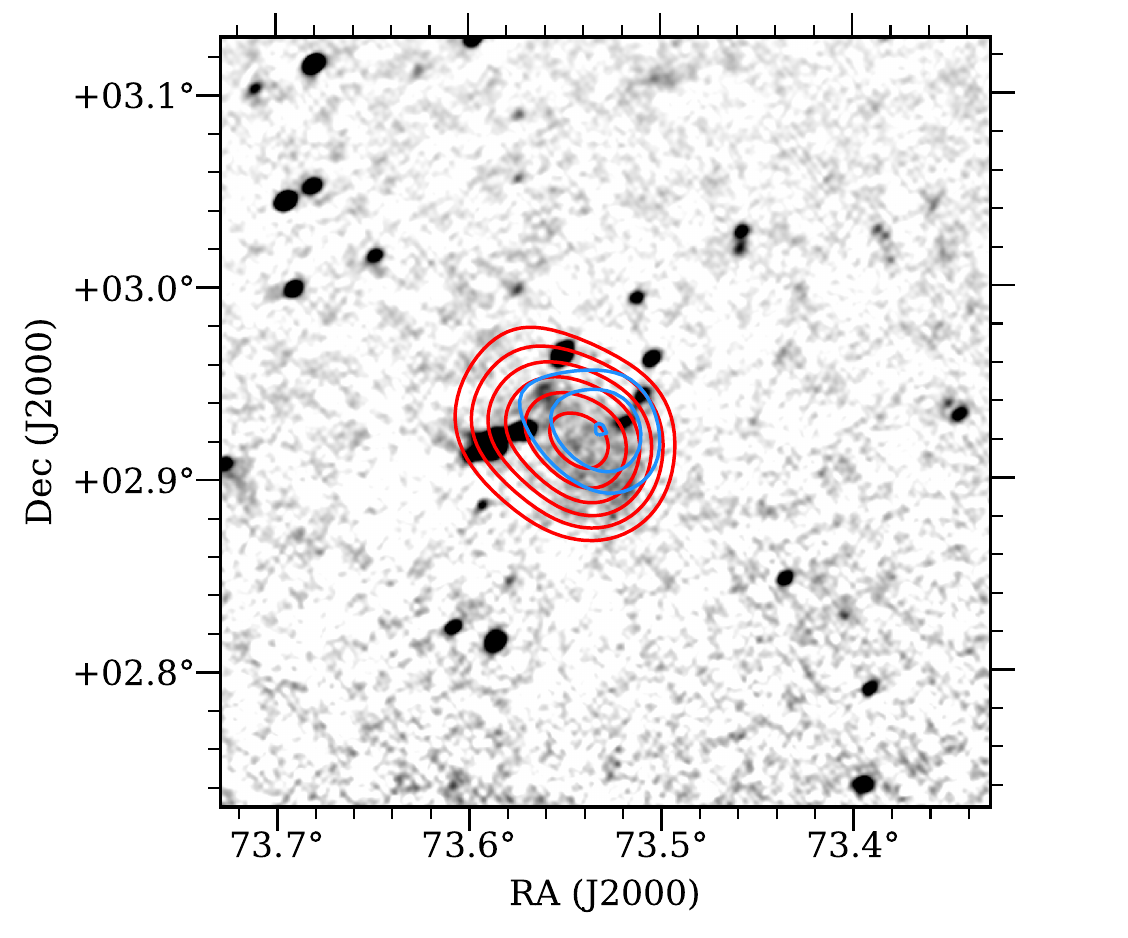}
    \end{subfigure}
    \begin{subfigure}{0.32\linewidth}
        \caption{I2}
        \includegraphics[width=1\linewidth,trim={0 -0.2cm 1cm 0.3cm},clip]{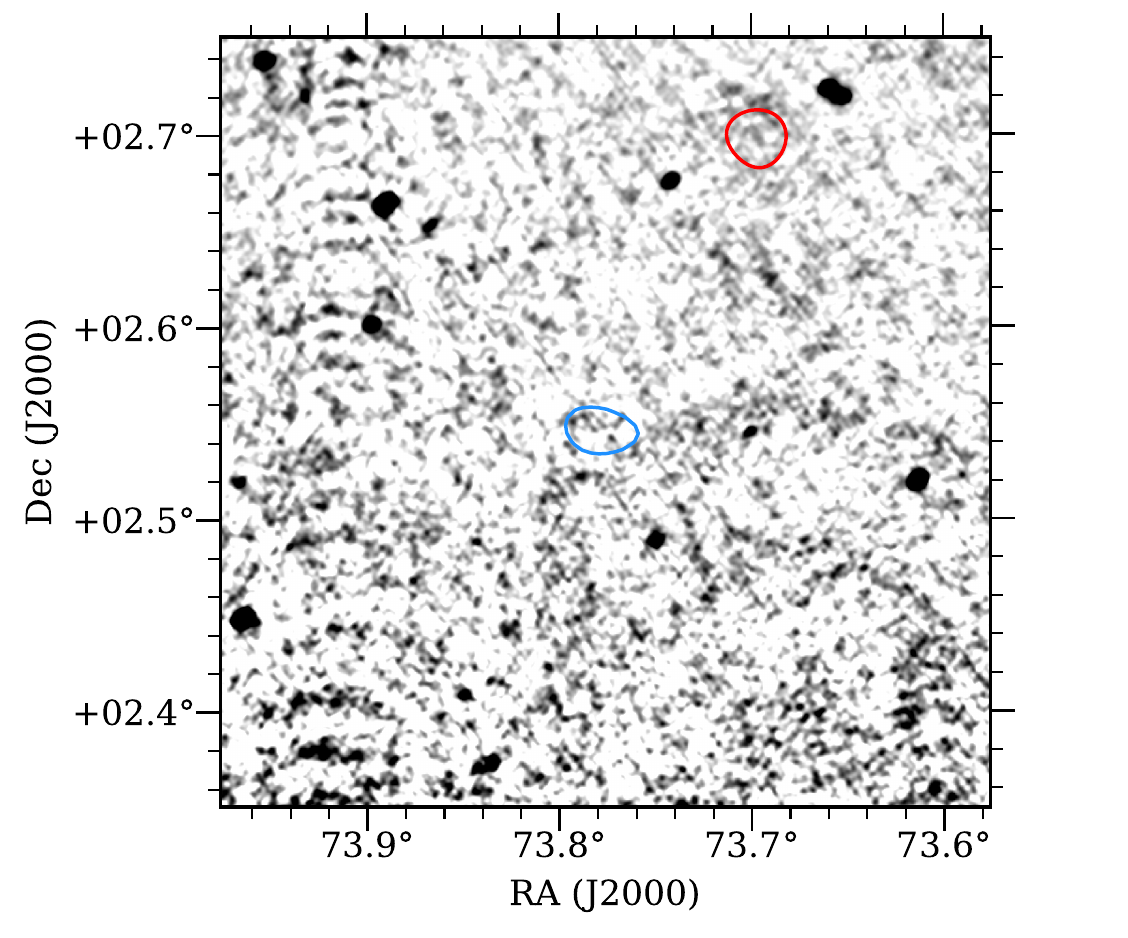}
    \end{subfigure}
    \begin{subfigure}{0.32\linewidth}
        \caption{I3}
        \includegraphics[width=1\linewidth,trim={0 -0.2cm 1cm 0.3cm},clip]{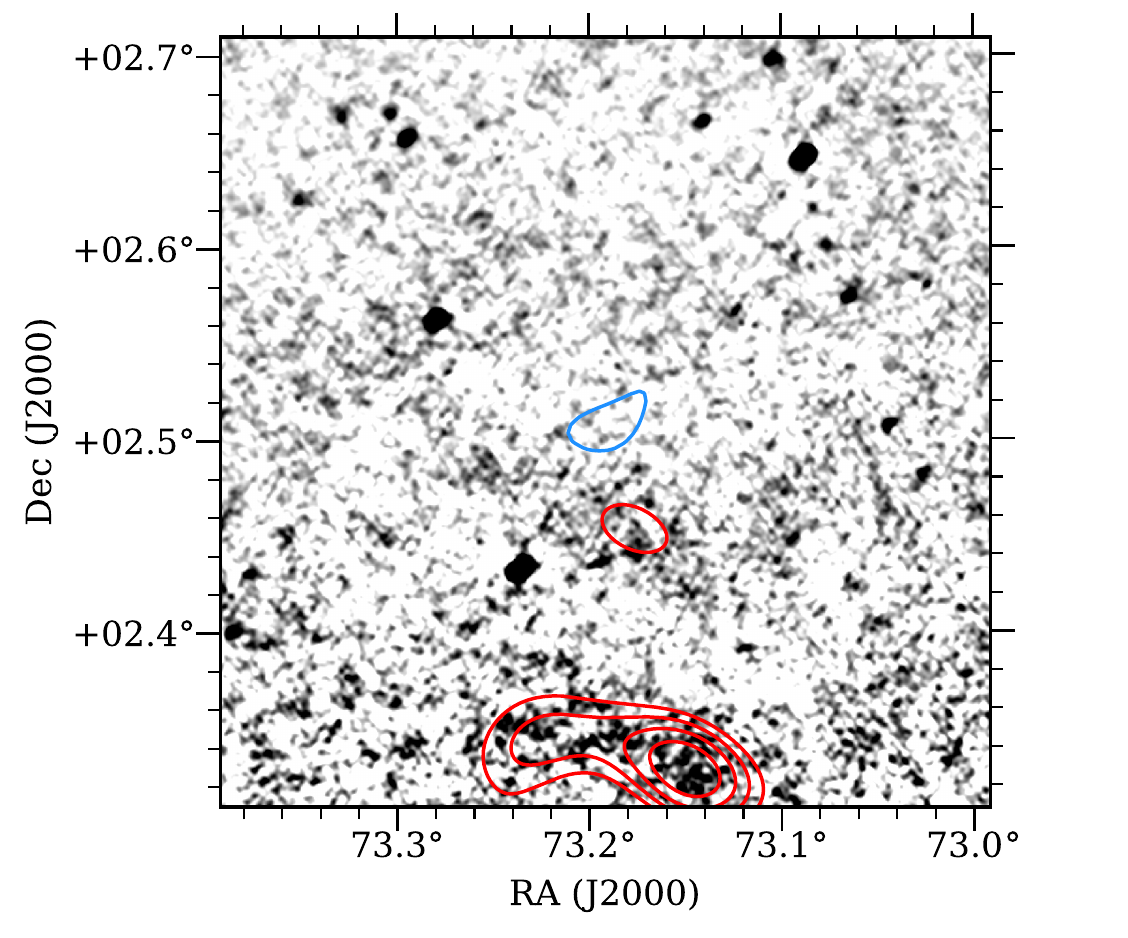}
    \end{subfigure}
    \begin{subfigure}{0.32\linewidth}
        \caption{J1}
        \includegraphics[width=1\linewidth,trim={0 -0.2cm 1cm 0.3cm} ,clip]{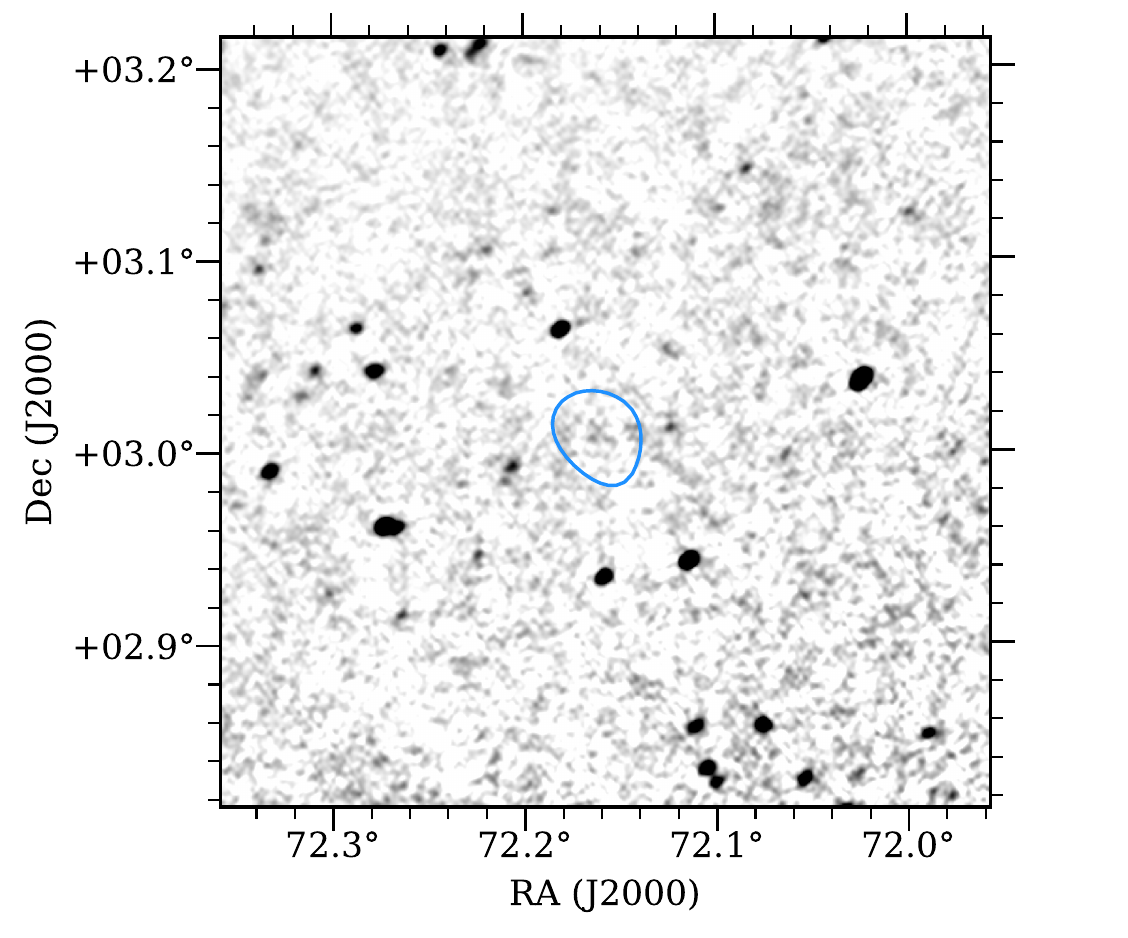}
    \end{subfigure}
\end{figure*}

\renewcommand{\thesubfigure}{\roman{subfigure}}
\begin{figure*}
    \centering
    \caption{MWA Phase II at 154 MHz with SRT+NVSS-diffuse contours (blue) and MWA-subtracted contours (red). Contours start at $3 \sigma$ of their respective map noises, and increase in increments of $1 \sigma$. Images are scaled linearly from -5 to \SI{50}{\milli \jansky \per \beam}.}
    \label{fig:mwa-2}
    
    \begin{subfigure}{0.45\linewidth}
        \caption{Region A}
        \includegraphics[width=1\linewidth,trim={0 -0.5cm 1cm 0.2cm},clip]{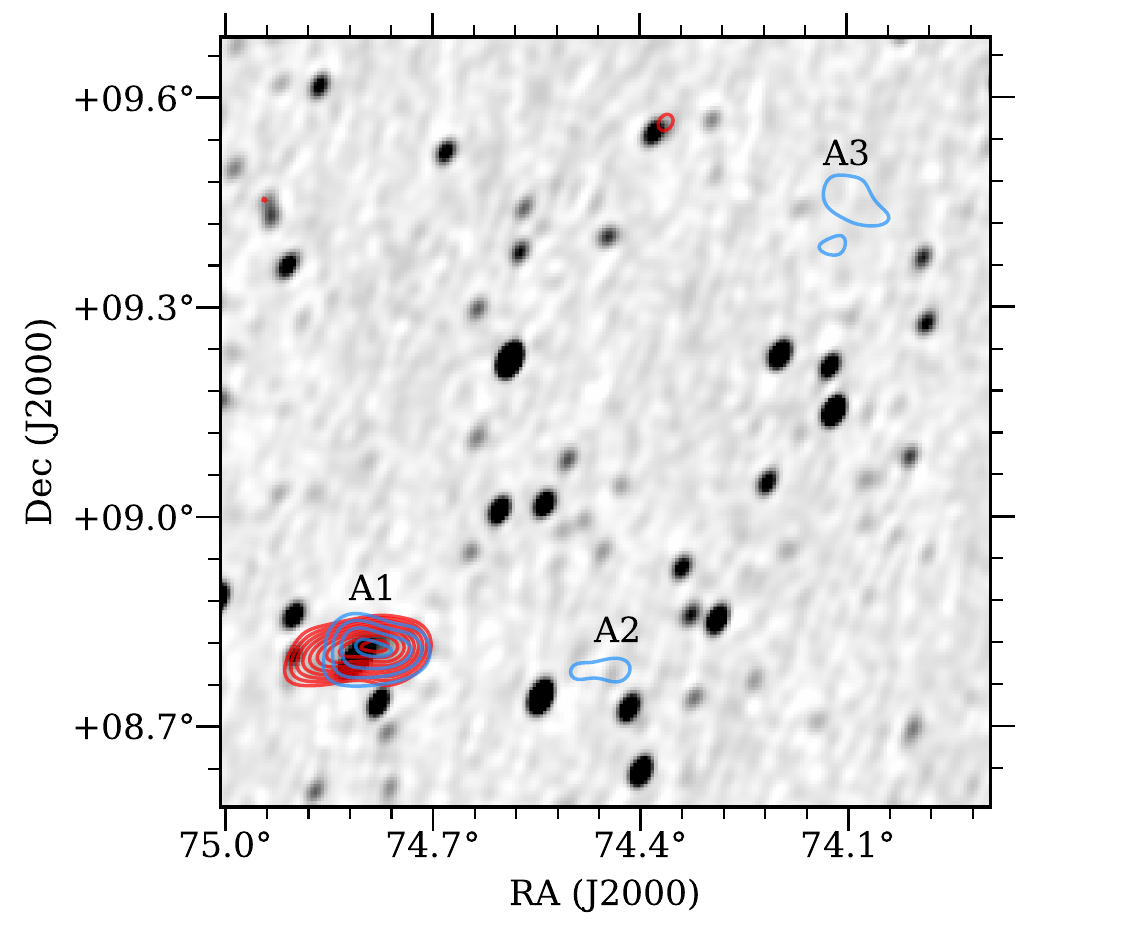}
    \end{subfigure}
    \begin{subfigure}{0.45\linewidth}
        \caption{Region B}
        \includegraphics[width=1\linewidth,trim={0 -0.5cm 1cm 0.2cm},clip]{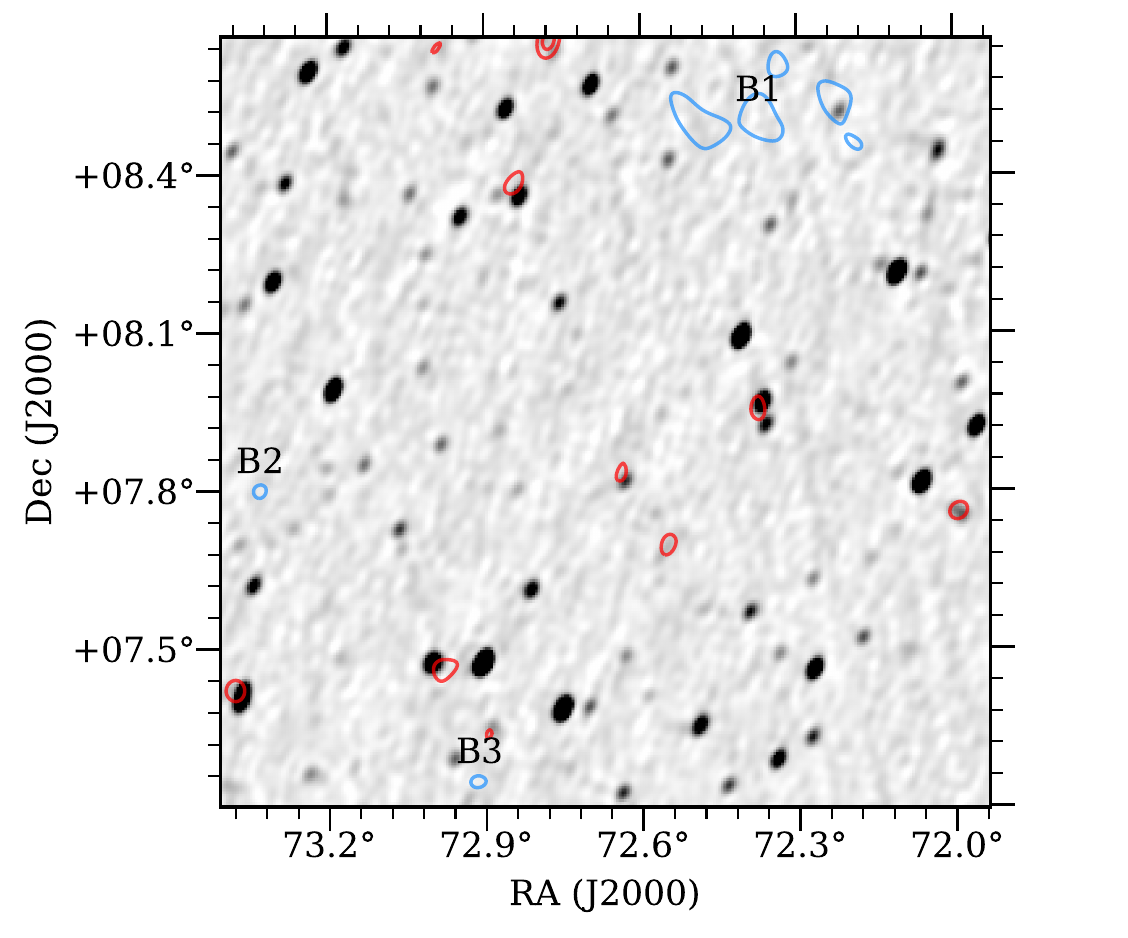}
    \end{subfigure}
    
    \begin{subfigure}{0.45\linewidth}
        \caption{Region C East}
        \includegraphics[width=1\linewidth,trim={0 -0.5cm 1cm 0.2cm},clip]{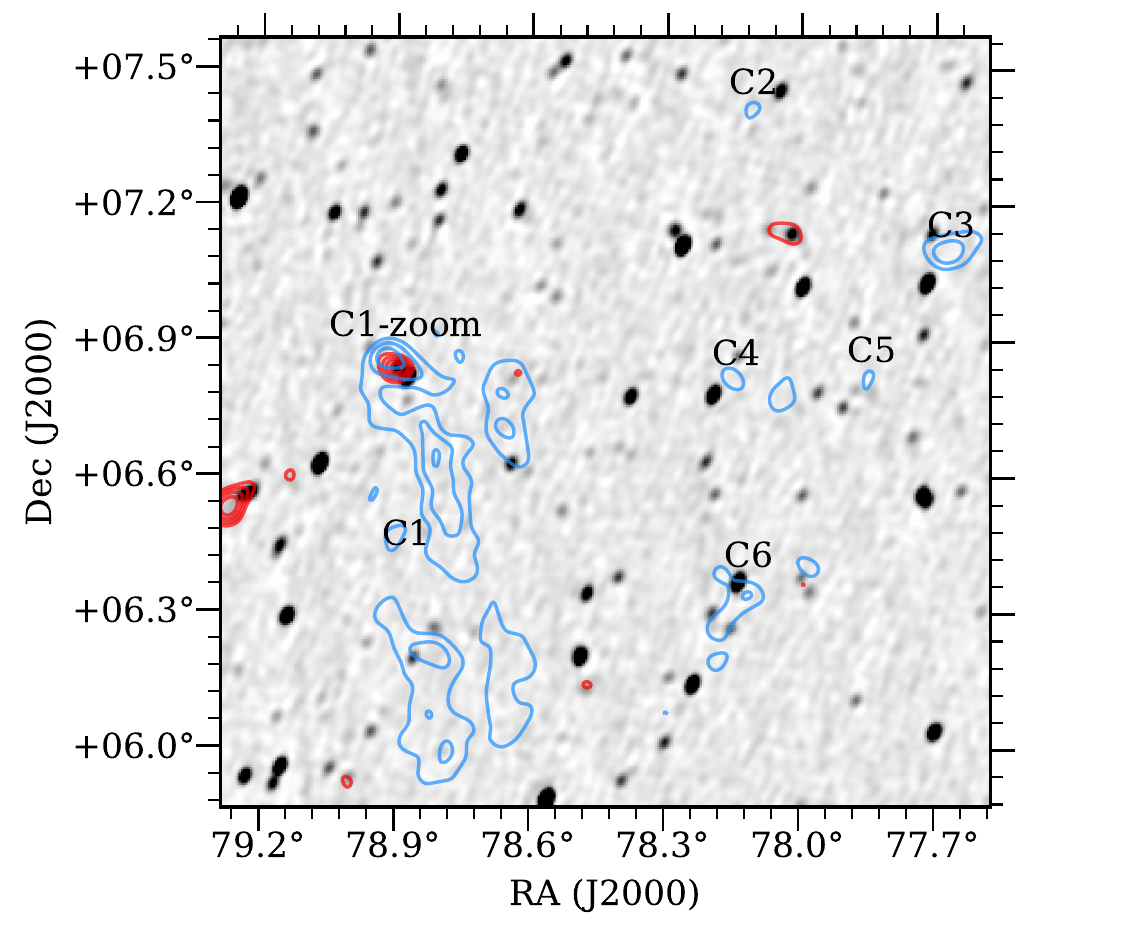}
    \end{subfigure}
    \begin{subfigure}{0.45\linewidth}
        \caption{Region C West \& D1}
        \includegraphics[width=1\linewidth,trim={0 -0.5cm 1cm 0.2cm},clip]{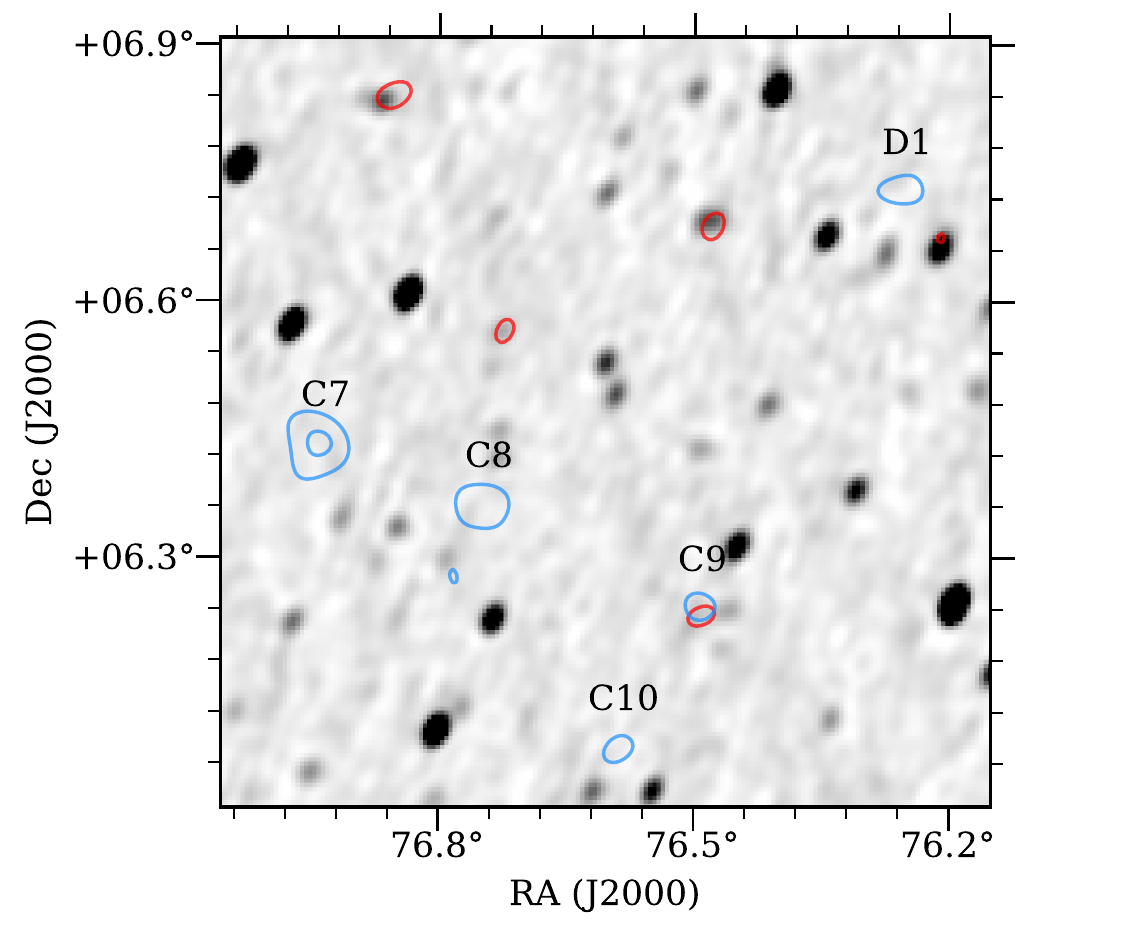}
    \end{subfigure}
    
    \begin{subfigure}{0.45\linewidth}
        \caption{Region D \& G1}
        \includegraphics[width=1\linewidth,trim={0 -0.5cm 1cm 0.2cm},clip]{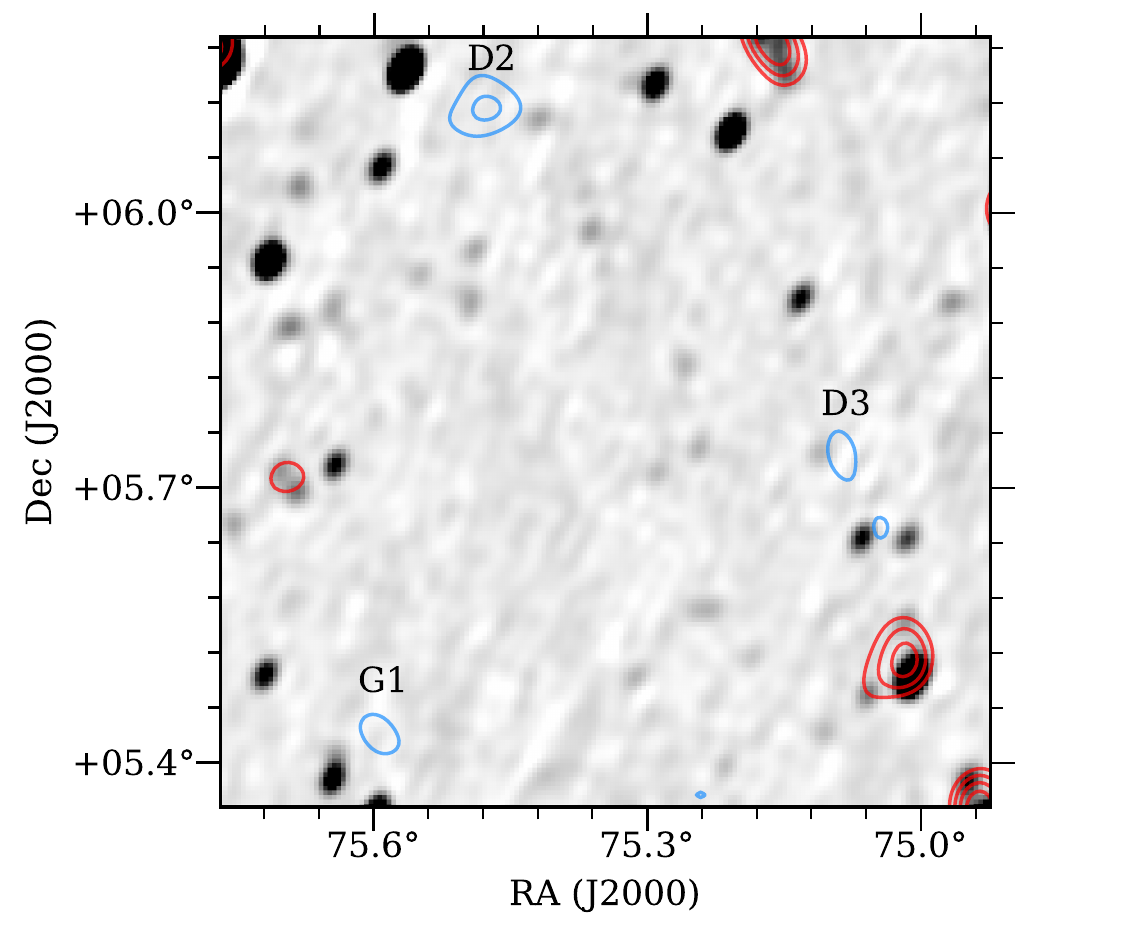}
    \end{subfigure}
    \begin{subfigure}{0.45\linewidth}
        \caption{Region E \& G2}
        \includegraphics[width=1\linewidth,trim={0 -0.5cm 1cm 0.2cm},clip]{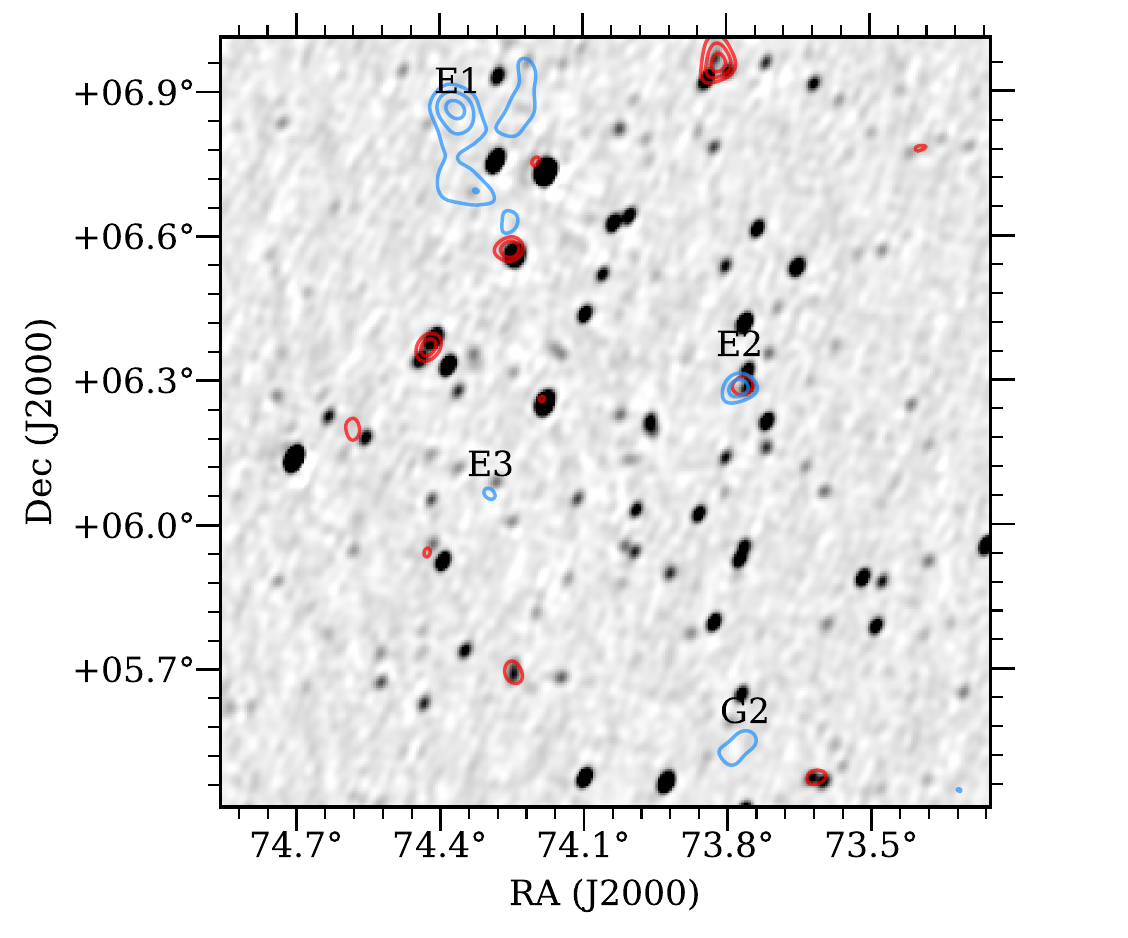}
    \end{subfigure}
\end{figure*}
\begin{figure*}\ContinuedFloat
    \centering
    \begin{subfigure}{0.45\linewidth}
        \caption{Region F}
        \includegraphics[width=1\linewidth,trim={0 -0.5cm 1cm 0.2cm},clip]{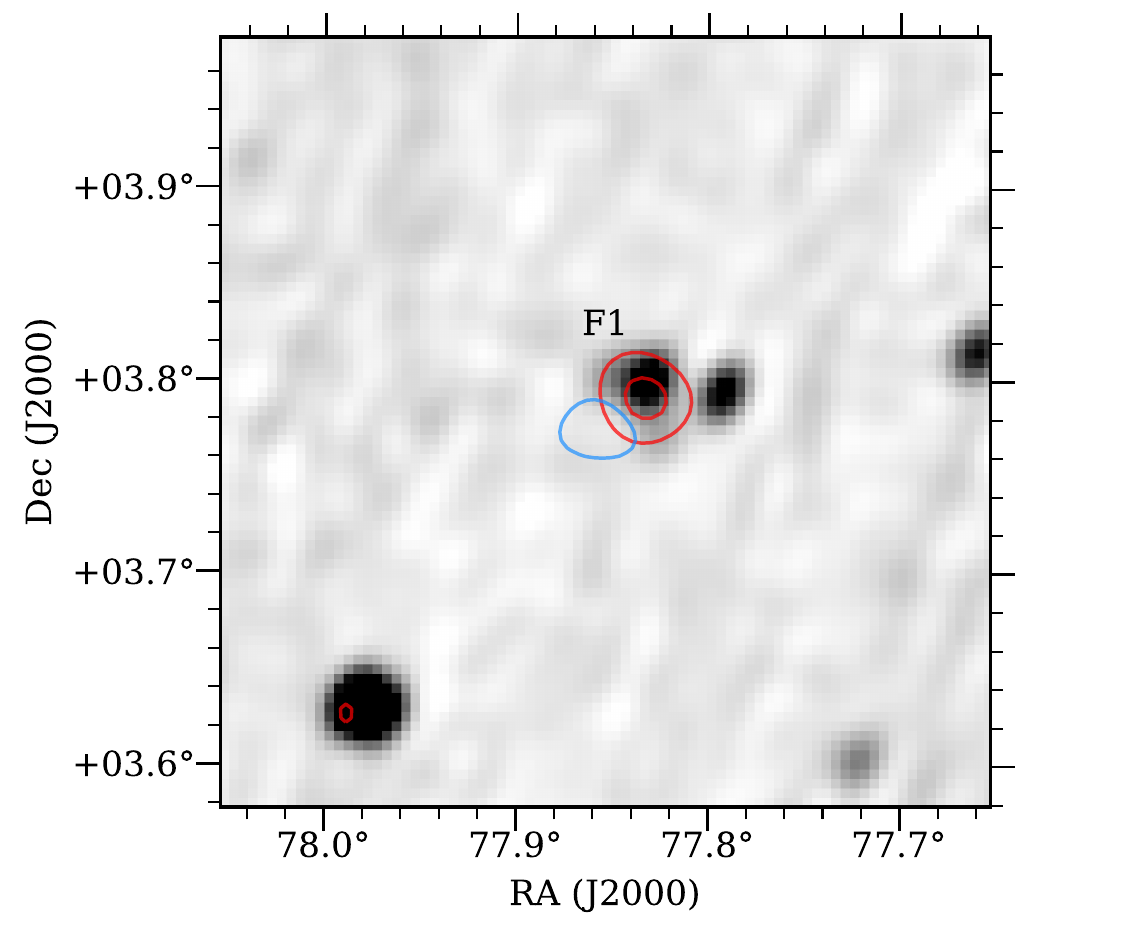}
    \end{subfigure}
    \begin{subfigure}{0.45\linewidth}
        \caption{Region G}
        \includegraphics[width=1\linewidth,trim={0 -0.5cm 1cm 0.2cm},clip]{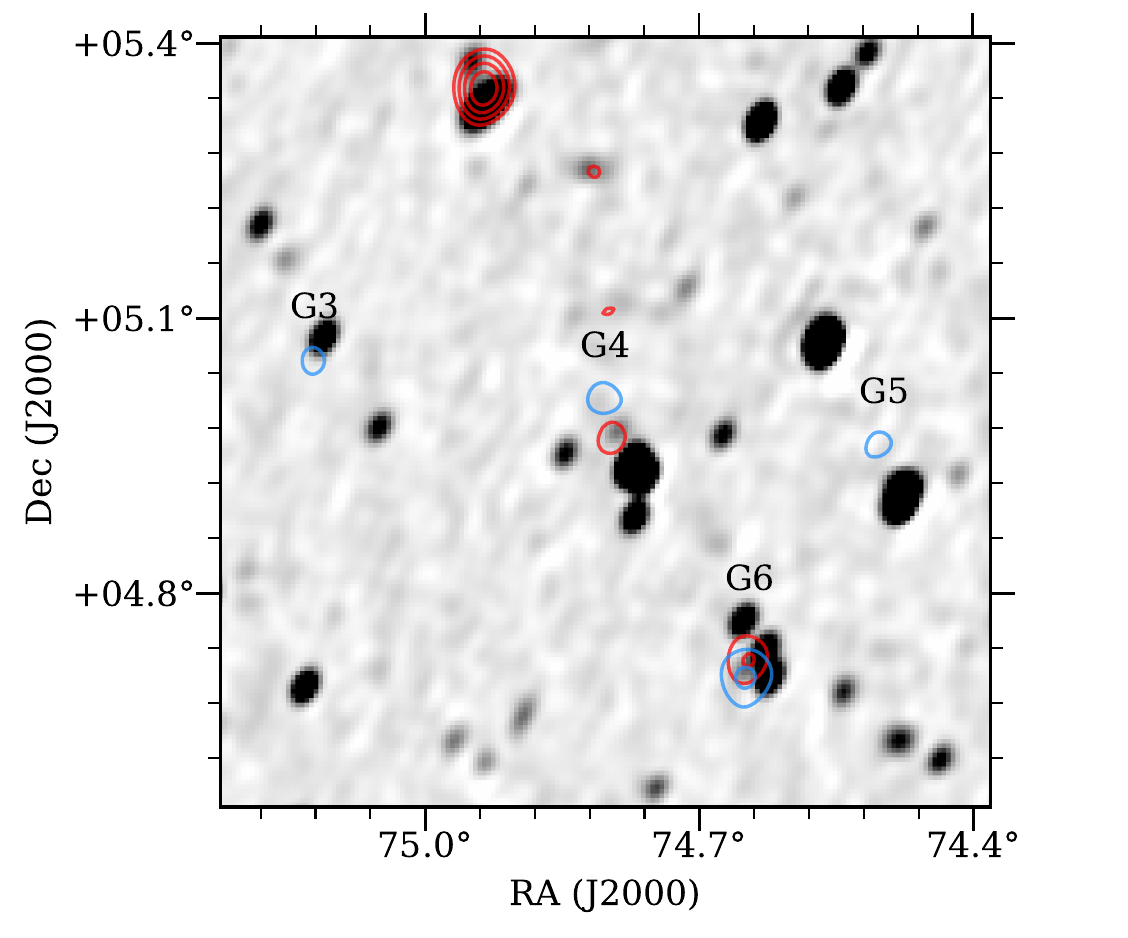}
    \end{subfigure}
    
    \begin{subfigure}{0.45\linewidth}
        \caption{Region H}
        \includegraphics[width=1\linewidth,trim={0 -0.5cm 1cm 0.2cm},clip]{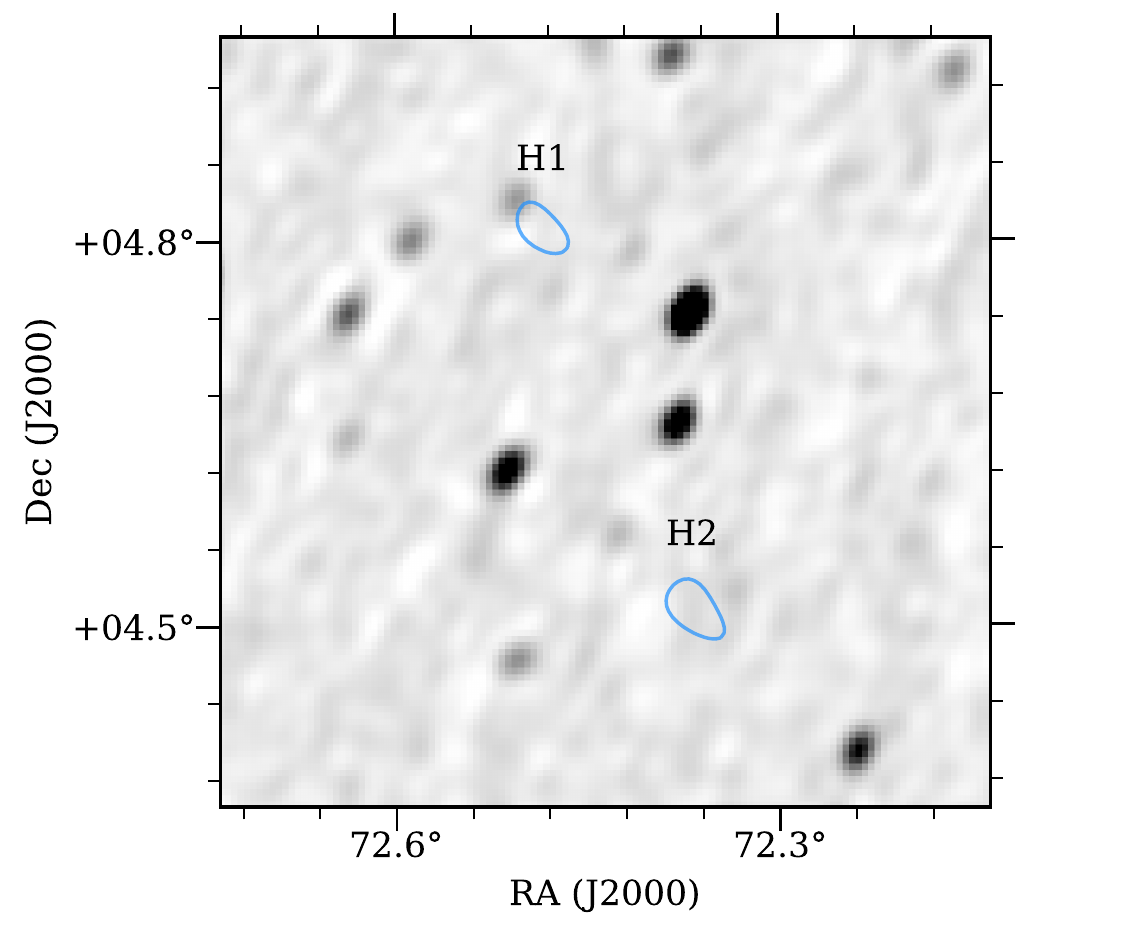}
    \end{subfigure}
    \begin{subfigure}{0.45\linewidth}
        \caption{Region I}
        \includegraphics[width=1\linewidth,trim={0 -0.5cm 1cm 0.2cm},clip]{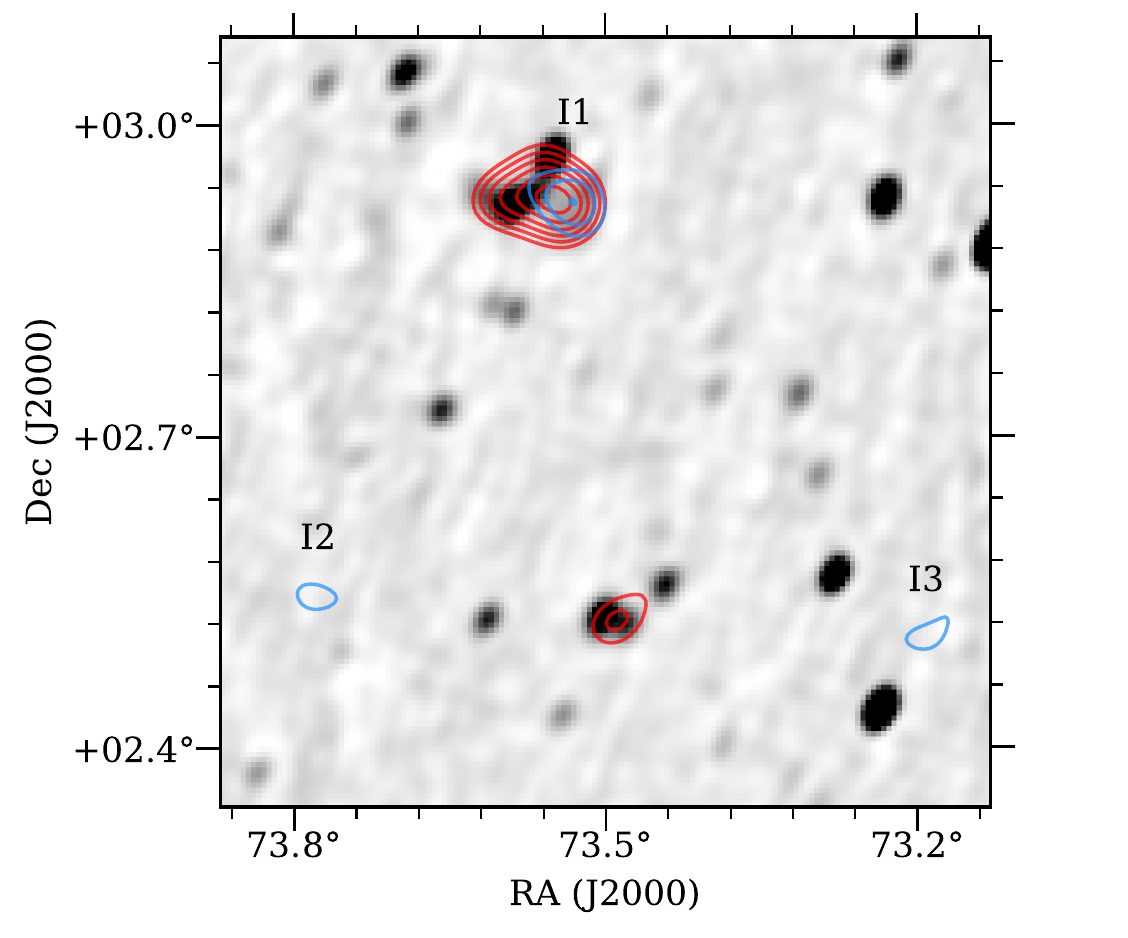}
    \end{subfigure}
    
    \begin{subfigure}{0.45\linewidth}
        \caption{Region J}
        \includegraphics[width=1\linewidth,trim={0 0 1cm 0.2cm},clip]{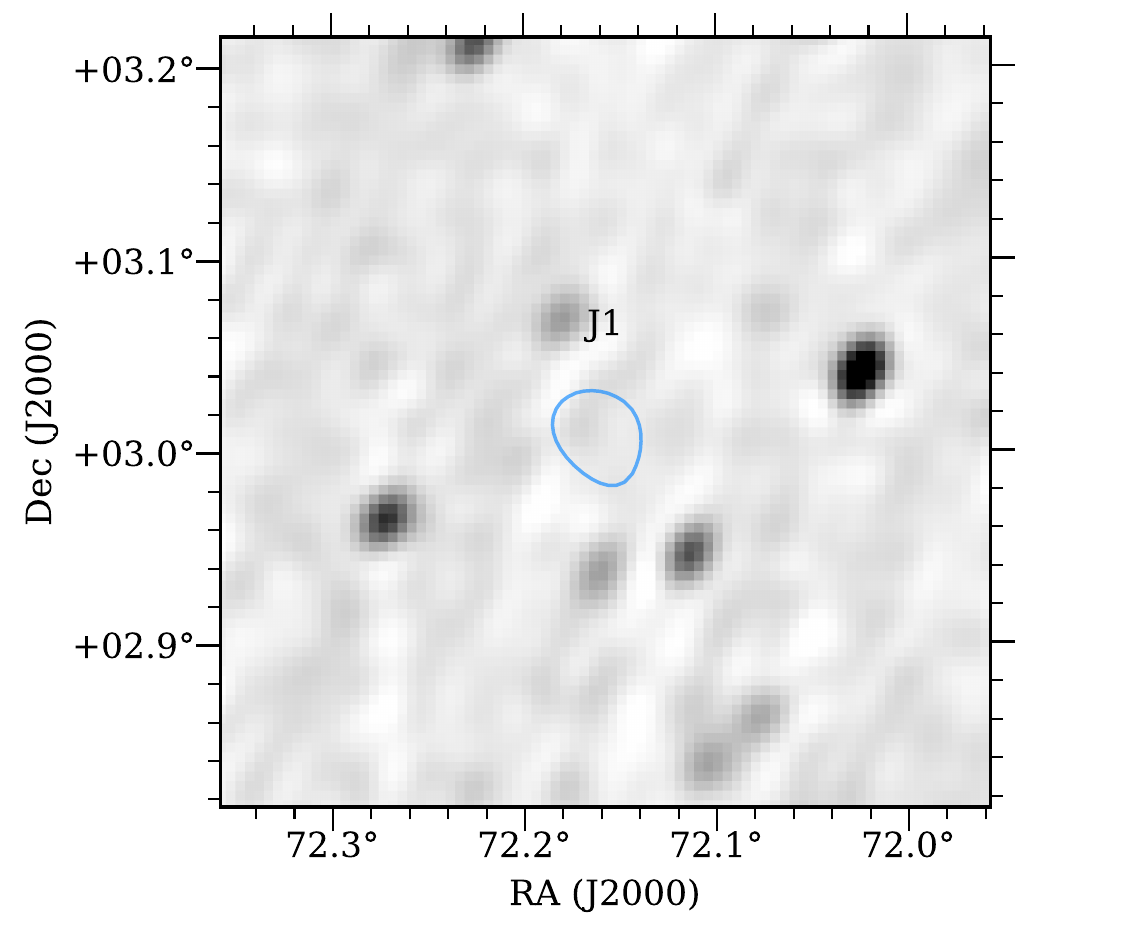}
    \end{subfigure}
\end{figure*}

\begin{figure*}
    \centering
    \includegraphics[width=1.0\linewidth,trim={0.2cm 0.2cm 0.2cm 0.2cm},clip]{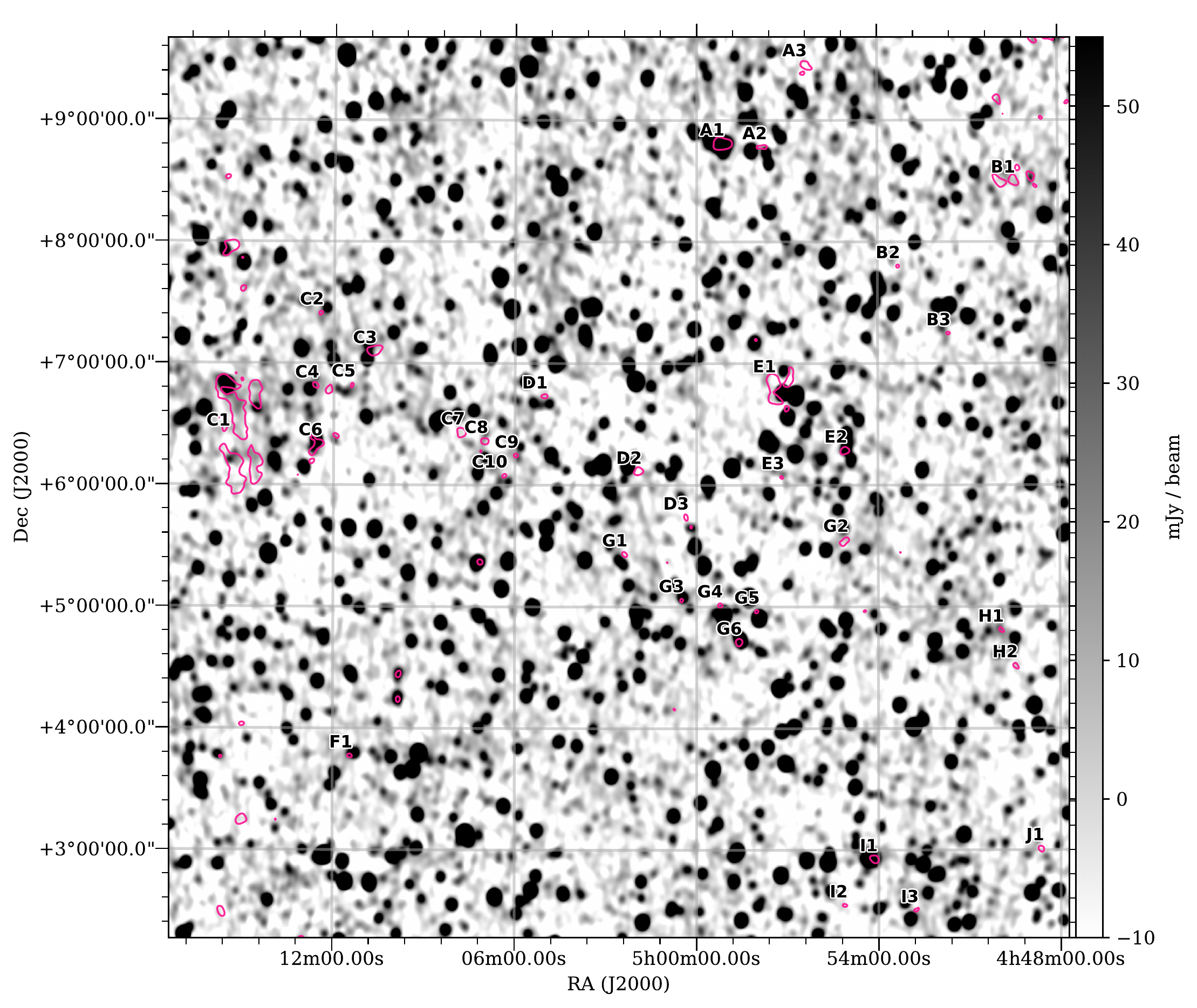}
    \caption{MWA Phase I at 154 MHz with $3\sigma$ SRT+NVSS-diffuse contours in magenta and named regions labelled above and to the left of the contour. The image is scaled linearly between -10 to \SI{55}{\milli \jansky \per \beam}  so that saturated black indicates a $5 \sigma$ detection.}
    \label{fig:mwa-1}
\end{figure*}

\end{document}